# Advancing Digital Precision Medicine for Chronic Fatigue Syndrome through Longitudinal Large-Scale Multi-Modal Biological Omics Modeling with Machine Learning and Artificial Intelligence

by
**Ruoyun Xiong**

A thesis submitted for the degree of
Doctor of Philosophy

December 2024

# Table of Contents

# Abstract

Myalgic Encephalomyelitis/Chronic Fatigue Syndrome (ME/CFS) is a complex debilitating disorder manifesting as severe fatigue and post-exertional malaise. The etiology of ME/CFS remains elusive. Here in chapter one, we present a deep metagenomic analysis of stool combined with plasma metabolomics and clinical phenotyping of two ME/CFS cohorts with short (<4y, n=75) or long-term disease (>10y, n=79) compared to healthy controls (n=79). First, we describe microbial and metabolomic dysbiosis in ME/CFS patients. Short-term patients showed significant microbial dysbiosis, while long-term patients had largely resolved microbial dysbiosis but had metabolic and clinical aberrations. Second, we identified phenotypic, microbial, and metabolic biomarkers specific to patient cohorts. These revealed potential functional mechanisms underlying disease onset and duration, including reduced microbial butyrate biosynthesis together with a reduction in plasma butyrate, bile acids, and benzoate. In addition to the insights derived, our data represent an important resource to facilitate mechanistic hypotheses of host-microbiome interactions in ME/CFS.

We then studied a more generalized question in chapter two: chronic diseases like ME/CFS and long COVID exhibit high heterogeneity with multifactorial etiology and progression, complicating diagnosis and treatment. To address this, we developed BioMapAI, an explainable Deep Learning framework using the richest longitudinal multi-'omics dataset for ME/CFS to date. This dataset includes gut metagenomics, plasma metabolome, immune profiling, blood labs, and clinical symptoms. By connecting multi-'omics to a symptom matrix, BioMapAI identified both disease- and symptom-specific biomarkers, reconstructed symptoms, and achieved state-of-the-art precision in disease classification. We also created the first connectivity map of these 'omics in both healthy and disease states and revealed how microbiome-immune-metabolome crosstalk shifted from healthy to ME/CFS. Thus, we proposed several innovative mechanistic hypotheses for ME/CFS: Disrupted microbial functions – SCFA (butyrate), BCAA (amino acid), tryptophan, benzoate - lost connection with plasma lipids and bile acids, and activated inflammatory and mucosal immune cells (MAIT, γδT cells) with INFγ and GzA secretion. These abnormal dynamics are linked to key disease symptoms, including gastrointestinal issues, fatigue, and sleep problems.

# Introduction

Myalgic Encephalomyelitis/Chronic Fatigue Syndrome (ME/CFS) is a complex, multi-system debilitating illness. The syndrome includes severe fatigue that is not alleviated by rest, post-exertional malaise (PEM), muscle and joint pain, headaches, sleep problems, hypersensitivity to sensory stimuli, and gastrointestinal symptoms.[1],[2],[3] In the US alone, ME/CFS affects up to 2.5 million people.[4] Our limited understanding of both the physiological changes associated with the syndrome and the underlying biological mechanisms are major impediments to identifying and developing both specific therapies and reliable biomarker-based diagnostics.[5],[6]

The human microbiome – the body's trillions of bacteria, fungi, and viruses – has recently emerged as an important potential contributor to, or biomarker of ME/CFS.[7] Patients have frequent gastrointestinal (GI) disturbances, and in lower-resolution studies based on 16S rRNA gene sequencing, altered gut microbiota.[7],[8],[9],[10],[11],[12],[13] Compared to healthy controls, the microbial dysbiosis observed in ME/CFS patients was characterized by decreased bacterial diversity, overrepresentation of putative pro-inflammatory species, and reductions in putative anti-inflammatory species.[11],[13] However, sample sizes for these studies were relatively small with limited taxonomic resolution. Much remains underexplored vis a vis the potential functional consequences of ME/CFS-associated microbial changes.

An important function of the intestinal microbiota is metabolism,[14],[15] feeding both microbial and host processes in its dynamic, symbiotic, and mutualistic relationship with the host. For example, the metabolic products of gut microbiota can feed into host pathways as energy sources[16] or function as immune regulators,[17],[18],[19] including short-chain fatty acids,[20] metabolites of bile acids,[21] or amino acid metabolites like tryptophan,[22] respectively. Thus, the microbiome can modulate host physiology via direct stimulatory effects [23], [24] or through secondary pathways coupled to metabolic processes [25]. Such host-microbe interactions can be identified both through mechanistic studies [26]. but also inferred by high resolution profiling [27], [28] and integrated analyses of the gut microbiome [29] – the microbial fingerprint, and the host metabolome – the collective chemical fingerprint.

Here, in chapter one, we performed a high-resolution characterization of the gut microbiome and the plasma metabolome in two ME/CFS cohorts compared to healthy controls. In an important departure from current studies, we profiled a 'short-term' cohort (diagnosed within the previous four years), vs. a 'long-term' cohort (patients who have been suffering from ME/CFS for more than ten years). Our goal was to gain an understanding of the baseline molecular mechanisms by which changes in the ME/CFS microbiome may be reflected in circulating metabolic markers, which could then potentiate further alterations in host physiology. In addition, we sought to identify potential molecular and biological markers of ME/CFS progression between the short- and long-term cohorts. Finally, we collected detailed clinical and lifestyle survey metadata for association analysis. Shotgun sequencing of the fecal microbiome of 149 ME/CFS patients (74 short- and 75 long-term) vs. 79 age- and sex-matched healthy controls, we found that short-term ME/CFS patients had more significant microbial and gastrointestinal abnormalities, and that long-term patients tended to establish a stable, but individualized gut microbiome. However, long-term patients had significantly more irreversible health problems and progressive metabolic aberrations. Finally, integrating detailed clinical and lifestyle survey metadata with these high resolution 'omics data allowed us to develop a highly accurate ME/CFS disease classifier. Taken together, our study presents a high resolution, multi-cohort and multi-'omics analysis, and provides a mechanistic hypotheses of host-microbiome interactions in ME/CFS.

We then noticed that chronic diseases, such as cancer[30], diabetes[31], rheumatoid arthritis (RA)[32], myalgic encephalomyelitis/chronic fatigue syndrome (ME/CFS)[33], and possibly long COVID[34],[35], the sequela of SARS-CoV-2 infection, can evolve over decades and exhibit diverse phenotypic and physiological manifestations across individuals. This heterogeneity is reflected in disease progression and treatment responses, complicating the establishment of standardized clinical protocols, and demanding personalized therapeutic strategies[36].

However, this heterogeneity has not been well studied, leaving substantial knowledge and technical gaps[37]. Current cohort studies often focus on identifying one or two key disease indicators, such as HbA1C levels for diabetes[38],[39] or survival rates for cancer[40], even with the advent of multi-'omics. This approach has difficulty accommodating the highly multifactorial etiology and progression of most chronic diseases, with different patients exhibiting varying symptoms and disease markers[41]. To address this challenge, methods must link a more complex matrix of disease-associated outcomes with a range of 'omics data types to enable precise targeting of biomarkers tailored to each patient's specific symptoms.

So in chapter two, we introduce BioMapAI, an explainable AI framework that we developed to integrate multi-'omics data to decode complex host symptomatology, specifically applied to ME/CFS. Affecting at least 10 million people globally, ME/CFS is a chronic, complex, multi-system illness characterized by impaired function and persistent fatigue, post-exertional malaise, multi-site pain, sleep disturbances, orthostatic intolerance, cognitive impairment, gastrointestinal issues, and other symptoms [42],[43],[44]. The pathogenesis of ME/CFS is not well understood, with triggers believed to include viral infections such as Epstein-Barr Virus (EBV)[45], enteroviruses[46] and SARS coronavirus[47]. As a chronic disease, ME/CFS can persist for years or even a lifetime, with each patient developing distinct illness patterns[42]. Therefore, a universal approach to clinical care and symptom management is insufficient, and a personalized approach is crucial for effectively addressing the complex nature of ME/CFS. Additionally, given similarities in causality and symptomatology to long COVID[48],[49], studying ME/CFS specifically can provide broader insights into post-viral syndromes, and more generally, our AI-driven approach can be applied to a range of diseases with complex symptomatology not readily explained by a single data type.

We generated a rich longitudinal, multi-'omics dataset of 153 ME/CFS patients and 96 age-gender-matched healthy controls, comprised of gut metagenomics, plasma metabolome, immune cell profiling, activation, and cytokines, together with blood labs, detailed clinical symptoms, and lifestyle survey data. We aimed to: 1) identify new disease biomarkers - not only for ME/CFS but also to specify biomarkers that could explain the complex symptomatology, and 2) define interactions between microbiome, immune system, and metabolome – rather than studying single data types in isolation, we created the first connectivity map of these 'omics. This map critically accounts for covariates such as age and gender, providing an important baseline in healthy individuals contrasted with aberrant connections identified in disease.

BioMapAI is a strategically designed Deep Neural Network (DNN) that connected the multi-'omics profiles to a matrix of clinical symptoms. Here, applying to ME/CFS, it identifies both disease- and symptom-specific biomarkers, accurately reconstructing key clinical symptoms, achieves state-of-the-art precision in disease classification, and generates several innovative mechanistic hypotheses for disease. By revealing microbiome-immune-metabolome crosstalk shifts from healthy to diseased states, we found depletion of microbial butyrate (SCFA) and amino acids (BCAA) in ME/CFS, linked with abnormal activation of inflammatory and mucosal immune cells – MAIT and γδT cells with INFγ and GzA. This altered dynamic correlated with clinical symptom scores, indicating deteriorated health perception and impaired social activity. Microbial metabolites, like tryptophan and benzoate, lost connections with plasma lipids in patients, in turn associated with fatigue, emotional and sleeping problems. This dataset is the richest multi-'omics dataset for ME/CFS, as well as for numerous other chronic diseases to date. It introduces a novel, generalizable, and explainable AI approach that captures the complexity of chronic disease and provides new hypotheses for host-microbiome interactions in both health and ME/CFS.

# Chapter 1: Multi-'omics of gut microbiome-host interactions in short- and long-term Myalgic Encephalomyelitis/Chronic Fatigue Syndrome (ME/CFS) patients

**Data characteristics.**

We enrolled 228 participants; 149 with ME/CFS (74 'short-term' and 75 'long-term') and 79 approximately age- and sex-matched healthy controls (Figure 1, Table S1-2). The short-term and long-term cohorts were designed to obtain a better understanding of the biological processes during the progression of ME/CFS. The cohort was 96.5% Caucasian, with an average age of 43 years and 67% female, characteristics consistent with epidemiological reports that women are 3-4X more susceptible to ME/CFS than men [50]. We collected detailed clinical metadata (Table S1-2), stool samples for shotgun metagenomics, and blood for targeted metabolomic analysis. Clinical metadata (n = 228) and blood samples (n = 184) were collected at time of enrollment, followed by self-collection of fecal samples (n = 224) within the following two weeks (n=180 complete datasets of metadata, blood, and stool). We established a workflow to integrate the 906 clinical features into major disease markers to optimize data dimensionality (Figure 1). Whole-genome shotgun metagenomic sequencing of the stool samples generated an average of 10,801,733 high-quality and classifiable reads per sample, which were then reconstructed to examine gut microbiome composition (species N=384, Table S3) and gene function (gene N=9652, Table S4). Plasma was fractionated from blood and sent for targeted LC-MS analysis, where 1278 metabolites were identified for host molecular 'omics profiling (Table S5). Finally, we analyzed each data type individually, then altogether to build multi-'omics models to describe and predict onset, stage of disorder, and associated microbial and metabolic features. This allowed us to target microbial pathways likely to affect host-microbiome interactions and alter the disease pathophysiology. For all datatypes, we performed two primary comparisons: 1) ME/CFS vs. healthy controls, to understand the broad differences inherent to the disease, and 2) short- vs. long-term ME/CFS vs. healthy controls, to understand disease progression.

**The host phenotype in ME/CFS**

To understand and interpret the host phenotype of ME/CFS compared to healthy controls, we collected comprehensive clinical metadata including detailed demographics and lifestyle information, an itemized dietary intake survey, medical history records, three general questionnaires regarding the physical and mental health of all participants, and five patient-specific surveys encompassing ME/CFS clinical symptoms and measurements (Table S1-2). We first excluded possible dietary biases and then established that the dietary habits were comparable among all groups (Figure S1, Table S1-2). We also found that the frequency of previous acute infections (Figure S1, Chi-square test, p < 0.001), which supports the potential association of infections with the onset of ME/CFS [51]. Our naïve Bayesian classification model (Figure S2A, area under the curve (a measure of classification accuracy), AUC = 0.85), which we used to identify clinical features that discriminate healthy controls vs. patients, showed that, besides some announced dysfunctions like orthostatic intolerance and fibromyalgia, most ME/CFS patients also suffered from additional complications, such as depression, headaches, constipation, and anxiety [52], [53] (Figure S2B). Altogether, these high-resolution data echo the known pathophysiologies of ME/CFS and established the clinical characteristics of our cohorts [54].

**ME/CFS patients have decreased gut microbial diversity and greater heterogeneity**

To begin to identify potential microbial mechanism(s) related to these symptoms, we first decoded the microbiota of ME/CFS patients. After classification of our shotgun metagenomic dataset to the species level (384 species passing quality cutoffs, see Methods, Table S3), we examined community-wide metrics to understand if broad dysbiosis was observed in patients compared to controls. Clustering using principal coordinate analysis (Bray-Curtis dissimilarity distance, which reflects the similarity of microbiome composition between each pairwise set of samples) showed 1) most of sample variation was explained by the onset of ME/CFS (Figure 2F, permutational analysis of variance (PERMANOVA), p = 0.002) and was not influenced by the age difference between the control and patient groups (Figure S3, Table S7), 2) patient samples had higher heterogeneity, as a population compared to controls (Figure 2G, p < 0.01, control N=79, ME/CFS N=149, heterogeneity was calculated over N=384 species,

Wilcoxon rank-sum test with Bonferroni correction). Interestingly, high heterogeneity was also observed in our recent study of frail older adults[55, 56], suggesting a non-uniform adjustment to the host's changing physiological conditions.

High microbial biodiversity has increasingly been associated with ecosystem health [57], with a less diverse (fewer members) and less even structure (i.e., more heavily weighted with fewer members) associated with decreased resilience and susceptibility to pathogenic colonization [58]. ME/CFS patients had fewer community members (Figure 2A, Chao 1 index; p <0.001, Control N=79, MECFS N=149, Chao 1 index was calculated over N=384 species, Wilcoxon rank-sum test with Bonferroni correction), and lower Evar evenness (Figure 2B, p <0.05, Wilcoxon rank-sum test with Bonferroni correction). To understand if there were specific species lacking in patients, we calculated a rarity and dominance index. A decrease of Chao 1 (Figure 2A) and rarity (Figure 2C, p <0.05, Wilcoxon rank-sum test with Bonferroni correction) indicated that ME/CFS patients had fewer low abundance members. The lower Gini index from those dominant species implied greater equality (Figure 2D, p <0.01, Wilcoxon rank-sum test with Bonferroni correction), with highly abundant commensal species. We also noted a modest change in the ratio of the overall relative abundance of Firmicutes to Bacteroidetes phyla (Figure 2E), a common metric for comparing broad compositional differences between microbiota, including IBD and other inflammatory diseases [59–61].

Finally, we predicted in-situ growth rate of individual microbes with our Growth Rate InDex (GRiD) algorithm, which leverages coverage differences over a microbe's genome to infer microbial growth rate, a proxy for metabolic activity in a community [62]. Interestingly, the species that grew slower in ME/CFS were primarily Firmicutes (Figure 2H), suggesting a further nuance to the already disproportionate Firmicutes:Bacteroides ratio. Taken together, the gut microbiome of ME/CFS, like in aging and other chronic inflammatory disorders, was characterized by modest but broad dysbiosis, including a less diverse and more uneven gut microbiome community with higher heterogeneity and altered Firmicutes:Bacteroidetes ratio, supported by a group of slower-replicating Firmicutes species.

**Microbial dysbiosis occurs in short-term ME/CFS and stabilizes in long-term disease**

Our analyses thus far investigated major differentiating features between ME/CFS patients and healthy controls. We wondered if microbial and metabolomic markers changed significantly during the progression of ME/CFS, or if early features could predict later severity. We analyzed our dataset comparing the short-term to the long-term group. First, we evaluated the effect of age as a confounder but found no significant differences (PERMANOVA with age as a variate, $p > 0.05$, Wilcoxon rank-sum test with Bonferroni correction, $p > 0.05$ ($\leq$50 years old (yo) vs. >50 yo subgroups, Figure S3, Table S7).

We then examined overall microbial composition differences between the patient cohorts as previously performed (Figure 3A). Interestingly, differences in the gut microbiome were more pronounced and variable during early stages of the disease compared to long-term and controls (Figure 3C-3F, $p < 0.05$, pairwise Wilcoxon rank-sum test with Bonferroni correction). As noted, heterogeneity was a feature we observed associated with aging, particularly frail aging . As we found no confounding effect of age, we conjectured that these differences were driven by ME/CFS disease duration. A significant shift in the Bacteroidetes (77.5%) to Firmicutes (19.4%) ratio was observed only in short-term but not long-term patients (Figure 3B, pairwise Wilcoxon rank-sum test with Bonferroni correction, $p < 0.05$ short-term; $p > 0.1$ long-term vs. controls), explaining the modest difference previously observed (Figure 2E).

To further resolve species-level differences between short-term and long-term patients, we performed a pairwise comparison for the mean relative abundance of every species in the top five most abundant phyla (Figure 3F). We discovered that species that had atypical relative abundance in short-term patients with respect to healthy controls (e.g., Bacteroidetes sp.) tended to stabilize in late-stage patients, i.e., become relatively closer to the relative abundance observed in controls.

Taken together, we found that microbial dysbiosis is most marked early in ME/CFS disease, characterized by a loss of diversity driven by low abundance bacteria with high heterogeneity. Over time, we conjecture that the microbiota then stabilizes and reverts to an ecosystem more characteristic of healthy controls, with the reacquisition of some low abundance species and normalization of diversity. These results highlight the importance of stratifying by disease duration to identify features relevant to disease progression.

**ME/CFS is distinguishable with microbiome and metabolomics features**

Currently, there are no approved laboratory diagnostics available for ME/CFS [6], likely due to the heterogeneity of the disease. We hypothesized that the combination of environmental and clinical factors may assist in a more comprehensive disease classification. We constructed multiple state-of-the-art classifiers for deep profiling. For metagenomic data, we used both microbial species as well as gene relative abundance (identified by KEGG gene profiling, see Methods, Table S4) for our classifier, as we hypothesized that both species relative abundance as well as gene-level differences could differ between cohorts. Thus, we constructed four models based on 1) species and 2) KEGG gene relative abundances, 3) normalized abundance of plasma metabolites or 4) a combination of all three (multi-'omics, see Methods), in addition to the clinical classifier described above. Irrespective of the model, results obtained using multi-'omics data (gradient boosting model, AUC=0.90) outperformed any individual dataset, followed by the metabolome (AUC = 0.82), KEGG gene profile (AUC = 0.73), and species relative abundance (AUC = 0.73, gradient boosting model, Figure 4B; LASSO logistic, SVM, and Random Forest, Figure S4). This improved performance of multi-'omics for differentiating ME/CFS patients from healthy controls underscores the complementarity of different 'omics in describing the molecular processes that can occur with shifts in the host physiological state.

We then examined the most discriminatory features for each of the individual 'omics models to identify potential biomarkers and to further biological interpretation (Figure 4A, S5). Low abundance microbes comprised the most discriminatory features, including microbes implicated in tryptophan, butyrate and propionic acid production that were largely depleted in ME/CFS (Wilcoxon rank-sum test with Bonferroni correction with over N=384 species, p < 0.01, Figure S4). Metabolites of both pathways are key immunomodulatory molecules that regulate metabolic and endocrine functions [22,63–66]. Microbial genes that discriminated ME/CFS and were decreased included genes involved in betaine production (*grdD*, *grdE*, *grdI*), an anti-inflammatory metabolite [67,68]. Besides betaine, sphingomyelin, serotonin, and cholesterol were highly discriminatory features, and have also been previously reported to be altered in ME/CFS patients [6], [69], [70], [71].

Finally, because of ME/CFS' inherent heterogeneity, it is possible that clinical or lifestyle factors could confound this analysis. We sought to determine whether biomarkers of interest were disease- or confounder- driven. We applied an unbiased general linear test (see Methods, Table S7) with an extensive array of metadata features. We found none of these biomarkers were significantly correlated with confounders. Additionally, from this general association study, we observed that broader microbiome features, such as alpha diversity, were not biased by metadata. However, a few metabolites (importantly, not those features selected in the ME/CFS classification model below) were influenced by gender, age, and IBS score (Figure S6). Despite this, we concluded our microbiome and metabolomic analyses are largely unconfounded by individual clinical and behavioral features.

**Severe phenotypic and metabolic abnormalities in long-term ME/CFS**

We then constructed individual and multi-'omics models to differentiate disease durations and controls. We followed this by over representation analysis (ORA) on metabolic pathways and Bayesian classification on phenotypic abnormalities to further pinpoint distinctive patterns in different stages of ME/CFS (see Methods). Overall classification accuracy was lower than for overall disease but still relatively accurate leveraging multi-'omics data (AUC=0.82, gradient boosting model, Figure S5). Low (<0.5%) relative abundance bacteria, including several putative butyrate producers (*Clostridium* sp.) [72], were identified as potential biomarkers. The metabolomics model also identified two cholesterol and several lipid metabolites as discriminatory (Figure 5A, Table S6), consistent with the phenotypic classifier which identified many metabolic-related abnormalities (Figure 5C) [73,74], [75], [76], [77], [78], [79], [80] to be more predictive in long-term ME/CFS. However, we note that some of these phenotypes also can increase with age, which is largely matched in our cohort but may contribute to these differences [81], [82]. In addition, we found that fibromyalgia [83] was a key feature distinguishing short- vs. long-term disease, as was a trend towards worsening sleep problems and more pronounced post-exertional malaise in the later stage of the disease (Figure 5B). Interestingly, we identified poor appetite [84] to be the most distinguishable phenotype in the short-term group, which is consistent with a trend of more GI disturbances in the early stage (Figure 5B, 3A).

ORA with the metabolomics profiles identified the most striking differences between short- and long-term patients and controls. Unlike trends observed in the microbiome data with the greatest dysbiosis observed in short-term patients, long-term patients had more metabolites differentiating them from healthy controls, especially in sphingolipids and diacylglycerol metabolites (Figure 5C), confirming previous metabolomic findings. Interestingly, most metabolic species either decreased across experimental groups (control>short-term>long-term, e.g., xanthine;), or increased (control<short-term<long-term; e.g. sphingomyelins, diacylglycerol, phosphatidylcholine, and ceramides, $p < 0.01$, pairwise Wilcoxon rank-sum test with Bonferroni correction with N=1278 metabolites, see Figure 5D), suggesting that metabolic irregularities associated with ME/CFS may gradually worsen over time (Figure 5D). Taken together, we postulate that long-term ME/CFS patients have developed a unique but stable pathophysiology characterized by more severe clinical symptoms and an array of altered host metabolic reactions.

**Gut and plasma butyrate is reduced in early-stage disease and is associated with host abnormal physiology**

We noted in our metagenomics classification model a particular refrain of a depletion of butyrate-synthesizing microbes, including *Roseburia* and *F. prausnitzii* in ME/CFS. Butyrate is a major energy source for colonic epithelial cells and one of the main intestinal anti-inflammatory metabolites [85], [63]. We thus performed a focused metagenomic and metabolomic analysis of the butyrate pathway to better understand its potential role in the crosstalk between the gut microbiota and host physiology in ME/CFS.

Strikingly, we found a depletion in plasma isobutyrate in the short-term group (Figure 6A, p = 0.03, Wilcoxon rank-sum test with Bonferroni correction with over N=1278 metabolites) [86]. Thus, we sought to link plasma metabolite abundance to gene-coding potential of the microbiome. We performed differential abundance analysis of KEGG gene matrix (p value computed by Wilcoxon rank-sum test with Bonferroni correction over N=9652 genes) and found that butanoate synthesis (KEGG map00650) differentiated between patients and controls (Figure 6C; fold change and p value of each gene in butyrate pathway is shown in Figure 6D, also in Table S4). Finally, we inferred gut butyrate abundance from metagenomic data using a Markov Chain Monte Carlo (MCMC) metabolite prediction algorithm (see Methods), in the absence of matched gut metabolomic data. Inferred concentrations of isobutyrate were significantly decreased in ME/CFS (Wilcoxon rank-sum test with Bonferroni correction, p = 0.05), especially in the short-term group (Figure 6B, pairwise Wilcoxon rank-sum test with Bonferroni correction, p < 0.01).

Finally, we performed a Spearman correlation analysis with the relative abundance of 1) butyrate producing microbes or 2) KEGG enzymes with plasma metabolites to identify the degree to which this pathway may influence circulating metabolite levels (Figure 6D, Table S4, p value computed by Spearman correlation and correlated by Holm's Method with N=18 species + N=113 butyrate genes ~ N=1278 metabolites, see Methods). From the thousands of plasma metabolites tested, we identified 24 positive correlations, including moieties from propionates, succinates, tryptophans, and hippurates, consistent with results of our differential abundance analysis, as well as 12 negative correlations, including sulfate and ursodeoxycholate moieties. The further suggests that changes in the ability of the gut microbiome to metabolize or synthesize short-chain-fatty acids (SFCA) is reflected in dysbiosis of these and related metabolites in plasma.

**Microbiome and metabolomic biomarkers are shared across additional ME/CFS cohorts**

To explore the generalizability of our findings, we compared our findings with those identified in 1) an additional timepoint for our current cohort: we collected metagenomics and metabolomics data again ~one year after their initial recruitment (n=107 patients, n=59 controls); 2) two independent cohorts: the microbiome cohort from Guo et al., 2023[87] with 197 long-term participants (n=106 patients, n=91 controls) and the metabolome cohort from Germain et al., 2020[69] with 56 female individuals (n=26 patients, n=26 controls). While there are notable biological differences in study design (e.g., long-term patients only vs. our short- and long-term cohort, gender), we found that the main conclusions and the identified biomarkers in ME/CFS were largely shared across all cohorts, supporting the robustness of our models and the validity of our findings.

First, the loss of gut microbial diversity in ME/CFS was observed in both the external microbiome cohort and our second timepoint. Guo et al. also found decreased evenness in the disease (Figure 2B). Most microbiome features that we at timepoint 1 were similar one year later (Figure 7A-B). Similarly, external cohorts and our second timepoint supported the depletion of butyrate producers and key butyrate-producing genes, indicating that the microbial butyrate biosynthesis capability was reduced in ME/CFS. Guo's cohort measured fecal SCFAs (acetate, propionate, butyrate) by GC-MS, and found significant decreases of acetate and butyrate in ME/CFS.(Figure 7G). We then validated our classifiers with plasma metabolomics data from our second timepoint (Figure 7C) and Germain's cohort (Figure 7D). The performance of the models was strong (accuracy > 70%) and numerous shared biomarkers among cohorts. Out of ten metabolic biomarkers identified from timepoint 1, six were shared among cohorts (Figure 7E).

Taken together, we found that our microbial and metabolic biomarkers were robust to cohort, supporting the reproducibility of our conclusions in ME/CFS.

**Conclusion**

Here, we performed a large-scale multi-'omics investigation integrating detailed clinical and lifestyle data, gut metagenomics, and plasma metabolomics in short- and long-term ME/CFS patients compared to healthy controls. Several studies have reported a disrupted gut microbiome in ME/CFS [11], [13] as well as changes in blood cytokine and metabolite levels. However, our cohort design differentiating short- and long-term patients was important to identify microbial and metabolic features that may contribute to disease progression.

Notably, the most significant microbial dysbiosis occurred in short-term ME/CFS. Compositional differences in short-term ME/CFS were consistent with previous studies using lower resolution sequencing [88] and an independent microbiome study from Guo et al. [87] co-published in this issue, identifying a broad reduction in microbial diversity [89], alterations in the ratio of Bacteroides:Firmicutes microbiota [10], and increased heterogeneity of low abundance organisms. This latter feature has recently been associated with the microbiome of frail older adults and supports its general association with reduced health outcomes [55]. There are several potential explanations for this early-stage dysbiosis. First, short-term patients suffer more GI disturbances, and gut microbiome changes may reflect these environmental changes. Second, it is possible that patients may try a range of interventions that impact their gut microbiome, which is dynamic and influenced by numerous intrinsic and extrinsic factors, including age and diet [90]. Diet can not only dramatically shift gut microbial community composition, but it can alter the metabolic potential of the microbes and production of immunomodulatory metabolites. We also analyzed the detailed dietary metadata, showing that most dietary habits are comparable among our cohorts, except infrequent sugar intake in our short-term patients, which might also contribute to the microbial differences observed in early stages of disease (Table S1, Figure S1).

The return of the gut microbiome of long-term patients to a configuration more similar to healthy controls (with notable differences nonetheless, in low abundance species and in heterogeneity) as well as the reduced occurrence of gastrointestinal illness in this cohort, suggests a return to a relative homeostasis. However, we conjecture that microbial dysbiosis seen in short-term patients may have cumulative and long-term effects, where damage may be caused by an initial trigger, resulting in cascading events. Long-term patients, despite their relatively more 'control-like' gut microbiome, have more severe clinical symptoms and metabolic dysbiosis. Thus, we hypothesize that ME/CFS progression may begin with loss of beneficial microbes, particularly SCFA producers, resulting in more pervasive gastrointestinal phenotypes that is later reflected in plasma metabolite levels. Individual-specific changes then lead to the irreversible metabolic and phenotypic changes and unrecoverable ME/CFS. Other possibilities for this 'normalization' observed in long-term ME/CFS include survival bias, as the cohort design for this study is cross-sectional, not truly longitudinal, and it is possible that the >10y cohort is comprised of individuals who have not seen significant improvement or decline in their symptoms, not that disease progression normalizes by default.

The abnormalities in the short-term cohort could result in potential increases in aberrant translocation of microbial metabolites that could affect host immune and metabolic processes. For example, one of the changes we noted in short-term patients was a reduction in potential immunomodulatory organisms (butyrate and tryptophan producers, e.g., *F. prausnitzii*), which we speculate could lead to long-term metabolic dysbiosis. The reduced prevalence of the butanoate synthesis pathway and the reduced relative abundance of butyrate-producing bacteria among all patients, but especially in the short-term group, suggested a loss of butyrate in the intestinal environment. This was consistent with the decrease of isobutyrate measured in the blood. We additionally note that some of these microbial signatures, such as these changes in SCFA producers, have been identified in other disease cohorts. For example, a particularly interesting study examining the gut microbiome and metabolome of cardiovascular disease [91] suggested that major alterations in both feature sets might begin significantly before clinical onset of disease, including a depletion of butyrate producers. One possibility is that the microbial and metabolic features identified here are less specific to ME/CFS but more generally biomarkers of future disease risk.

Butyrate, tryptophan, and other microbial metabolites have been linked to mucosal immune regulation, and our team previously showed a striking immune dysbiosis in different blood immune markers, including changes in the functional capacity of mucosal associated invariant T (MAIT) cells and Th17 cells, and a decrease in the frequency of CD8+ T cells and natural killer cells in long-term ME/CFS patients [92]. This is an exciting association because each of these cell types have been linked to bacterial or fungal infections, respond to microbial metabolites, and have been linked to the pathogenesis of autoimmune or chronic inflammatory diseases. Thus, it is possible that the microbiome primes or sustains an aberrant immune response following disease onset. This is supported by an observed shift from a predominantly Th1 to Th2 immune response in ME/CFS [93].

There is currently no standard diagnostic test for ME/CFS because of many phenotypes of ME/CFS are shared with other disorders [5,6], such as fibromyalgia. Here, our integration of multiple 'omics data significantly increased classification accuracy and identified microbial and metabolic features that could pinpoint potential hypotheses for further investigation and therapeutic strategies. Longitudinal sampling of short-term patients, particularly as they progress to long-term disease, would help to untangle directionality of microbial dysbiosis and potential effects on the blood metabolome. For example, recent studies performing large-scale associations with the gut microbiome and blood metabolome have identified that a significant fraction (upwards of 15%) of blood metabolites can be predicted by gut microbiome composition [94]. However, understanding of the temporal nature of this association is limited. Here, our 'omics workflows could be one of the guiding frameworks to intergrading microbiome, metabolome, and host phenotypes, and thus, bring a more understanding to host-microbiome interactions.

Finally, we believe that recent potential associations between the chronic immune dysfunctions in ME/CFS patients and 'long COVID' increase the relevance of the results reported here. 'Long COVID' refers to phenotypes suffered by numerous patients infected by SARS-CoV-2 (COVID-19) that have 'recovered', but did not return to full health. Notably, it manifests as numerous phenotypical abnormalities shared with ME/CFS, including lingering chronic fatigue and myalgias. In ME/CFS, key symptoms might also be triggered by acute infections including SARS coronavirus, MERS [95], the Epstein-Barr Virus (EBV) [96], or other agents, and such infections were reported to be more frequent in the medical histories of our patient cohort, even preceding the onset of ME/CFS symptoms [97], [98] Understanding the biological mechanisms underlying ME/CFS may now have further urgency and generalizability in the worldwide COVID-19 pandemic. Taken together, we have established a framework to study host-microbiome interactions leveraging 'omics to identify host and microbial metabolites and functions implicit in ME/CFS, presenting a rich clinical and 'omics dataset to further mechanistic hypotheses to better understand this debilitating disease.

This study was designed to recruit age-, sex- and diet- matched healthy cohort and patient cohort. However, we note that our long-term cohort is slightly but significantly older than the short-term cohort (though age-matched in our healthy controls), which could contribute to some of the phenotypes, particularly clinical abnormalities, observed. Additionally, we do not have matched cohorts with other disorders with overlapping phenotypes, such as IBS, neuroinflammatory disorders, and others. This is particularly relevant because ME/CFS is often misdiagnosed. While our results support that we can very accurately discriminate healthy individuals from ME/CFS patients, future studies would include cohorts with matched individuals with differing diagnoses to pinpoint microbial/metabolomic changes that are specific to ME/CFS rather than general disease risk. In addition, a rigorous control of exposures and co-morbidities could more effectively narrow down on microbial and metabolomic features that are ME/CFS-specific vs. a generalizable risk or protective factors for inflammatory disease. In advance of these data, the use of these features as potential diagnostic biomarkers may be premature.

**Figure**

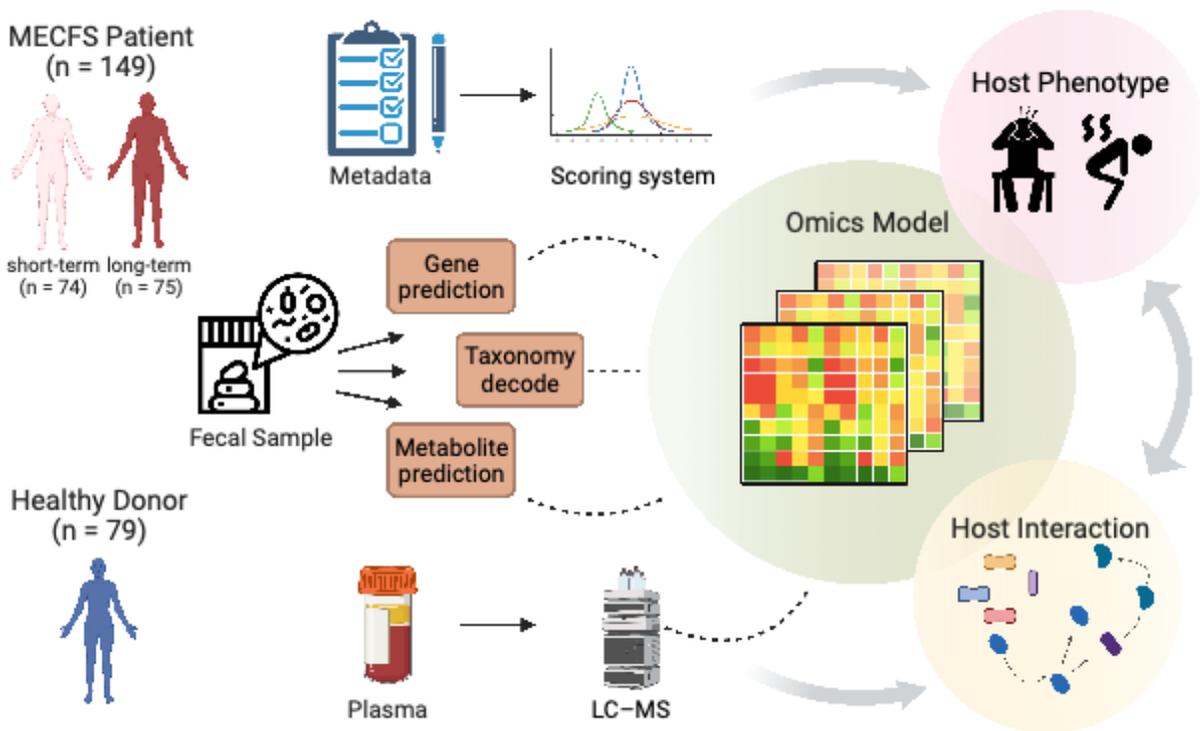

**Figure 1. Summary of study design and analytical pipeline.** We collected detailed clinical metadata, fecal samples, and blood samples for 228 individuals in three cohorts: healthy controls, patients with short-term (<4y) or long-term (>10y) ME/CFS. A comprehensive 'omics workflow was constructed with the multi-data types (phenotypic, metagenomics, and metabolomics, respectively) and multi-computational models to understand potential host-microbe interactions. LC-MS, Liquid chromatography-mass spectrometry.

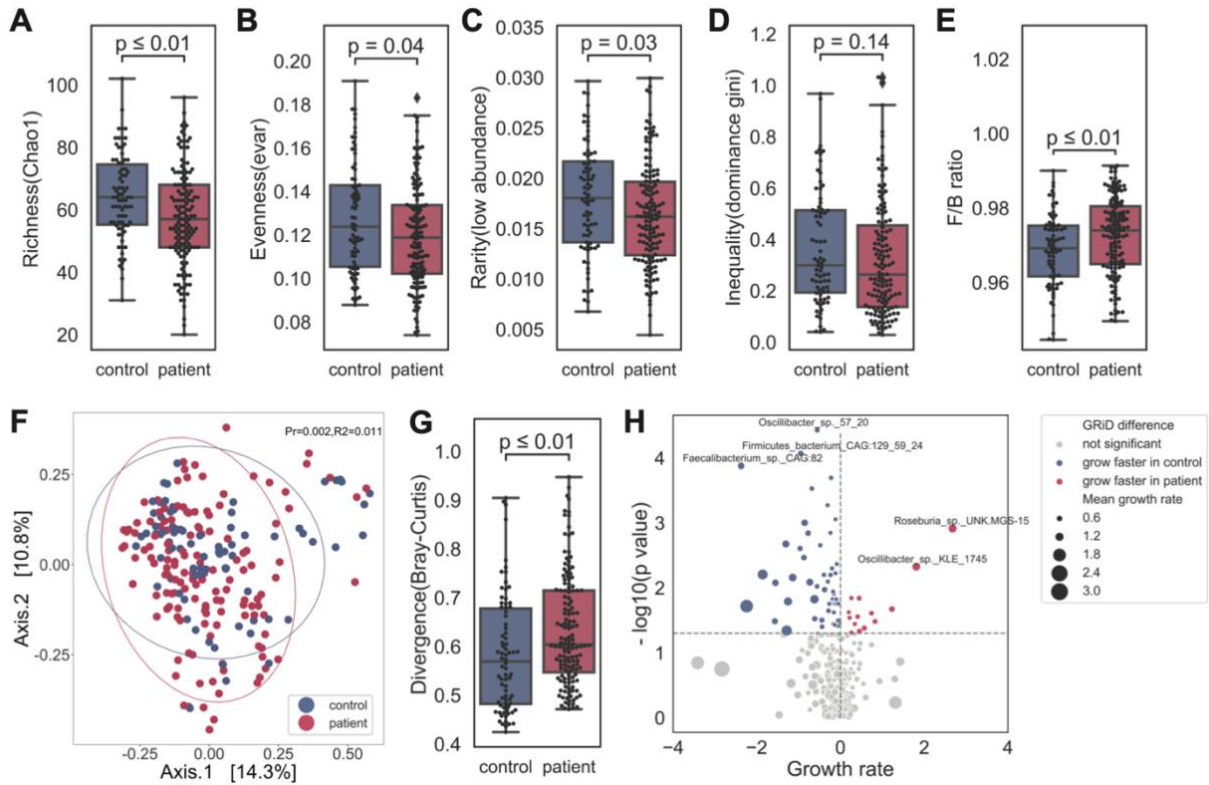

**Figure 2. Microbial dysbiosis in ME/CFS is characterized by decreased diversity and greater heterogeneity.** Comparing controls vs. ME/CFS patients (irrespective of disease stage), community structure differed in ME/CFS with A) decreased richness (Chao 1 index, which measures the number of observed species); B) decreased evenness (lower values of Smith and Wilson's Evar index); C) decreased rarity (smaller proportion of the least abundant species (<0.2% relative abundance); D) decreased inequality (smaller Gini index of the dominant species (>0.2% relative abundance); E) decreased Firmicutes/Bacteroidetes ratio. p-values were computed by Wilcoxon rank-sum test. F) First and second principal coordinates of dimensionality reduction for Bray-Curtis dissimilarity distances, which measures pairwise similarity of two given samples). Values in brackets indicate the amount of total variability explained by each principal coordinates. p-value and $R^2$ were calculated by permutational multivariate analysis of variance (PERMANOVA) test with patient/control as a variable. G) Increased heterogeneity observed in ME/CFS as measured by divergence, or Bray-Curtis dissimilarity. p-value was computed by Wilcoxon rank-sum test. H) Volcano plot showing differences in predicted growth rate in select species in ME/CFS. Each dot indicates a microbe, sized by the value of its inferred growth rate. The x-axis shows the absolute difference (mean growth rate in patient – mean growth rate in control) and the y-axis is the log10(p-value, Wilcoxon rank-sum test). Species that were predicted to grow faster in patients were colored red and slower in blue. p-value > 0.05 was considered not significant (gray). See also Table S3.

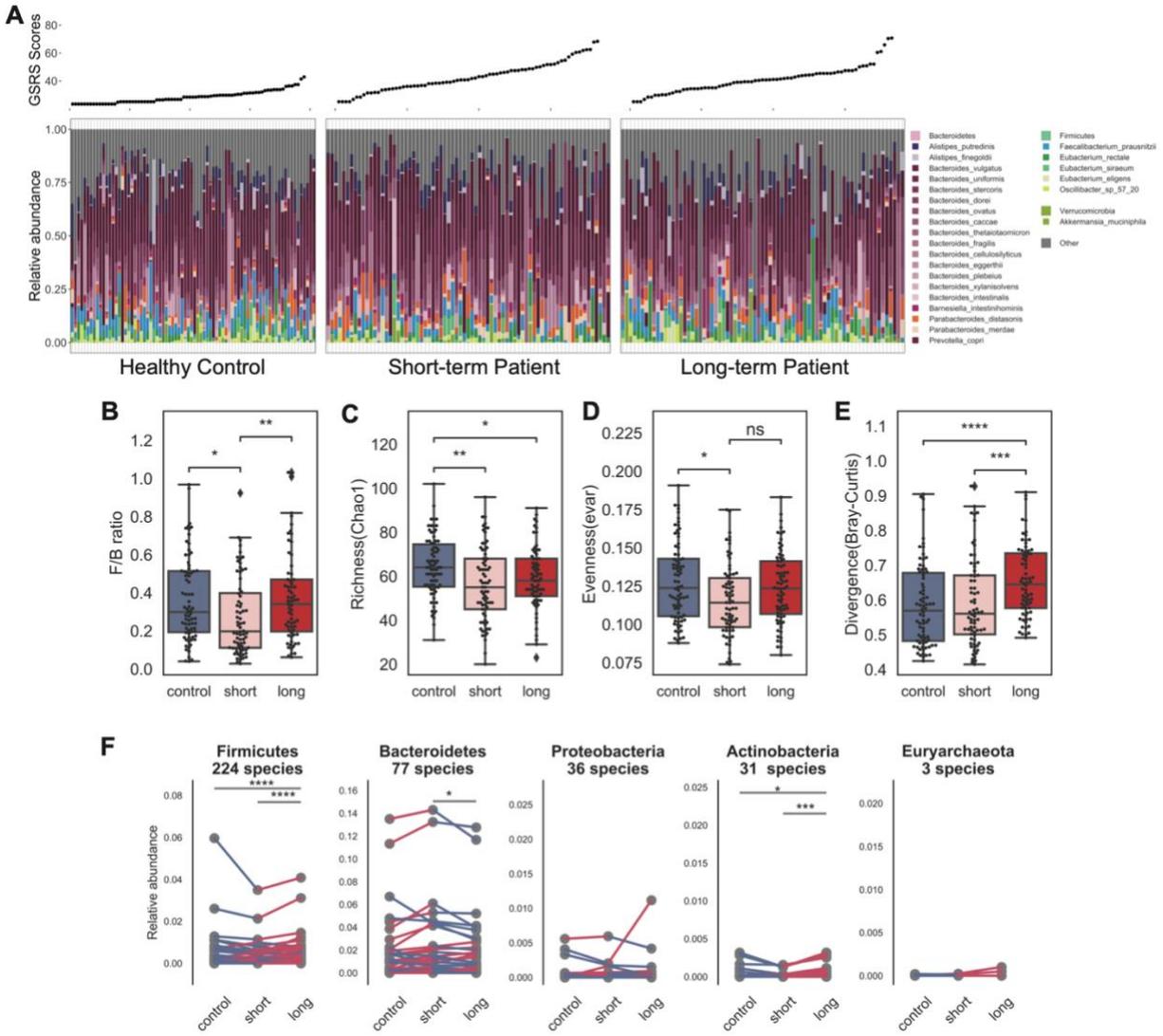

**Figure 3. Significant microbial dysbiosis in observed in short-term ME/CFS.** A) Taxonomic classification at species-level resolution for all individuals in the three cohorts: healthy controls, short-term patients, and long-term patients. Relative abundances of the most abundant gut species (top 25) are presented with gray representing the aggregate relative abundance of the remaining species. Gastrointestinal Symptom Rating Scale (GSRS) score, indicating the scale of gastrointestinal abnormality, is shown above for each individual. Similarly to Figure 2, we showed differences in the microbial community structures in short-term and long-term patients in: B) composition at phylum-level – decreased Firmicutes/Bacteroidetes ratio in short-term patients; C) reduced richness in short- and long-term (Chao 1 index, which measures the number of observed species); D) decreased evenness in short-

term (lower values of Smith and Wilson's Evar index). p-values were computed by Wilcoxon rank-sum test. E) Increased heterogeneity observed in long-term cohort as measured by divergence, or Bray-Curtis dissimilarity. p-value was computed by Wilcoxon rank-sum test. F) Each species in the five most abundant phyla were compared among three groups to observe the dynamics of gut community with respect to progression of disease. Here, each point represents the average relative abundance for a given species, connected with a line that is colored by increase (red) or decrease (blue). p-values were computed by Wilcoxon signed-rank test. p-value annotation legend: ns: $p > 0.05$, *: $0.01 < p <= 0.05$, **: $0.001 < p <= 0.01$, ***: $1e\text{-}04 < p <= 0.001$, ****: $p <= 1e\text{-}04$. See also Table S3.

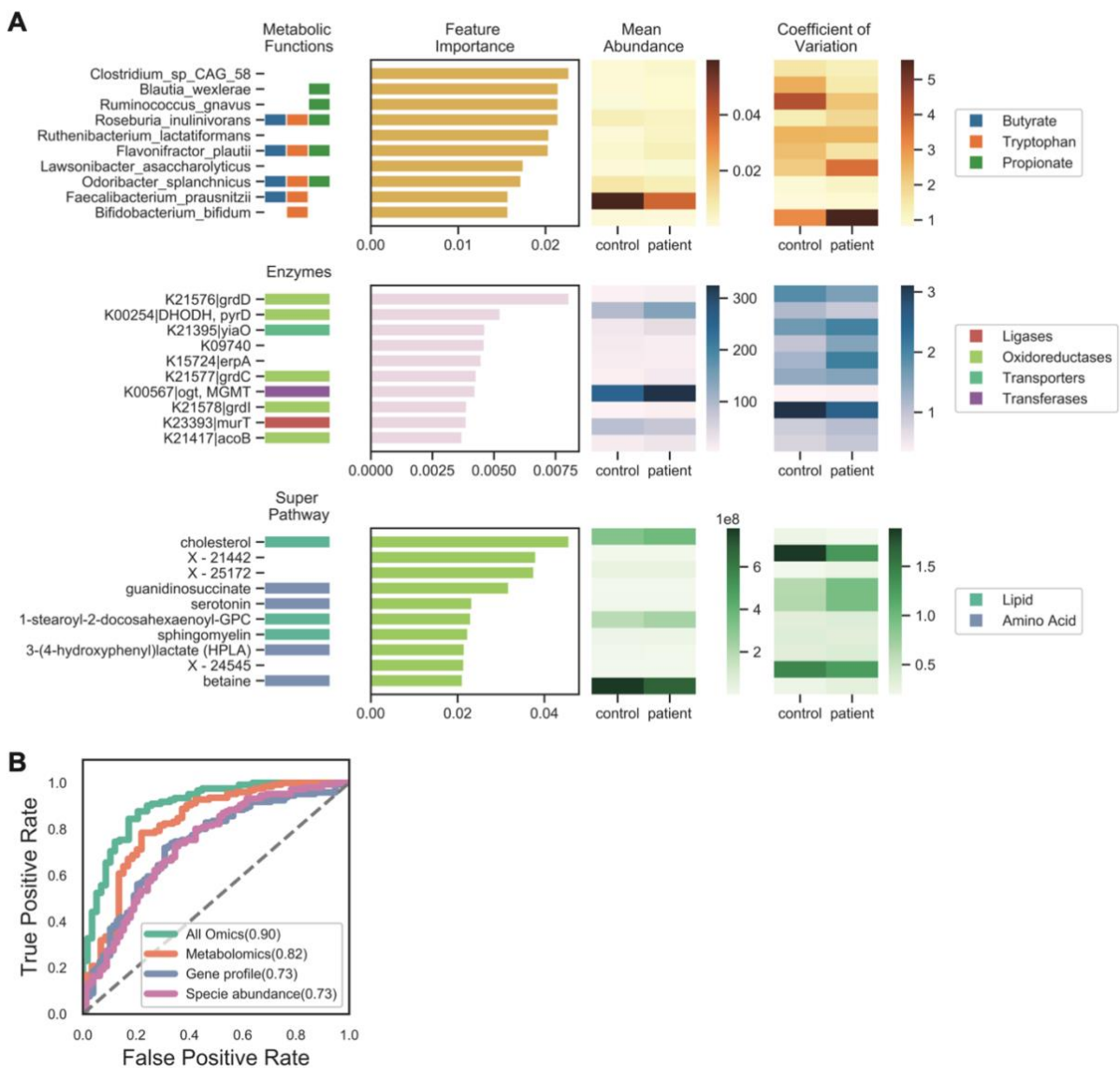

**Figure 4. Out-performing multi-'omics model identifies microbial, metagenomic, and metabolic biomarkers for ME/CFS compared to controls.** A) Biomarkers from three supervised Gradient Boosting (GDBT) models are shown. Models from top to bottom: species relative abundance, relative abundance of KEGG gene profile, normalized abundance of plasma metabolomics. The top ten most important features in each model are shown together with their general functional class, raw abundance, and variance. From left to right: 1. Functional annotations: species relative abundance model - the metabolic function (capacity of butyrate, tryptophan, and propionate pathway); KEGG gene profile model - the class identification of the enzyme; metabolomics models - the superfamily for the metabolite; 2. Feature importance:

features were ranked by their contribution to the model on the y-axis; the x-axis indicates the feature importance value from each model; 3. Average feature abundance in control and patient groups (Figure S4); 4. Variation in mean relative abundance in control and patient groups with coefficient of variation. B) Performance of the classifiers using area under the curve (AUC) was evaluated using 10 randomized and 10-fold cross-validations for each model: species relative abundance (pink), KEGG gene profile (blue) or metabolites (orange) alone, or taken altogether ('omics, green), which used the combination of the top 30 features from three models. See also Figure S4 and Table S3-5.

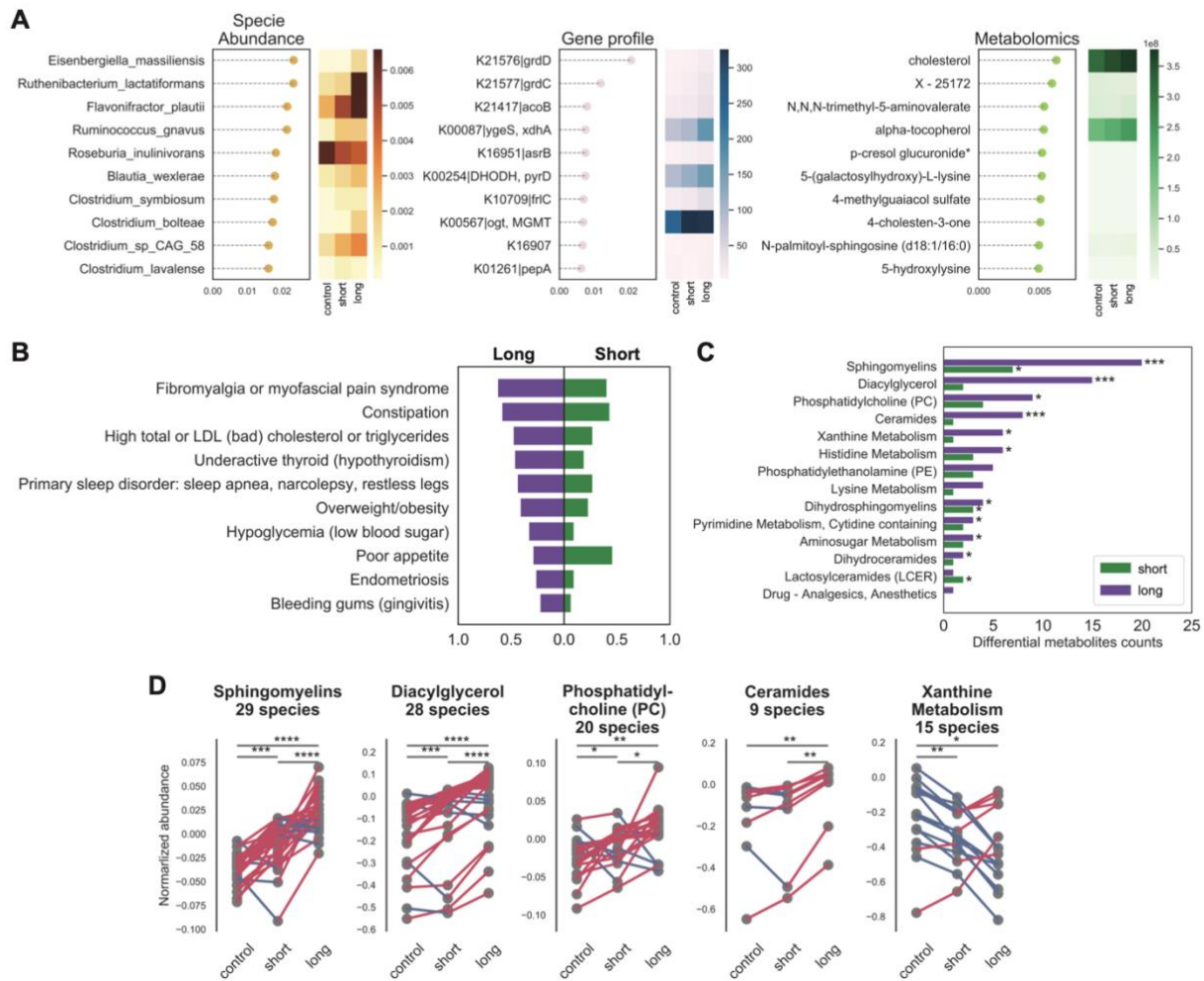

**Figure 5. Phenotypical and metabolic abnormality are most pronounced in long-term patients.** A) Multi gradient boosting models identified most important species, genes and metabolites differentiating controls, short-, or long-term ME/CFS. In each model, the top ten features are ranked by their contribution to the model on the y-axis, and the x-axis indicates the feature importance value. The heatmap shows the average feature abundance or relative abundance in each group. For full classification model performance (AUCs), see also Figure S5. B) Naïve Bayesian model based on medical history records classified the stage of disease and identified nine significant clinical phenotypes in the long-term cohort and one significant phenotype in the short-term cohort. For each feature, the probability of experiencing the symptom in the long-term patients was presented to the left on the x-axis and the probability in the short-term patients was presented to the right. C) Overrepresentation analysis (ORA) on the plasma metabolome identified the most differential metabolites and pathways in the long-term group. For each pathway, two comparisons were conducted, control vs. short-term and control vs. long-term. P-values were computed by linear global t-test and the counts of differential metabolites presented on the x-axis. D) The trend of gradually changing metabolic irregularities along with the progression of disease are indicated by the difference between control, short-term and long-term cohorts. Here, each point represents the average normalized abundance for a given metabolite in the top five most abundant superfamilies, connected with a line that is colored by if increasing (red) or decreasing (blue). p-values were computed by Wilcoxon signed-rank test. p-value annotation legend: *: $0.01 < p <= 0.05$, **: $0.001 < p <= 0.01$, ***: $1e-04 < p <= 0.001$, ****: $p <= 1e-4$. See also Table S5-6.

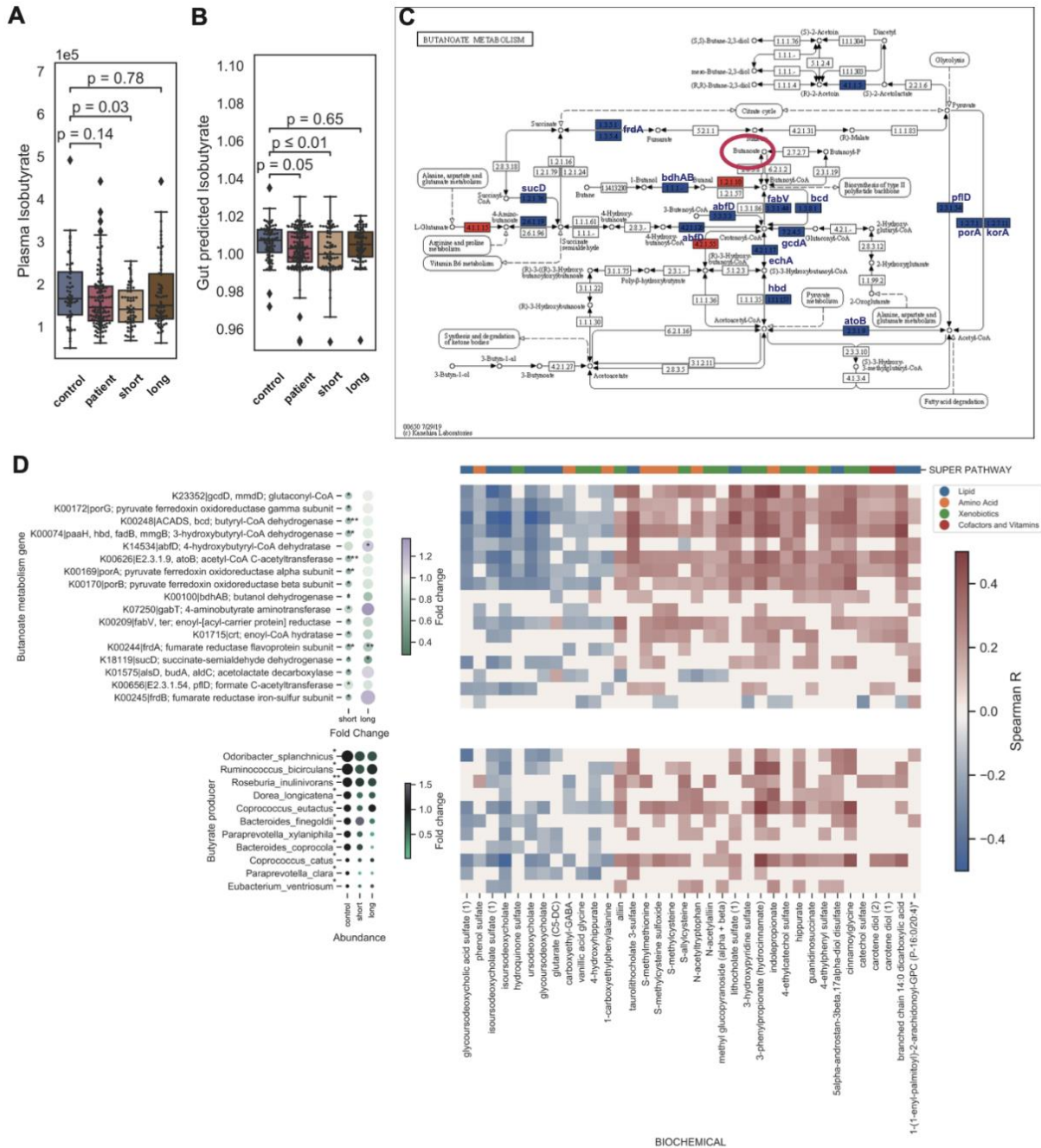

**Figure 6. Limited microbial butyrate biosynthesis capacity associates with reduced plasma isobutyrate and multiple blood metabolites.** In the blood and gut environment, decreased butyrate abundance in the short-term patient was indicated by: A) significantly reduced plasma isobutyrate normalized abundance; B) decreased predicted gut isobutyrate in patients, especially in short-term patients. Boxes show median relative abundance and interquartile ranges (IQR); whiskers specify ±1.5∗IQR

from the box's quartile. P-values were computed by Wilcoxon rank-sum test. C) The reduced abundance of most key enzymes in the butanoate mechanism (KEGG pathway map00650) indicated a more limited microbial butyrate biosynthesis capacity in ME/CFS. Differentiating enzymes were colored and annotated on the map (decreased in blue and increased in red). D) Correlation of plasma metabolite normalized abundance and relative abundance of microbial butyrate biosynthesis features, with fold changes. Heatmap shows significant correlations (Spearman, $p < 0.05$) with the top bar indicating the metabolite superfamily. The top half shows the key enzymes in the KEGG butanoate pathway. On the left, different fold changes between the two patient cohorts (short-term vs. control and long-term vs. control, respectively) indicated a significant decrease in butyrate biosynthetic capacity in the early stages of ME/CFS. P-values were calculated in each group with Wilcoxon rank-sum test. Finally, the bottom half shows the correlation between the relative abundance of predicted butyrate producers and plasma metabolites. Microbes were ordered by relative abundance. For each microbe, the size of the dot indicates the mean abundance in each group and the color indicated fold change over the control group. P-value was computed by Kruskal–Wallis H test. p-value annotation legend: *: $0.01 < p <= 0.05$, **: $0.001 < p <= 0.01$, ***: $1e\text{-}04 < p <= 0.001$. See also Table S6.

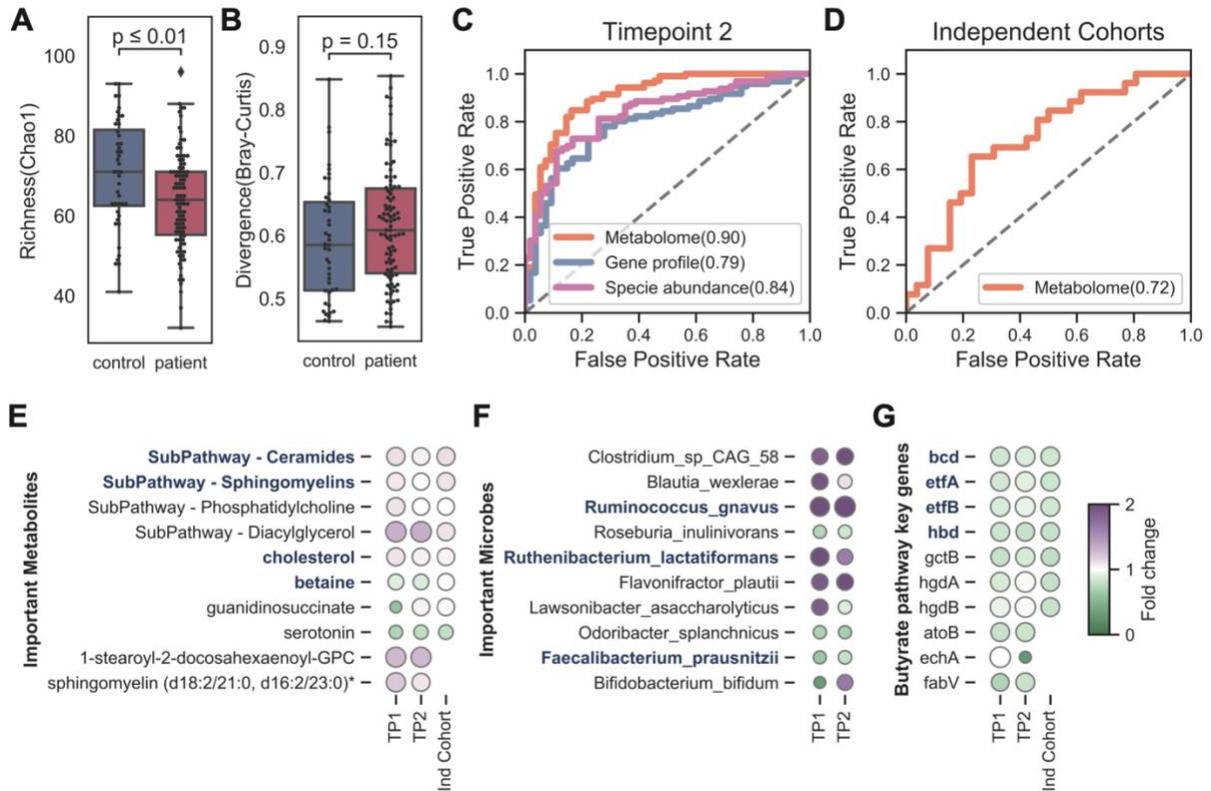

**Figure 7. Microbial and metabolomic features of ME/CFS identified in additional cohorts.** A) Decreased richness and B) increased heterogeneity (Bray-Curtis dissimilarity) in the gut microbiota were consistent in our second timepoint (Wilcoxon rank-sum test). B) The performance of our classifiers was assessed using ROC curves in C) our second timepoint and D) an independent cohort (Germain et al., 2020). The models were colored by species relative abundance (pink), KEGG gene profile (blue), and metabolites (orange). E) The fold change of important metabolic subpathways and biomarkers are shown from three cohorts: timepoint 1, 2, and an independent cohort. The shared features with Germain's study were highlighted in blue. F) The fold change of microbial biomarkers from the classifier in timepoints 1 and 2. The shared biomarkers with Guo's study are highlighted in blue. G) The fold change of key genes in the butyrate pathway across three cohorts. The shared essential genes from the Acetyl-CoA branch are highlighted in blue. See also Figure S7.

**Methods**

Detailed methods are provided in the online version of this paper and include the following:

- KEY RESOURCE TABLE
- RESOURCE AVAILABILITY
    - Materials availability
    - Data and code availability
- EXPERIMENTAL MODEL AND SUBJECT DETAILS
    - Cohort and study design
    - Participants
- METHOD DETAILS
    - Clinical metadata collection and preprocess
    - Plasma sample collection and preparation
    - Plasma untargeted metabolome by UPLC-MS/MS
    - Fecal sample collection and DNA extraction
    - Metagenomic shotgun sequencing
- QUANTIFICATION AND STATISTICAL ANALYSIS
    - Metabolome enrichment study
    - Taxonomic and KEGG gene profiling of metagenomics samples
    - Microbial community structure analysis
    - Gut metabolic status prediction
    - Multi-'omics classification models
    - Confounder analysis
    - Interaction study and targeted pathway analysis
    - Statistical Analysis

# KEY RESOURCE TABLE

| REAGENT or RESOURCE | SOURCE | IDENTIFIER |
|---|---|---|
| Biological samples | | |
| Stool samples from ME/CFS patients and healthy donors | This study | N/A |
| Plasma samples from ME/CFS patients and healthy donors | This study | N/A |
| Critical commercial assays | | |
| QIAamp 96 DNA QIAcube HT Kit | QIAGEN | #51331 |
| BioCollector fecal collection kit | The BioCollective, Denver, CO | https://www.thebiocollective.com/products-services |
| OMNIgene•GUT tube | DNA Genotek | OMR-200 |
| Deposited data | | |
| Metagenomics sequencing raw data | This study | NCBI-SRA: Bioproject PRJNA878603 |
| Software and algorithms | | |
| R version (v4.0.4) | | |
| Python (v3.10.4) | | |
| scythe (v0.994) | Buffalo, 2014 | https://github.com/vsbuffalo/scythe |
| sickle (v1.33) | Joshi and Fass, 2011 | https://github.com/najoshi/sickle |
| MetaPhlAn3 | Beghini et al., 2021 | https://huttenhower.sph.harvard.edu/metaphlan/ |
| Bowtie2 (v2.3.1) | Langmead and Salzberg, 2012 | https://github.com/BenLangmead/bowtie2 |
| GRiD (v1.3) | Emiola et al., 2018 | https://github.com/ohlab/GRiD |
| USEARCH (v8.0.1517) | Edgar, 2010 | https://www.drive5.com/usearch/ |
| dplyr_1.0.5 | R package | https://dplyr.tidyverse.org/index.html |
| microbiome_1.12.0 | R package | https://microbiome.github.io/tutorials/ |
| phyloseq_1.34.0 | R package | https://joey711.github.io/phyloseq/ |
| maaslin2_1.8.0 | R package | https://huttenhower.sph.harvard.edu/maaslin/ |
| globaltest_5.44.0 | R package | https://www.bioconductor.org/packages/release/bioc/html/globaltest.html |
| bcdstats_0.0.0.9005 | R package | https://github.com/bcdudek/bcdstats/ |

| lme4_1.1-29 | R package | https://cran.r-project.org/web/packages/lme4/index.html |
| vegan_2.6-2 | R package | https://cran.r-project.org/web/packages/vegan/index.html |
| deseq2_v1.18.1 | R package | https://bioconductor.org/packages/release/bioc/html/DESeq2.html |
| scipy_1.8.0 | python package | https://scipy.org/ |
| jupyter_4.9.2 | python package | https://jupyter.org/ |
| numpy_1.22.3 | python package | https://numpy.org/ |
| pandas_1.4.2 | python package | https://pandas.pydata.org/ |
| scikit-learn_1.0.2 | python package | https://scikit-learn.org/stable/index.html |

## RESOURCE AVAILABILITY

## Materials availability

This study did not generate new unique reagents.

## Data and code availability

Metagenomics data have been deposited at PRJNA878603 and are publicly available as of the date of publication. Accession numbers are listed in the key resources table.

All original code has been deposited at https://github.com/ohlab/MECFS_2021 and is publicly available as of the date of publication.

Any additional information required to reanalyze the data reported in this paper is available from the lead contact upon request.

**EXPERIMENTAL MODEL AND SUBJECT DETAILS**

**Cohort and study design** All subjects were recruited at Bateman Horne Center, Salt Lake City, UT, based on who met the 1994 CDC Fukuda (Fukuda et al., 1994) and/or Canadian consensus criteria for ME/CFS (Carruthers, 2007). Healthy controls were frequency-matched to cases on age, sex, race/ethnicity, geographic/clinical site, and season of sampling. Patients or controls taking antibiotics or who had any infections in the prior one month, or who were taking any immunomodulatory medications were excluded from the study. The study was approved by The Jackson Laboratory IRB (Study number 17-JGM-13) and written informed consent and verbal assent when appropriate were obtained from all participants in this study. We enrolled a total of 149 ME/CFS patients (of which 74 had been diagnosed with ME/CFS <4 years before recruitment and 75 had been diagnosed with ME/CFS >10 years before recruitment) and 79 healthy controls. Among them, 107 patients and 59 healthy controls were followed one year after the recruitment as timepoint 2. Subject characteristics are shown in Supplemental Table 1.

**Participants** 149 ME/CFS patients who had been seen at Bateman Horne Center (Salt Lake City, UT) for routine clinical care between February 2018 and September 2019 and 79 matched HCs were recruited for the study. The 149 MECFS subjects included 74 sick with ME/CFS for <4 years and 75 sick for greater than 10 years. The age range of ME/CFS participants and HCs was 18-65 years at the time of informed consent. HCs were matched with <4 ME/CFS participants by age (± 5 years), gender and ethnicity. Enrolled ME/CFS participants were required to fulfill the International Chronic Fatigue Syndrome Study Group research criteria [99], the Canadian Consensus Criteria [100], and the IOM clinical diagnostic criteria [101]. HCs were recruited from the Salt Lake City metropolitan area using advertisements posted on social media, the clinic webpage or by phone contact with a volunteer pool from previous studies. HCs were considered generally healthy and between 18 to 65 years of age. HCs were excluded if they fulfilled ME/CFS diagnostic criteria or had a history of illness, had a BMI>40 or had been treated with long-term (longer than 2 weeks) antiviral medication or immune modulatory medications within the past 6 months or had been treated with short-term (less than 2 weeks) antiviral or antibiotic medication within the past 30 days.

**METHOD DETAILS**

**Clinical metadata collection and preprocess** Clinical symptoms and baseline health status was assessed on the day of physical examination and biological sample collection from both case and control subjects. For each participant, we collected demographic information (including age, gender, diet, race, family, work, and education), medical histories, and three questionnaires regarding the general physical and mental health condition (RAND-36 form), sleep quality (PSQI form) and gastrointestinal health (GSRS form). The summary of analyzed clinical features and questionnaires are shown in supplemental Table 2. Age and diet were analyzed and discussed as potential confounders (Figure S1 and S4). Medical histories were simplified into binary features (0 - no records, 1, had/having the disease) and further constructed naïve Bayesian classification models. Every questionnaire was transformed into a 0–100 scale to facilitate combination and comparison wherein a score of 100 is equivalent to maximum disability or severity and a score of zero is equivalent to no disability or disturbance.

**Plasma sample collection and preparation** Healthy and patient blood samples were obtained from Bateman Horne Center, Salt Lake City, UT and approved by JAX IRB. One 4 mL lavender top tube (K2EDTA) was collected, and tube slowly inverted 8-10 times immediately after collection. Blood was centrifuged within 30 minutes of collection at 1000 x g with low brake for 10 minutes. 250 uL of plasma was transferred into three 1 mL cryovial tubes, and tubes were frozen upright at -80°C. Frozen plasma samples were batch shipped overnight on dry ice to The Jackson Laboratory, Farmington, CT, and stored at -80°C.

**Plasma untargeted metabolome by UPLC-MS/MS** Plasma samples were sent to Metabolon platform and processed by Ultrahigh Performance Liquid Chromatography-Tandem Mass Spectroscopy (UPLC-MS/MS) following the CFS cohort pipeline. In brief, samples were prepared using the automated MicroLab STAR® system from Hamilton Company. The extract was divided into five fractions: two for analysis by two separate reverse phases (RP)/UPLC-MS/MS methods with positive ion mode electrospray ionization (ESI), one for analysis by RP/UPLC-MS/MS with negative ion mode ESI, one for analysis by HILIC/UPLC-MS/MS with negative ion mode ESI, and one sample was reserved for backup. QA/QC were analyzed with several types of controls were analyzed including a pooled matrix sample generated by taking a small volume of each experimental sample (or alternatively, use of a pool of well-characterized human plasma), extracted water samples, and a cocktail of QC standards that were carefully chosen not to interfere with the measurement of endogenous compounds were spiked into every analyzed sample, allowed instrument performance monitoring, and aided chromatographic alignment. Compounds were identified by comparison to Metabolon library entries of purified standards or recurrent unknown entities. The output raw data included the annotations and the value of peaks quantified using area-under-the-curve for metabolites.

**Fecal sample collection and DNA extraction** Stool was self-collected at home by volunteers using a BioCollector fecal collection kit (The BioCollective, Denver, CO) according to manufacturer instructions for preservation for sequencing prior to sending the sample in a provided Styrofoam container with a cold pack. Upon receipt, stool and OMNIgene samples were immediately aliquoted and frozen at –80°C for storage. Prior to aliquoting, OMNIgene stool samples were homogenized by vortexing (using the metal bead inside the OMNIgene tube), then divided into 2 microfuge tubes, one with 100µL aliquot and one with 1mL. DNA was extracted using the Qiagen (Germantown, MD, USA) QIAamp 96 DNA QIAcube HT Kit with the following modifications: enzymatic digestion with 50µg of lysozyme (Sigma, St. Louis, MO, USA) and 5U each of lysostaphin and mutanolysin (Sigma) for 30 min at 37 °C followed by bead-beating with 50 µg 0.1 mm of zirconium beads for 6 min on the Tissuelyzer II (Qiagen) prior to loading onto the Qiacube HT. DNA concentration was measured using the Qubit high sensitivity dsDNA kit (Invitrogen, Carlsbad, CA, USA).

**Metagenomic shotgun sequencing** Approximately 50µL of thawed OMNIgene preserved stool sample was added to a microfuge tube containing 350 µL Tissue and Cell lysis buffer and 100 µg 0.1 mm zirconia beads. Metagenomic DNA was extracted using the QiaAmp 96 DNA QiaCube HT kit (Qiagen, 5331) with the following modifications: each sample was digested with 5µL of Lysozyme (10 mg/mL, Sigma-Aldrich, L6876), 1µL Lysostaphin (5000U/mL, Sigma-Aldrich, L9043) and 1µL oh Mutanolysin (5000U/mL, Sigma-Aldrich, M9901) were added to each sample to digest at 37°C for 30 minutes prior to the bead-beating in the in the TissueLyser II (Qiagen) for 2 x 3 minutes at 30 Hz. Each sample was centrifuged for 1 minute at 15000 x g prior to loading 200µl into an S-block (Qiagen, 19585) Negative (environmental) controls and positive (in-house mock community of 26 unique species) controls were extracted and sequenced with each extraction and library preparation batch to ensure sample integrity.

Sequencing adapters and low-quality bases were removed from the metagenomic reads using scythe (v0.994) and sickle (v1.33), respectively, with default parameters. Host reads were removed by mapping all sequencing reads to the hg19 human reference genome using Bowtie2 (v2.3.1), under 'very-sensitive' mode. Unmapped reads (i.e., microbial reads) were used to estimate the relative abundance profiles of the microbial species in the samples using MetaPhlAn3.

**QUANTIFICATION AND STATISTICAL ANALYSIS**

**Metabolome enrichment study** From the raw data (peaks area-under-the-curve), we first kept metabolic features present in >50% of the samples for further analysis, and missing values were imputed with the minimum value. To remove batch variability, for each metabolite, the values in the experimental samples are divided by the median of those samples in each instrument batch, giving each batch and thus the metabolite a median of one. For each metabolite, the minimum value across all batches in the median scaled data is imputed for the missing values. We first applied qualitative enrichment analysis (Over Representation Analysis ORA) in two patient cohorts (short-term vs. control, and long-term vs. control). We conducted Wilcoxon rank-sum test on all metabolites and counted the significantly differential metabolites in every sub pathway. Fisher test with Benjamini-Hochberg adjustment was followed to identify the distributions of the over-represented genes in the pathway. For every sub pathway, we also applied a global quantitative enrichment analysis with linear *globaltest* in R to compute the association between a group of metabolites from that pathway and the duration of disease (control, short-term, and long-term).

**Taxonomic and KEGG gene profiling of metagenomics samples** Taxonomic compositions were profiled using Metaphlan3.0 [102] and the species whose average relative abundance > 1e-5 were kept for further analysis, giving 384 species. The gene profiling was computed with USEARCH(v8.0.15) [103] (with parameters: evalue 1e-9, accel 0.5, top_hits_only) to KEGG Orthology (KO) database v54, giving a total of 9452 annotated KEGG genes. The reads count profile was normalized by DeSeq2 in R [104]. Metagenomics read depth was considered as potential confounders and analyzed as below (See Method Confounder analysis).

**Microbial community structure analysis** The overall community structures was examined by Correspondence Analysis (PCoA) and PERMANOVA were performed in R with the *adonis* function in the R package vegan to analyze the partitioning of variation giving potential confounders including age and gender. The heterogeneity index (Inter-individual divergence) and community indexes including chao1, evenness(evar), rarity(low_abundance) and inequality(dominance_gini) were computed by R package microbiome. The species replication rate was predicted using GRiD with default settings [62]. Metagenomics read depth was considered as potential confounders and analyzed as below (See Method Confounder analysis).

**Gut metabolic status prediction** Here, we adapted from the MAMBO [105] pipeline and predicted 224 metabolites status of the microbial community in our cohort. We first constrained the Genome-scale metabolic models (GSMMs) community of all 384 microbes identified in the species profile. We started with the same gut metabolic environment for all samples giving randomized 224 metabolites initialization and using a Markov chain Monte Carlo (MCMC) to sample metabolites. In each step, we sampled one metabolite and used the Flux balance analysis to model reaction flux with a slight change of the target metabolites and accepted the step only if the probed growth rate is correlated with the species relative abundance. Samples were first subjected to 100,000 search steps, and 100,000 steps were subsequently added until a high Pearson correlation ($\rho > 0.6$) with the target metagenomic abundance profile was achieved. Finally, the 10% time points with the highest Pearson correlation scores between the biomass profile and the metagenomic abundance profile were averaged, yielding a robustly predicted metabolome.

**Multi-'omics classification models** To identify phenotypic, metagenomic, and metabolomic markers of the onset(control/patient) of the disease, we constructed a naïve Bayesian classification model with medical history records and three individual classification models based on the species abundance, normalized KEGG gene abundance, and normalized metabolite profile and one combination multi-omics model with all top ten features collected from each model. We also tested four different classification methods, LASSO logistic regression, Support vector machine (SVM), Random Forest (RF), and Gradient Boosting (GDBT). The same Multi-'omics classification model system was also applied to classify the duration(control/short-term/long-term) of the disease.

All analyses were carried out using the Python package 'scikit-learn'. Normalized KEGG gene and normalized plasma metabolome were standardized (by centering to mean 0 and dividing by the standard deviation of each feature) before fitting into the models. The models were optimized by five-fold RandomizedSearchCV to probe the best parameters giving lists of candidates. Models were then validated by 10-fold stratified cross-validation testing (we resampled dataset partitions 10 times). In each test, the accuracy of the model was examined using ROC (area under the curve). The comparison among models showed that GDBT outcompeted the rest three and reached the best performance. In the Gradient boosting model, two steps were carried out. In the first step, the model was constructed using each of the three profiles (species abundance, KEGG gene abundance, and metabolite profile) individually to compute the feature importance as the feature contributions to the classification. In the second step, the collective model was constructed using a combination of top ten important features determined from the species relative abundance model + top ten important features determined from the KEGG gene model + top ten important features determined from the plasma metabolite model.

**Confounder analysis** Confounder analysis was done by R package MaAsLin2[106]. The demographic features (including age, gender, ethnicity, and race), diet records, antivirals, antifungals, antibiotics, and probiotics medications, and self-reported IBS scores are considered potential confounders. We ran general linear tests to determine the multivariable association between each meta feature and each microbial and metabolomic features (including microbiome diversity indexes, specie abundance, KEGG gene abundance, plasma metabolites). The three formulas for fixed effects: 1) Demographic model: expr ~ age + gender + ethnic + race + IBS; 2) Diet model: expr ~ diet_meat + diet_sugar + diet_veg + diet_grains + diet_fruit; 3) expr ~ antifungals + antibiotics + probiotics + antivirals. The *expr* in these formulas have combined microbiome diversity indexes and 3 'omics in the model training and covered all features in the differentiation analysis. Metagenomics reads depth was also considered as another confounder. The general linear tests were computed with all microbiome features by the formula: reads_count ~ microbiome diversity indexes + specie abundance + KEGG gene abundance. No microbiome features were significantly selected by the model. All significant features identified by the association models are shared in supplemental table 10. The codes and run log files are shared on our GitHub.

**Interaction study and targeted pathway analysis** We computed the Spearman correlation for the butyrate pathway genes (N = 113) and butyrate producers (N = 18) with metabolites (N = 1278) along with the fold change of the elements. P-value was correlated by Holm's method.

**Statistical Analysis** The dimensionality reduction analysis was conducted by Principal Correspondence Analysis (PCoA) using sklearn.manifold.MDS function for both gut microbiome Bray-Curtis dissimilarity distance matrix and normalized plasma metabolome profile. The statistically significant differences among independent groups (healthy/patient/short-term/long-term) were determined by nonparametric test using pairwise Wilcoxon rank-sum test two-sided with Bonferroni correction. The average abundances of each species, genes and metabolites were determined to be significantly elevated or depleted in short-term or long-term groups by pairwise nonparametric comparison using Wilcoxon signed-rank test with Bonferroni correction over N=384 species, N=9652 genes and N=1278 metabolites, respectively. Chi-squared test was used to compare the infection frequencies between healthy and patient groups. The correlation analysis was done by Spearman's rank correlation, and the p-values were correlated by Holm's method. P value annotations: ns: $p > 0.05$, *: $0.01 < p <= 0.05$, **: $0.001 < p <= 0.01$, ***: $1e-04 < p <= 0.001$, ****: $p <= 1e-04$.

**Supplemental Figure**

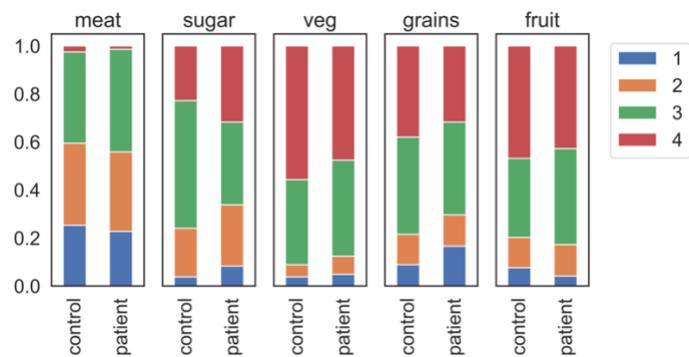

**Supplemental Figure 1. The dietary distributions of control and patient cohorts.** The dietary questionnaire was summarized into five categories and their frequency in the last week. Categories: Meat - red meat; sugar - desserts, sweets, soda, or juice; veg – fresh vegetables; grains, whole grains (e.g., oatmeal, quinoa, whole wheat products); fruit - fresh fruit. Frequency: 1, Never; 2, Once; 3, 2 to 5 times; 4, Daily.

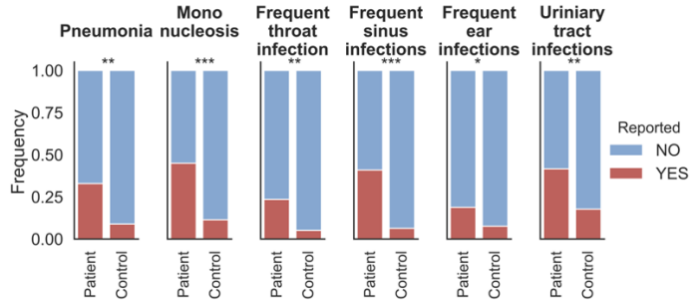

**Supplemental Figure 2. ME/CFS patients reported significantly more infection-related histories.** More than 30% ME/CFS patients reported pneumonia, mononucleosis, frequent sinus infections and urinary tract history. The p value was computed by Chi-squared test. p-values annotations: *: 0.01 < p <= 0.05, **: 0.001 < p <= 0.01, ***: 1e-04 < p <= 0.001

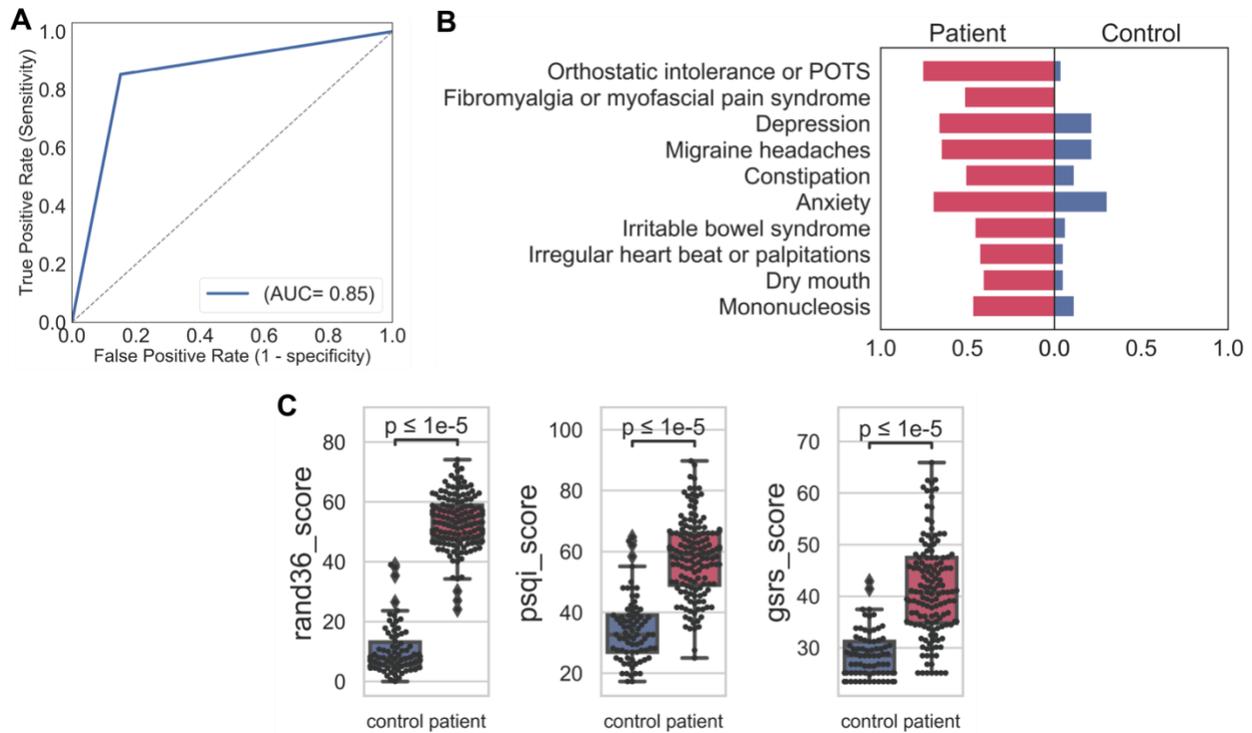

**Supplemental Figure 3. The host phenotype in ME/CFS.** A) The performance of the naïve Bayesian classification model based on medical history records to identify clinical features that discriminate healthy controls vs. patients. The area under the curve, AUC = 0.85. B) Top ten predicted features and their probabilities to discriminate both cohorts were presented in separate directions on the x-axis. For each feature, the probability of experiencing the symptom in the patients was presented to the left on the x-axis and the probability in the controls was presented to the right. C) Based on our scoring system, patients had significantly more anomalous mental and physical health conditions identified by higher rand36, Pittsburgh sleep quality, and gastrointestinal symptom rating scale scores (Table S2). p-values were computed by Wilcoxon rank-sum test.

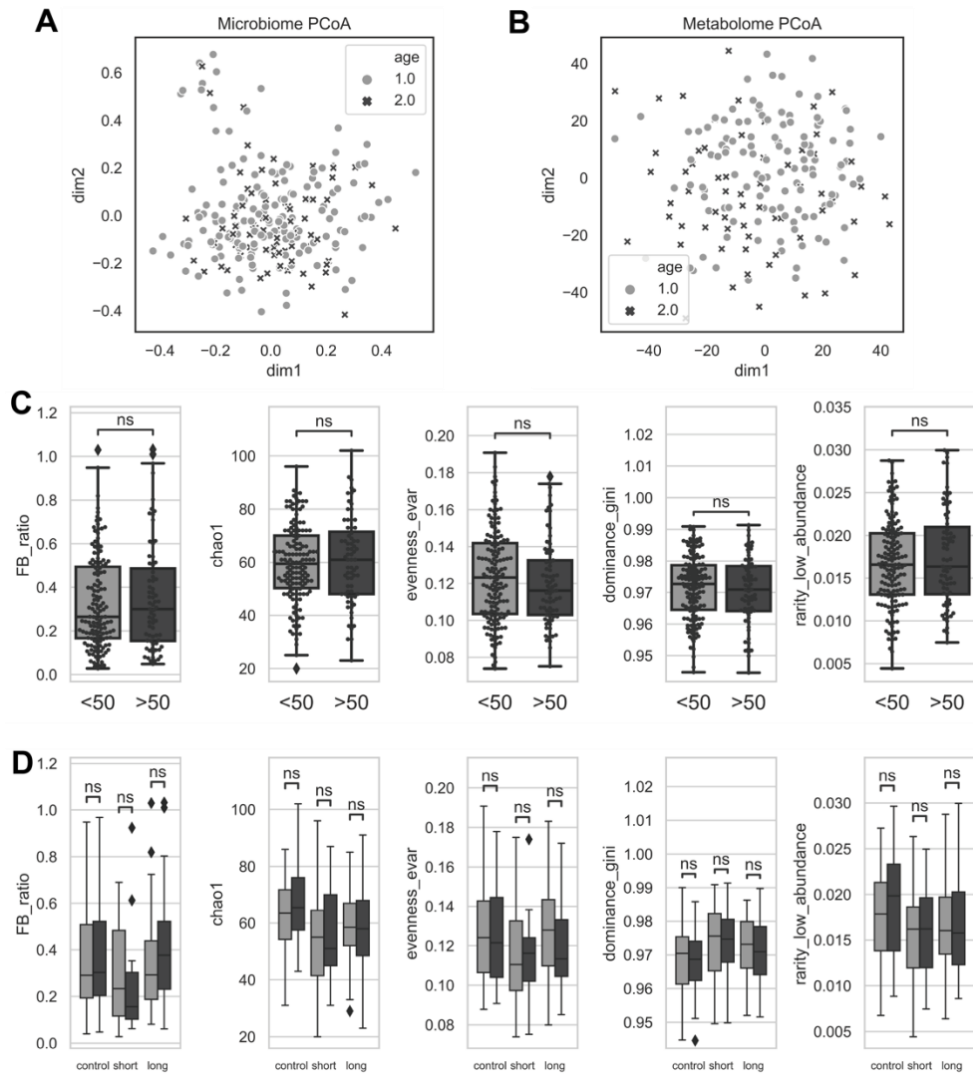

**Supplemental Figure 4. Age is not a significant confounder of the gut microbiome and plasma metabolome.** A) The Principal Correspondence Analysis (PCoA) based on gut microbiome-derived Bray-Curtis dissimilarity distance. The contribution of age was not significant with a p-value > 0.05 (PERMANOVA, see Methods). B) Principal Correspondence Analysis (PCoA) based on normalized plasma metabolome profile. In C) and D), microbial community structure was not significantly different between young (<50 years old) and old (>=50 years old) in C) all cohorts and D) cohort by disease stage (control, short-term, and long-term), respectively. Community structure was indicated by FB_ratio (Firmicutes:Bacteroides ratio), Chao 1 index (richness), evenness_evar (Smith and Wilson's Evar index), dominance_gini (Gini index of the dominant species >0.2% relative abundance), and rarity_low_abundance (proportion of the least abundant species <0.2% relative abundance). p-values were computed by Wilcoxon rank-sum test. ns, not significant.

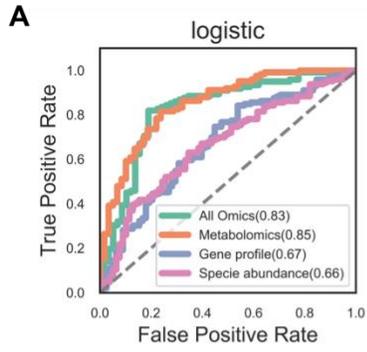

**A** logistic

True Positive Rate vs False Positive Rate

All Omics(0.83)
Metabolomics(0.85)
Gene profile(0.67)
Specie abundance(0.66)

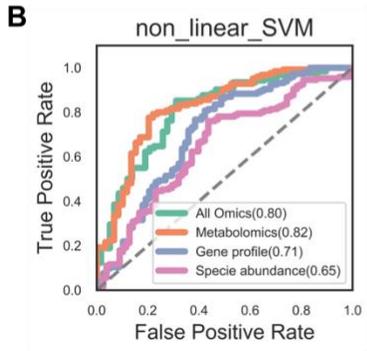

**B** non_linear_SVM

All Omics(0.80)
Metabolomics(0.82)
Gene profile(0.71)
Specie abundance(0.65)

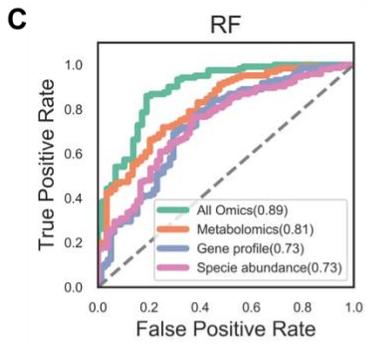

**C** RF

All Omics(0.89)
Metabolomics(0.81)
Gene profile(0.73)
Specie abundance(0.73)

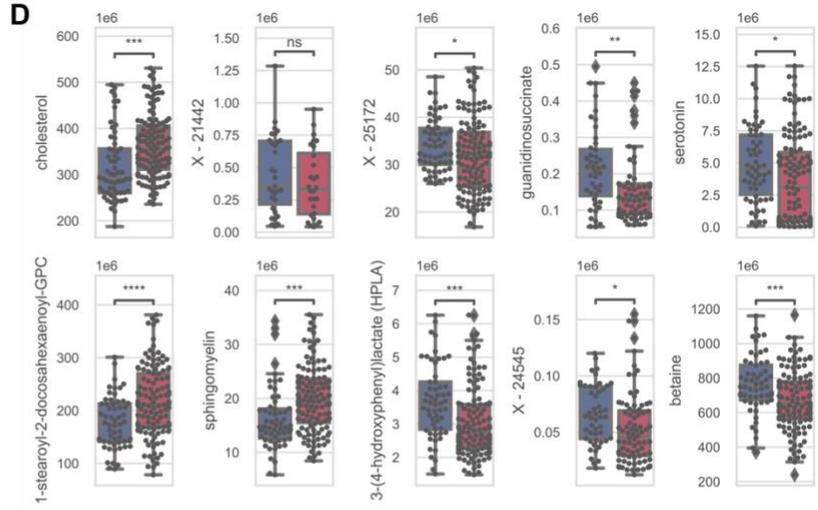

**D**

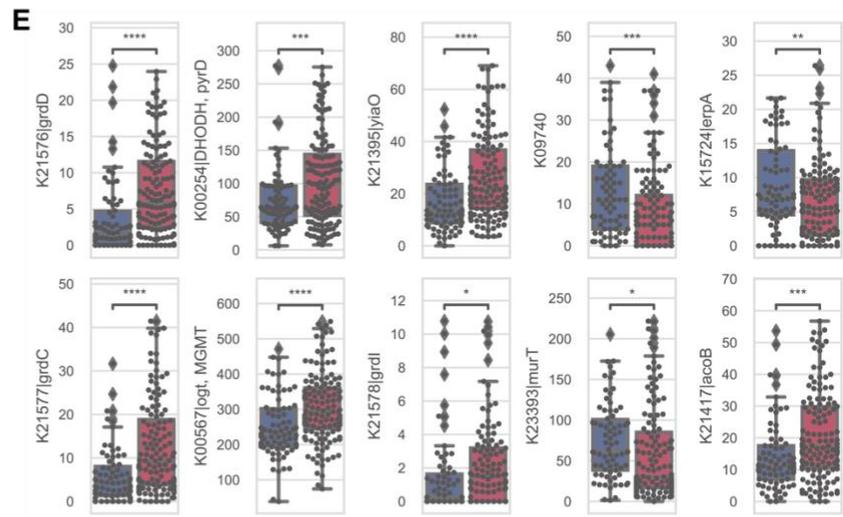

**E**

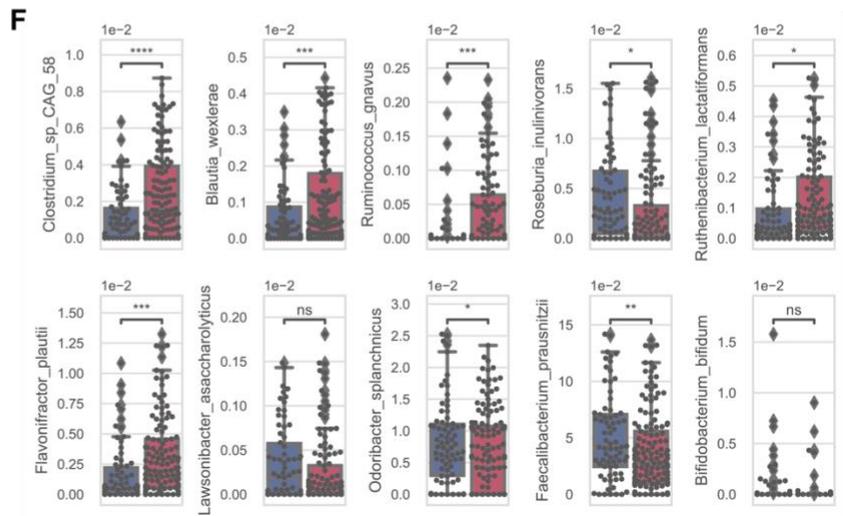

**F**

**Supplemental Figure 5. Multi-'omics models to classify the onset of ME/CFS (control vs. patient).** A-C) Performance of the classifiers using area under the curve (AUC) was evaluated using 10 randomized and 10-fold cross-validations for each model: LASSO logistic regression, Support vector machine (SVM) and random-forest (RF) models. Models were designed based on species relative abundance (pink), KEGG gene profile (blue) or plasma metabolome (orange) individually, or a taken altogether ('omics green) with top 30 features from three individual models (see Methods). D-F) Discriminant features identified from gradient boosting classifiers significantly changed in the patient cohort compared to the healthy individuals. p-values were computed by Wilcoxon signed-rank test. p-value annotation legend: ns: p > 0.05, *: 0.01 < p <= 0.05, **: 0.001 < p <= 0.01, ***: 1e-04 < p <= 0.001, ****: p <= 1e-04.

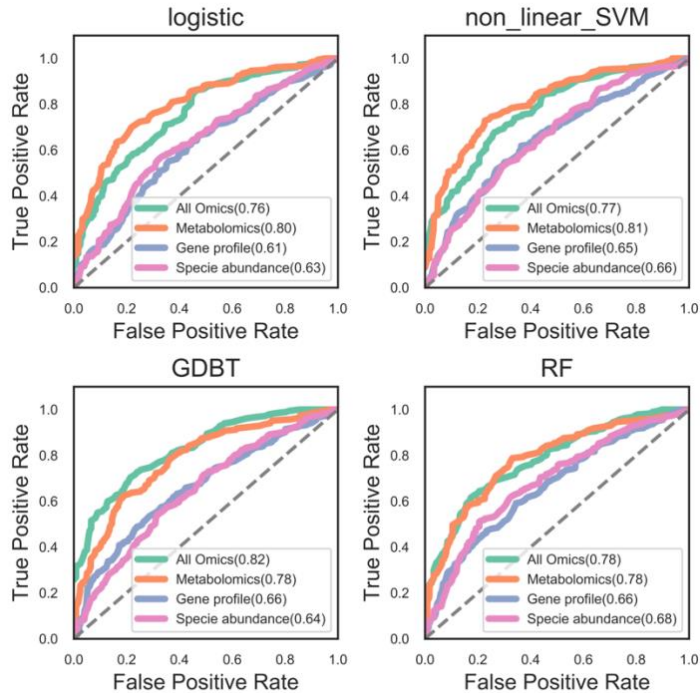

**Supplemental Figure 6. Multi-'omics models to classify the duration of ME/CFS (control vs. short-term vs. long-term).** As above, the performance of the classifiers using the area under the curve (AUC) was evaluated using 10 randomized and 10-fold cross-validations for each model: LASSO logistic regression, support vector machine (SVM), and random forest (RF) models. Models were designed based on species relative abundance (pink), KEGG gene profile (blue), or plasma metabolome (orange) individually, or taken all together ('omics green) with top 30 features from three individual models shown.

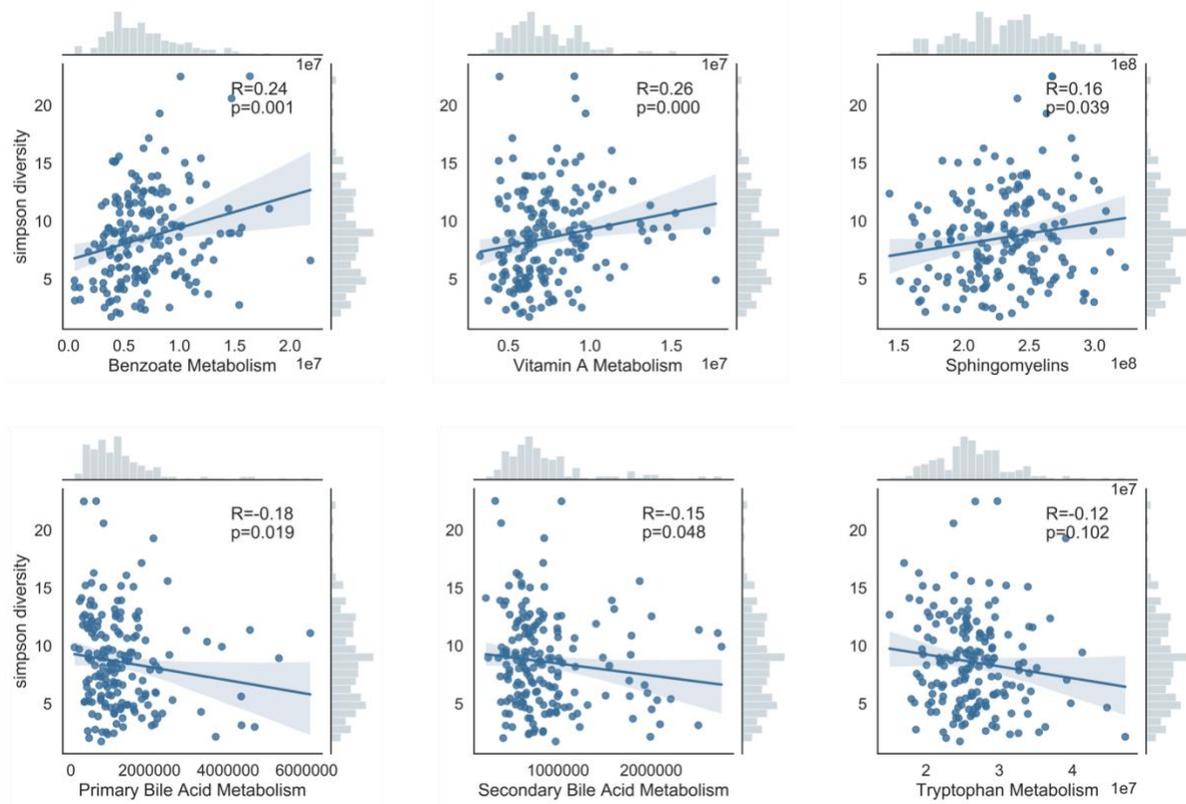

**Supplemental Figure 7. Plasma metabolic pathways are correlated with gut community structure in the cohort.** The gut microbiome community structure index was indicated by Simpson diversity and the relative abundance of the secondary metabolic pathway was indicated by the mean of all metabolites in the pathway. Each dot indicates one sample, and regression line and the confidence intervals shaded are shown. The data distributions are displayed in the top and right, respectively. The rho value and p-value were computed by the Spearman correlation.

**Supplemental Tables**

**Supplementary Table 1** Demographic information of cohorts and Clinical metadata summary. Related to STAR Methods.

**Supplementary Table 2** Metadata. Related to STAR Methods.

**Supplementary Table 3** Species relative abundances derived from MetaPhlAn3. Related to Figure 2, 3 and 4.

**Supplementary Table 4** KEGG gene reads count from metagenomic data. Related to Figure 4.

**Supplementary Table 5** Plasma metabolome raw data (peaks area-under-the-curve). Related to Figure 4 and 5.

**Supplementary Table 6** Feature importance and annotation from multi-'omics classification model and Butyrate pathway: the statistics of key microbes and key enzymes in KEGG butanoate metabolism, and their correlations with key plasma metabolites. Related to Figure 5 and 6.

**Supplementary Table 7** Confounder driven features identified by MaAsLin2, Related to STAR Methods

# Chapter 2: BioMapAI: Artificial Intelligence Multi-Omics Modeling of Myalgic Encephalomyelitis / Chronic Fatigue Syndrome

**Cohort Overview**

We tracked 249 participants over 3-4 years, including 153 ME/CFS patients (75 'short-term' with disease symptoms < 4 years and 78 'long-term' with disease symptoms > 10 years) and 96 healthy controls (Fig 1A; Supplemental Table 1). The cohort is 68% female and 32% male, aligning with the epidemiological data showing that women are 3-4 times more likely to develop ME/CFS[107],[108]. Participants ranged in age from 19 to 68 years with body mass indexes (BMI) from 16 to 43 kg/m². Throughout the study, we collected detailed clinical metadata, blood samples, and fecal samples. In total, 1471 biological samples were collected across all participants at 515 timepoints (Methods, Supplemental Figure 1A, Supplemental Table 1).

Blood samples were 1) sent for clinical testing at Quest Laboratory (48 features measured, N=503 samples), 2) fractionated into peripheral blood mononuclear cells (PBMCs), which were examined via flow cytometry, yielding data on 443 immune cells and cytokines (N=489), 3) plasma and serum, for untargeted liquid chromatography with tandem mass spectrometry (LC-MS/MS), identifying 958 metabolites (N=414). Detailed demographic documentation and questionnaires covering medication use, medical history, and key ME/CFS symptoms were collected (Methods). Finally, whole-genome shotgun metagenomic sequencing of stool samples (N=479) produced an average of 12,302,079 high-quality, classifiable reads per sample, detailing gut microbiome composition (1293 species detected) and KEGG gene function (9993 genes reconstructed)

**Heterogeneity and Non-linear Progression of ME/CFS**

First, we demonstrated the phenotypic complexity and heterogeneity of ME/CFS. Collaborating with clinical experts, we consolidated detailed questionnaires and clinical metadata, foundational to diagnosing ME/CFS, into twelve essential clinical scores (Methods). These scores covered core symptoms including physical and mental health, fatigue, pain levels, cognitive efficiency, sleep disturbances, orthostatic intolerance, and gastrointestinal issues (Supplemental Table 1).

While healthy individuals consistently presented low symptom scores (Supplemental Figure 1D), ME/CFS patients exhibited significant variability in symptom severity, with each individual showing different predominant symptoms (Figure 1B). Principal coordinates analysis (PCoA) of the 'omics matrices highlighted the difficulty in distinguishing patients from controls, emphasizing the complex symptomatology of ME/CFS and the challenges in developing predictive models (Supplemental Figure 1E). Additionally, over time, in contrast to the stable patterns typical of healthy individuals (Supplemental Figure 1B), ME/CFS patients demonstrated distinctly varied patterns each year, as evidenced by the diversity in symptom severity and noticeable separation on the 'omics PCoA (Figure 1B, Supplemental Figure 1C). Despite employing multiple longitudinal models (Methods), we found no consistent temporal signals, confirming the non-linear progression of ME/CFS.

This individualized, multifaceted, and dynamic nature of ME/CFS that intensifies with disease progression necessitates new approaches that extend beyond simple disease versus control comparisons. Here, we created and implemented an AI-driven model that integrates the multi-'omics profiles to learn host phenotypes. This allowed us not only to develop a state-of-the-art classifier for disease, but for the first time, to identify biomarker sets for each clinical symptom as well as unique interaction networks that differed between patients and controls.

## BioMapAI, an Explainable Neural Network Connecting 'Omics to Multi-Type Outcomes

To connect multi-'omics data to clinical symptoms, a model must accommodate the learning of multiple different outcomes within a single framework. However, traditional machine learning models are generally designed to predict a single categorical outcome or continuous variable[109],[110],[111]. This simplified disease classification and conventional biomarker identification typically fails to encapsulate the heterogeneity of complex diseases[112],[113].

We developed an AI-powered multi-'omics framework, BioMapAI, a fully connected deep neural network that inputs 'omics matrices ($X$), and outputs a mixed-type outcome matrix ($Y$), thereby mapping multiple 'omics features to multiple clinical indicators (Figure 2A). By assigning specific loss functions for each output, BioMapAI aims to comprehensively learn every $y$ (i.e., each of the 12 continuous or categorical clinical scores in this study), using the 'omics data inputs. Between the input layer $X$ and the output layer $Y = [y_1, y_2, \dots, y_n]$, the model consists of two shared hidden layers ($Z^1$ with 64 nodes, and $Z^2$ with 32 nodes) for general pattern learning, followed by a parallel hidden layer ($Z^3 = [z_1^3, z_2^3, \dots, z_n^3]$), with sub-layers ($z_n^3$, each with 8 nodes) tailored for each outcome ($y_n$), to capture outcome-specific patterns (Figure 2A). This unique architecture – two shared and one specific hidden layer – allows the model to capture both general and output-specific patterns. This model is made 1) explainable by incorporating a SHAP (SHapley Additive exPlanations) explainer, which quantifies the feature importance of each predictions, providing both local (symptom-level) and global (disease-level) interpretability, and 2) flexible by automatically finding appropriate learning goals and loss functions for each type of outcomes (without need of format refinement), facilitating BioMapAI's adaptability to broader research applications.

## BioMapAI Reconstructed Clinical Symptoms and Achieved State-of-the-Art Performance in Discriminating ME/CFS from Healthy Controls

BioMapAI is a versatile AI framework connecting a biological 'omics matrix to multiple phenotypic outputs. It does not have a specific disease focus and is designed to be applicable to a range of applications. Here, we trained and validated its usage with our ME/CFS datasets, employing a five-fold cross-validation. This trained model, nicknamed DeepMECFS for the ME/CFS community, accurately represented the structure of diverse clinical symptom score types and discriminated between healthy individuals and patients (Figure 2, Supplemental Figure 2, Supplemental Table 2-3). For example, it effectively differentiated the physical health scores, where patients exhibited more severe conditions compared to healthy controls (category datatype 4 vs. 0, respectively, Figure 2B, Supplemental Table 2) and pain scores (continuous datatype ranging from 1(highest)- 0(lowest), mean 0.52±0.24 vs. 0.11±0.12 for patients vs. controls). Though compressing some inherent variance, BioMapAI accurately reconstructed key statistical measures such as the mean and interquartile range (25%-75%), and highlighted the distinctions between healthy and disease. (Figure 2B, Supplemental Figure 2A-B, Supplemental Table 2).

To determine the accuracy of reconstructed clinical scores by BioMapAI's integration of 'omics data, we compared their ability to discriminate ME/CFS patients from controls with the original clinical scores. We used one additional fully connected layer to regress the 12 predicted clinical scores $\hat{Y}(12,)$ into a binary outcome of patient vs. control $\hat{y}(1,)$. Because the diagnosis of ME/CFS relies on clinical interpretation of key symptoms (i.e., the original clinical scores), the original clinical scores have near-perfect accuracy in classification as expected (AUC, Area Under the Curve >99%, Supplemental Figure 2C). Notably, BioMapAI's predicted scores based on the 'omics data achieved a 91% AUC, highlighting its leading-edge accuracy in disease vs. healthy classification (Figure 2D, Supplemental Figure 2D), which was also superior to the performance of three ML models - linear regression (LR), support vector machine (SVM), and gradient boosting (GDBT) - and one deep learning model (DNN) without the hidden 3, 'spread out' layer (Supplemental Table 3). BioMapAI particularly excelled utilizing immune features (AUC = 80%), KEGG genes (78%), blood measure models (71%) and combined 'omics (91%). GDBT, however, led in the microbial species (75%) and metabolome (74%) models, likely due to its emphasis on specific features.

Finally, to assess the robustness of our BioMapAI model, we validated it with independent, published ME/CFS cohorts (Figure 2E, Supplemental Table 4). Using data from two microbiome cohorts, Guo, Cheng et al., 2023 (US)[114] and Raijmakers, Ruud et al., 2020 (Netherlands)[115], BioMapAI achieved 72% and 63% accuracy in species relative abundance and 58% and 60% accuracy in microbial KEGG gene abundance. When applied to two metabolome cohorts, Germain, Arnaud et al., 2022 (US)[116] and Che, Xiaoyu et al., 2022 (US)[117], BioMapAI attained 68% and 59% accuracy. These results were strong given that the metabolomic features only overlap by 79% and 19%, respectively, due to methodological variations.

Importantly, BioMapAI significantly surpassed GDBT and DNN in external cohort validation, supporting our theory that while commonly used models, such as tree-based GDBT, may be effective within a single study, their overemphasis on specific key features can limit its generalizability across different studies, which may not share the same biomarkers. BioMapAI's effectiveness also highlighted the value of incorporating clinical symptoms into a predictive model, proving that connecting 'omics features to clinical symptoms improves disease classification. Given the limitations of using external cohorts – which often have significant methodological differences and cohort characteristics – to validate traditional microbiome and metabolite ML models[118,119,120], BioMapAI represents a breakthrough as a far more adaptable and broadly applicable model.

**'Omics' Strengths Varied in Symptom Prediction; Immune is the Most Predictive**

A major innovation of BioMapAI is its ability to leverage different 'omics data to predict individual clinical scores in addition to disease vs. healthy classification. We evaluated the predictive accuracy by calculating the mean squared error between actual ($y$) and predicted ($\hat{y}$) scores and observed that the different 'omics showed varying strengths in predicting clinical scores (Figure 2C). Immune profiling consistently excelled in forecasting a wide range of symptoms, including pain, fatigue, orthostatic intolerance, and general health perception, underscoring the immune system's crucial role in health regulation. In contrast, blood measurements demonstrated limited predictive ability, except for cognitive efficiency, likely owing to their limited focus on 48 specific blood bioactives. Plasma metabolomics, which encompasses nearly a thousand measurements, performed significantly better with notable correlations with facets of physical health and social activity. These findings corroborate published metabolites and mortality[121,122], longevity[123,124], cognitive function[125], and social interactions[126,127,128]. Microbiome profiles surpassed other 'omics in predicting gastrointestinal abnormalities (as expected[129,130]), emotional well-being, and sleep problems, supporting recently established links in gut-brain health[131,132,133].

**BioMapAI is Explainable, Identifying Disease- and Symptom-Specific Biomarkers**

Deep learning (DL) models are often referred to as 'black box', with limited ability to identify and evaluate specific features that influence the model's predictions. BioMapAI is made explainable by incorporating SHAP values, which quantify how each feature influenced the model's predictions. BioMapAI's architecture – two shared layers ($Z^1$ and $Z^2$) for general disease pattern learning and one parallel layer for each clinical score ($Z^3 = [z_1^3, z_2^3, ..., z_{12}^3]$) – allowed us to identify both disease-specific biomarkers, which are shared across symptoms and models (Supplemental Figure 3, Supplemental Table 5), and symptom-specific biomarkers, which are tailored to each clinical symptom (Figure 3, Supplemental Figure 4-5, Supplemental Table 6).

Disease-specific biomarkers are important features across symptoms and models (Methods, Supplemental Figure 3). Increased B cells (CD19+CD3-), CCR6+ CD8 memory T cells (mCD8+CCR6+CXCR3-), and CD4 naïve T cells (nCD4+FOXP3+) in patients were pivotal for most symptoms, indicating a systemic dysregulation of the adaptive immune response. The species model highlighted the importance of *Dysosmobacteria welbionis*, a gut microbe previously reported in obesity and diabetes, with a critical role in bile acid and butyrate metabolism[134,135]. The metabolome model categorized increased levels of glycodeoxycholate 3-sulfate, a bile acid, and decreased vanillylmandelate (VMA), a catecholamine breakdown product[136]. These critical features for all symptoms were consistently validated across ML and DL models, demonstrating the efficacy of BioMapAI (Supplemental Table 5).

More uniquely, BioMapAI linked 'omics profiles to clinical symptoms and thus enabled the identification of symptom-specific biomarkers (Figure 3A). Certain 'omics data, like species-gastrointestinal and immune-pain associations, were especially effective in predicting specific clinical phenotypes (Figure 2C). Utilizing SHAP, BioMapAI identified distinct sets of biomarkers for each symptom (Supplemental Table 6, Supplemental Figure 5). We found that while disease-specific biomarkers accounted for a substantial portion of the variance, symptom-specific biomarkers crucially refined the predictions, aligned predicted scores – consistently across age and gender – more closely with actual values (Figure 3A-B, Supplemental Figure 4B-D). For example, in the case of pain, CD4 memory and CD1c+ dendritic cells (DC) were particularly important features, and *Faecalibacterium prausnitzii* was uniquely linked as well with varying impact across individual (Figure 3B). Similar to pain, each clinical score in ME/CFS was characterized by its unique 'omics features, distinct from those common across other symptoms (Supplemental Table 6).

In addition, we observed a spectrum of interaction types (linear, biphasic, and dispersed) extending beyond conventional linear interactions, underscoring the heterogeneity inherent in ME/CFS (Figure 3C). High-abundance species and immune cells often had a biphasic relationship with symptoms, showing dual effects, while low-abundance species and metabolites displayed a linear relationship with positive or negative associations with clinical scores (Supplemental Figure 5).

An example of a relatively straightforward monotonic (linear) relationship was observed between CD4 memory (CD4 M) cells, CD1c+ DCs and pain, with positive contributions of CD4 M cells to pain intensity severity. Conversely,  CD1c+ DCs contributed negatively to pain severity in both patients and control (Figure 3C, E). These variations suggest alterations in inflammatory responses and specific pathogenic processes in ME/CFS, which may be virally triggered and is marked by prolonged infection symptoms. Many microbial biomarkers demonstrated linear contributions to symptoms, evidenced by numerous negative peaks indicating their beneficial role in symptom reduction (Figure 3A). For example, *Dysosmobacteria welbionis*, a disease-specific biomarker, exacerbated sleeping and gastrointestinal issues (Supplemental Figure 3), whereas *Clostridium sp.* and *Alistipes communis* alleviated these issues (Figure 3A, Supplemental Figure 5B).

A more complex, biphasic relationship was observed in the interaction of *Faecalibacterium prausnitzii* with pain, whose saddle curve (Figure 3C) and mixture of positive and negative contribution peaks (Figure 3B) revealed how abnormal low and high abundances could be associated with amplified pain. In disease, *F. prausnitzii* was associated with exacerbated pain, while in healthy individuals, it appeared to mitigate pain (Figure 3D). *F. prausnitzii* was identified as a biomarker in several ME/CFS cohorts[114],[115],[137], but also has been implicated in numerous anti-inflammatory effects[138],[139],[140],[141]. Here notably, BioMapAI elaborated its role at ME/CFS by recognizing its potential dual contribution to symptom severity. Similar biphasic relationships were observed for plasma metabolomics biomarkers, glucuronide and glutamine, in relation to pain (Figure 3C).

Distinct from other 'omics features, KEGG genes exhibited sparse and dispersed contributions (Figure 3C, Supplemental Figure 4C). The vast feature matrix of KEGG models complicated the identification of a universal biomarker for any single symptom, as individuals possessed distinct symptom-specific KEGG biomarkers. For example, the gene FNR, an anaerobic regulatory protein transcription factor, negatively impacted pain but was active in only a small portion of patients, with the majority showing no significant impact (Figure 3C). This pattern was consistent for other KEGG biomarkers, which contributed sparsely to symptom severity (Supplemental Figures 4C).

Taken together, BioMapAI achieved a comprehensive mapping of the intricate nature of symptom-specific biomarkers to clinical phenotypes that has been inaccessible to single models to date. Our models unveil a nuanced and precise correlation between 'omics features and disease symptomology, emphasizing ME/CFS' complex etiology and consequent disease management approaches.

**Healthy Microbiome-Immune-Metabolome Networks are Dysbiotic in ME/CFS**

BioMapAI elucidated that each 'omics layer provided distinct insights into the disease symptoms and influenced host phenotypes in a dynamic and complex manner. To examine crosstalk between 'omics layers, we modeled co-expression modules for each 'omics using weighted gene co-expression network analysis (WGCNA), identifying seven microbial species, six microbial gene set, nine metabolome, and nine immune clusters (Methods, Supplemental Table 7). Observing significant associations of these modules with disease classification (microbial modules), age and gender (immune and metabolome modules) (Supplemental Figure 6A), we first established baseline networks of inter-'omics interactions in healthy individuals as a function of these and other clinical covariates such as age, weight, and gender (Figure 4A), and then examined how these interactions were altered in patient populations (Figure 4B, Supplemental Figure 6B-C).

Healthy control-derived host-microbiome interactions, such as the microbial pyruvate module interacting with multiple immune modules, and connections between commensal gut microbes (*Prevotella*, *Clostridia* sp., *Ruminococcaceae*) with Th17 memory cells, plasma steroids, phospholipids, and tocopherol (vitamin E) (Figure 4A), were disrupted in ME/CFS patients. Increased interactions between gut microbiome and mucosal/inflammatory immune modules, including CD8+ MAIT, and INFg+ CD4 memory cells, suggested a microbiome-mediated intensified inflammatory in ME/CFS (Supplemental Figure 6D). Young, female, and normal-weight patients shared those changes, while male patients showed more distinct alterations in the interplay between microbial and plasma metabolites. Elderly and overweight patients had more interaction abnormalities than other subgroups, with specific increases between *Blautia*, *Flavonifractor*, *Firmicutes* sp. linked with TNFα cytotoxic T cells and plasma plasmalogen, and decreased interactions between *Lachnospiraceae* sp. with Th17 cells (Figure 4B).

Further examining the pyruvate hub as well as several other key microbial modules whose networks were dysbiotic in patients, we mapped the interactions of their metabolic subpathways to plasma metabolites and immune cells and detailed the collective contributions to host phenotypes (Figure 4C, Supplemental Table 8). We further validated these findings with two independent cohorts (Guo 2023[114] and Raijmakers 2020[115]). For example, increased tryptophan metabolism, linked to gastrointestinal issues, lost its inhibitory effect on Th22 cells, and gained interactions with γδ T cells and the secretion of INFg and GzA from CD8 and CD8+ MAIT cells. Several networks linked with emotional dysregulation and fatigue – again underscoring the gut-brain axis[133] – differed significantly in patients vs. controls, including decreased butyrate production - especially from the pyruvate[142] and glutarate[143] sub-pathways- and branched-chain amino acid (BCAA) biosynthesis, which lost or reversed their interactions with Th17, Treg cells, and plasma lipids while gaining interactions with inflammatory immune cells including γδ T and CD8+ MAIT cells in patients; and increased microbial benzoate, synthesized by *Clostridia* sp.[144],[145] then converted to hippurate in the liver[146],[147], showed a strong positive correlation with plasma hippurate in long-term ME/CFS patients, supporting enhanced pathway activity in later stages of the disease. This change altered its interactions with numerous plasma metabolites, including steroids, phenols, BCAAs, fatty acids, and vitamins B5 and B6. Finally, we noted that connections of short-term patients often resembled a transitional phase, with dysbiotic health-associated networks and emergent pathological connections that solidified in long-term ME/CFS patients.

Based on BioMapAI's outputs and network analyses, we propose that the shift in disease pathology in ME/CFS is linked to the topological interaction of the gut microbiome, immune function, and metabolome. (Figure 5). A decrease in key microbes, including *Faecalibacterium prausnitzii*, and resultant dysfunction of microbial metabolic pathways such as butyrate, tryptophan, and BCAA, contributed to critical ME/CFS phenotypes, particularly pain and gastrointestinal abnormalities. In healthy individuals, these microbial metabolites regulate mucosal immune cells, including Th17, Th22, and Treg cells, an interaction that is dysfunctional in ME/CFS resulting in elevated pro-inflammatory interactions via elevated activation of γδ T cells and CD8 MAIT cells with the secretion of INFg and GzA, particularly impacting health perception and social activities. Additional health-associated networks between gut microbial metabolites, particularly benzoate, with plasma metabolites such as lipids, GPE, fatty acids, and bile acids, were weakened or reversed in ME/CFS. This breakdown in the host metabolic-microbiome balance were collectively associated with fatigue, emotional and sleeping problems, supporting recent findings underscoring microbial mechanisms in the gut-brain axis that occur via modulation of plasma metabolites[148],[149],[150].

**Figure**

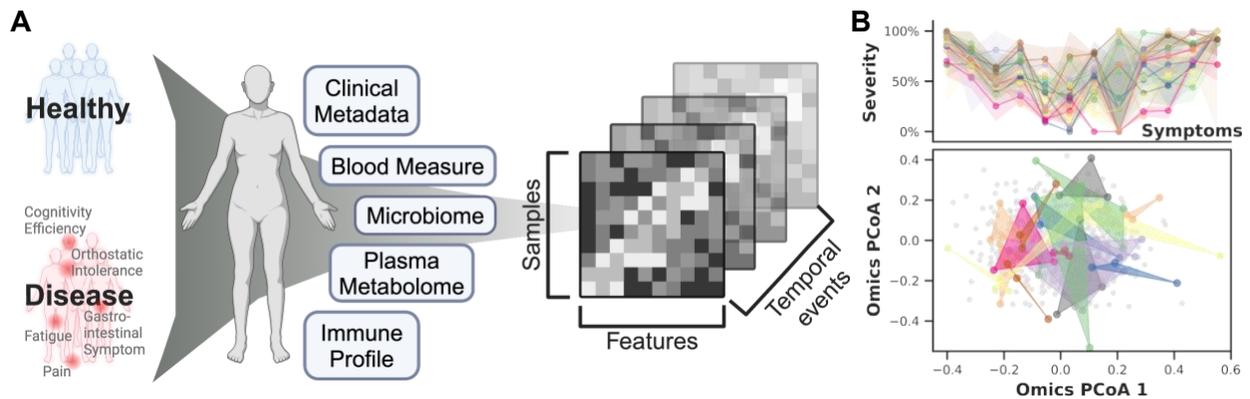

**Figure 1: Cohort Summary and Heterogeneity of ME/CFS. A) Cohort Design and 'Omics Profiling.** 96 healthy donors and 153 ME/CFS patients were followed over 3-4 years with yearly sampling. Clinical metadata including lifestyle and dietary surveys, blood clinical laboratory measures (N=503), gut microbiome (N=479), plasma metabolome (N=414), and immune profiles (N=489) were collected (Supplemental Table 1 and Supplemental Figure 1A). **B) Heterogeneity and Non-Linear Progression of ME/CFS in Symptom Severity and 'Omics Profiles.** Variability in symptom severity (top) and 'omics profiles (bottom) for 20 representative ME/CFS patients over 3-4 time points. For symptom severity, the 12 major clinical symptoms (x-axis) vs. severity (scaled from 0% to 100%, y-axis) is shown for each patient (each color), with lines showing average severity and shaded areas showing severity range over their timepoints. The widespread highlights the lack of consistent temporal patterns and unique symptomatology of ME/CFS (controls shown in Supplemental Figure 1C). Bottom, PCoA of integrated 'omics data with color dots matching patient timepoints in the symptom plot and grey dots representing the entire cohort. Again, the spread and overlap of the colored space reflect the diversity in 'omics signatures vs. the more consistent pattern typical of controls (Supplemental Figure 1B). **Abbreviations:** ME/CFS, Myalgic Encephalomyelitis/Chronic Fatigue Syndrome; PCoA, Principal Coordinates Analysis. **Supporting Materials:** Supplemental Table 1, Supplemental Figure 1.

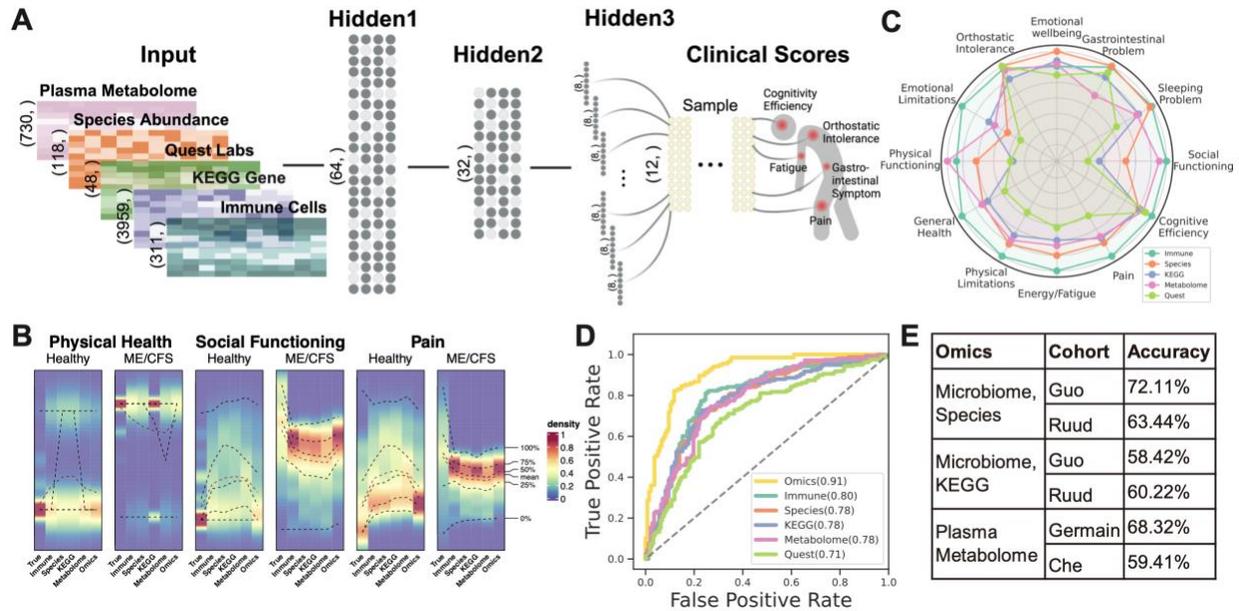

**Figure 2: BioMapAI's Model Structure and Performance. A) Structure of BioMapAI.** BioMapAI is a fully connected deep neural network comprised of an input layer ($X$), a normalization layer (not shown), three sequential hidden layers ($Z^1, Z^2, Z^3$), and one output layer ($Y$). Hidden layer 1 ($Z^1$, 64 nodes) and hidden layer 2 ($Z^2$, 32 nodes), both feature a dropout ratio of 50% to prevent overfitting (visually represented by dark and light gray nodes). Hidden layer 3 has 12 parallel sub-layers each with 8 nodes ($Z^3 = [z_1^3, z_2^3, \ldots, z_{12}^3]$) to learn 12 objects in the output layer ($Y = [y_1, y_2, \ldots, y_{12}]$) representing key clinical symptoms of ME/CFS. **B) True vs. Predicted Clinical Scores highlight BioMapAI's accuracy.** Three example density maps (full set, Supplemental Figure 2A) compare the true score, $y$ (Column 1) against BioMapAI's predictions generated from different 'omics profiles - $\hat{y}_{immune}$, $\hat{y}_{species}$, $\hat{y}_{KEGG}$, $\hat{y}_{metabolome}$, $\hat{y}_{omics}$ (Columns 2-6). The color gradient from blue (lower density) to red (higher density) illustrates the occurrence frequency (e.g., true scores for ~100% of healthy controls' physical health ~ 0 = red), with dashed lines indicating key statistical percentiles (100%, 75%, 50%, 25%, and 0%). Note that model's predicted scores a preserve differences between healthy controls and patients for these three examples, irrespective of 'omics type. **C) 'Omics' Strengths in Symptom Prediction.** Radar plot shows BioMapAI's performance in predicting the 12 clinical outcomes for each 'omics datatype. Each of the 12 axes represents a clinical score output ($Y = [y_1, y_2, \ldots, y_{12}]$), with five colors denoting the 'omics datasets used for model training. The spread of each color along an axis reflects the normalized mean square error (MSE, Supplemental Table 2) between the actual, $y$, and the predicted, $\hat{y}$, outputs, illustrating the predictive strength or weakness of each 'omics for specific clinical scores. For instance, species abundance predicted gastrointestinal, emotional, and sleep issues effectively, while the immune profile was broadly accurate across most scores. **D) BioMapAI's Performance in Healthy vs. Disease Classification.** ROC curves show BioMapAI's performance in disease classification using each 'omics dataset separately or combined ('Omics'), with the AUC in parentheses showing prediction accuracy (full report in Supplemental Table 3). **E) Validation of BioMapAI with External Cohorts.** External cohorts with microbiome data (Guo et al.[114], Ruud et al.[115]) and metabolome data (Germain et al.[116], Che et al.[32]) were used to test BioMapAI's model, underscoring its generalizability (detailed classification matrix, Supplemental Table 4). **Abbreviations:** KEGG, Kyoto Encyclopedia of Genes and Genomes; 'Omics' refers to the combined multi-'omics matrix; MSE, Mean Square Error; ROC curve, Receiver Operating Characteristic curve; AUC**,** Area Under the Curve; $y$, True Score; $\hat{y}$, Predicted Score. **Supporting Materials:** Supplemental Tables 2-4, Supplemental Figures 1-2.

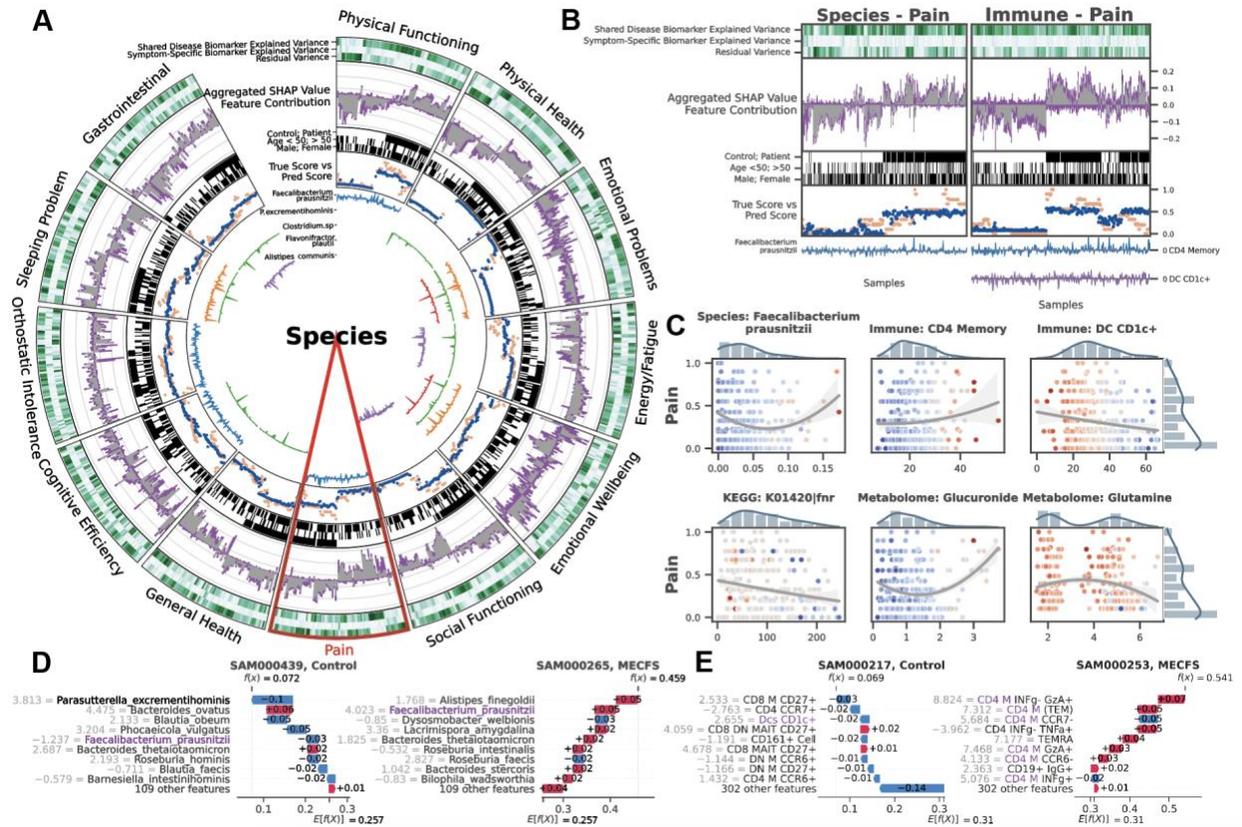

**Figure 3: BioMapAI Identifies both Disease- and Symptom-Specific Biomarkers. For Symptom-Specific Biomarkers, A) Circularized Diagram of Species Model with B) Zoomed Segment for Pain.** Each circular panel illustrates how the model predicts each of the 12 symptom-specific biomarkers derived from one type of 'omics data (all datatypes shown in Supplemental Figure 4). The x-axis for each panel represents an individual's values for each of the following contributors to the model's performance (from top to bottom): *1. Variance Explained by Biomarker Categories:* Gradients of dark green (100%) to white (0%) show variance explained by the model. For many biomarkers, disease-specific biomarkers account for the greatest proportion of variance, and symptom-specific biomarkers provide additional tailored explanations, with residual accounting for the remaining variance; *2. Aggregated SHAP Values* quantify the contribution of each feature to the model's predictions, with disease-specific biomarkers in grey and symptom-specific in purple. *3. Demography and Cohort Classification:* cohort (controls, white vs. patients, black); age <50 (white) vs. >50 years

old (black); sex (male, white vs. female, black); *4. True vs. Predicted Scores* show BioMapAI's predictive performance at the individual sample level, with true in blue and model-predicted scores in orange; *5. Examples of Symptom-Specific Biomarkers:* Line graphs show the contribution of select symptom-specific biomarkers to the model across individuals, e.g., 5 gut species in A). In B), the three features most specific to the pain model include gut microbe *F. prausnitzii,* CD4 memory T, and DC CD1c+ cells. Peaks above 0 (middle line) indicate a positive contribution and below 0 for a negative contribution. For example, the mixed positive and negative contribution peaks of *F. prausnitzii* indicated a biphasic contribution to pain intensity. Disease-Specific Biomarkers are shown in Supplemental Figure 3. **C) Different Correlation Patterns of Biomarkers to Symptoms:** For pain (other symptoms in Supplemental Figure 5), correlation analysis of raw abundance (x-axis) of each biomarker with pain score (y-axis) show monotonic (e.g., CD4 memory and DC CD1c+ markers), biphasic (microbial and metabolomic markers), or sparse (KEGG genes) contribution patterns for those features. Dots represent an individual color-coded to SHAP value, where the color spectrum indicates negative (blue) to neutral (grey) to positive (red) contributions to pain prediction. Superimposed trend lines with shaded error bands represents the predicted correlation trends between biomarkers and pain intensity. Adjacent bar plots represent the data distribution. **D-E) Examples of Pain-Specific Biomarkers' Contributions.** SHAP waterfall plots (colors corresponding to gradient in C) illustrate the contribution of individual features to a model's predictive output. The top 10 features for two pairs of controls and patients are shown here, illustrating the species and the immune model (additional examples in Supplemental Figure 4A). The contribution of each feature is shown as a step (SHAP values provided adjacent), and the cumulative effect of all the steps provides the final prediction value, $E[f(X)]$. Our example of *F. prausnitzii* exhibits a protective role (negative SHAP) in controls but exacerbates pain (positive SHAP) in patients – consistent with the biphasic relationship observed in C). As a second example, all CD4 memory cells in this model have positive SHAP values, reinforcing the positive monotonic relationship with pain severity observed in C). Conversely, DC CD1c+ cells contribute negatively and thus may have a protective role. **Abbreviation:** SHAP, SHapley Additive exPlanations; DNN, Deep Neuron Network; GBDT, Gradient

Boosting Decision Tree; KEGG, Kyoto Encyclopedia of Genes and Genomes.

**Supporting Materials:** Supplemental Table 5-6, Supplemental Figure 3-5.

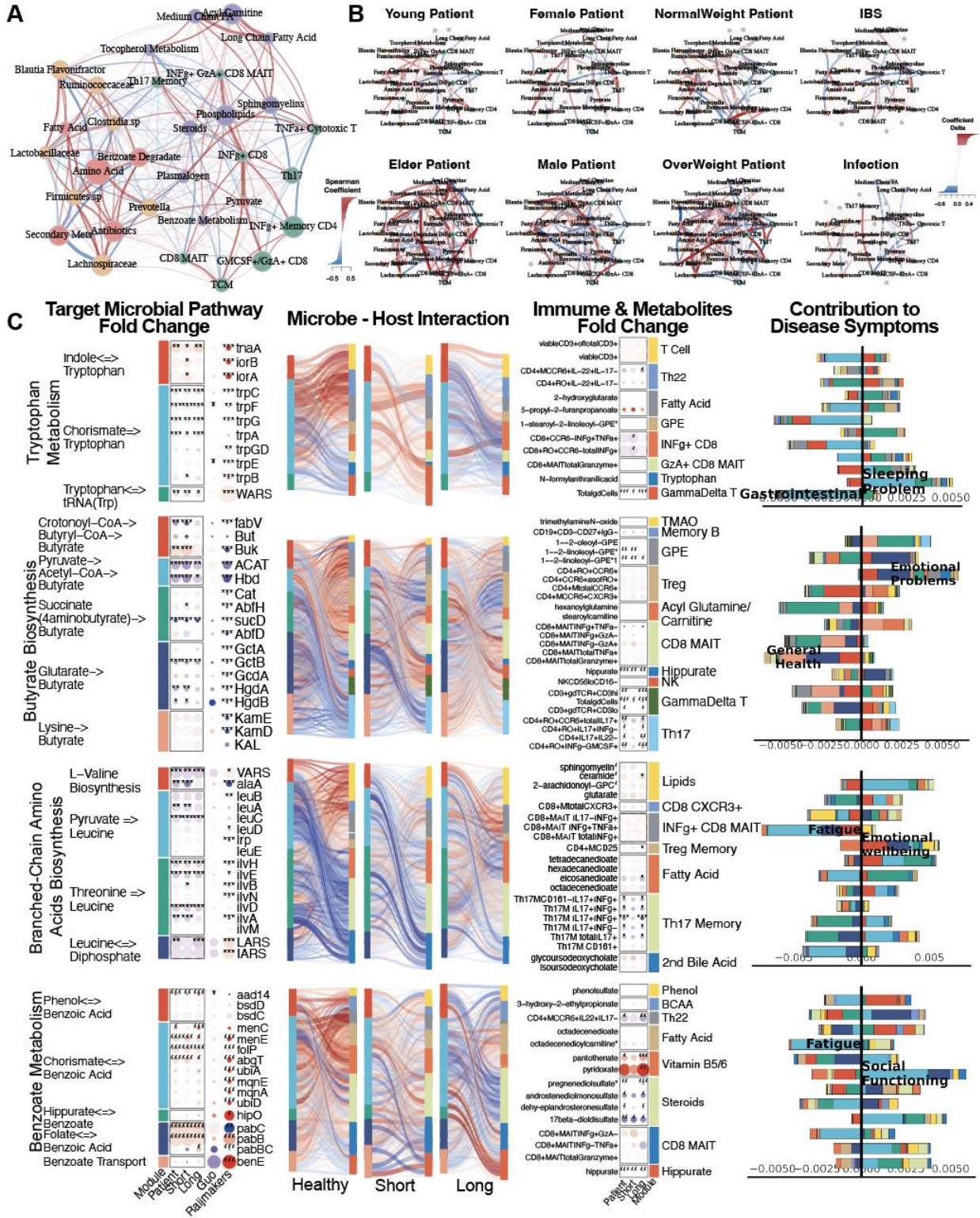

**Figure 4: Microbiome-Immune-Metabolome Crosstalk is Dysbiotic in ME/CFS. A-B) Microbiome-Immune-Metabolome Network in A) Healthy and B) Patient Subgroups.** A baseline network was established with 200+ healthy control samples (**A**), bifurcating into two segments: the gut microbiome (species in yellow, genetic modules in orange) and blood elements (immune modules in green, metabolome modules in purple). Nodes: modules; size: # of members; colors: 'omics type; edges: interactions between modules, with Spearman coefficient (adjusted) represented by thickness, transparency, and color - positive (red) and negative (blue). Here, key microbial pathways (pyruvate, amino acid, and benzoate) interact with immune and metabolome modules in healthy individuals. Specifically, these correlations were disrupted in patient subgroups (**B**), as a function of gender, age (young <26 years old vs. older >50), BMI (normal <26 vs. overweight >26), and health status (individuals with IBS or infections). Correlations significantly shifted from healthy counterparts (Supplemental Figure 6C) are highlighted with colored nodes and edges indicating increased (red) or decreased (blue) interactions. **C) Targeted Microbial Pathways and Host Interactions.** Four important microbial metabolic mechanisms (tryptophan, butyrate, BCAA, benzoate) were further analyzed to compare control, short and long-term ME/CFS patients, and external cohorts for validation (Guo[114] and Raijmakers[115]). *1. Microbial Pathway Fold Change:* Key genes were grouped and annotated in subpathways. Circle size: fold change over control; color: increase (red) or decrease (blue), p-values (adjusted Wilcoxon) marked. *2. Microbiome-Host Interactions:* Sankey diagrams visualize interactions between microbial pathways and host immune cells/metabolites. Line thickness and transparency: Spearman coefficient (adjusted); color: red (positive), blue (negative). *3. Immune & Metabolites Fold Change:* Pathway-correlated immune cells and metabolites are grouped by category. *4. Contribution to Disease Symptoms:* Stacked bar plots show accumulated SHAP values (contributions to symptom severity) for each disease symptom (1-12, as in Supplemental Table 1). Colors: microbial subpathways and immune/metabolome categories match module color in fold change maps. X-axis: accumulated SHAP values (contributions) from negative to positive, with the most contributed symptoms highlighted. **P-values:** *p < 0.05, **p < 0.01, ***p < 0.001. **Abbreviations:** IBS, Irritable Bowel Syndrome; BMI, Body Mass Index; BCAA,

Branched-Chain Amino Acids; MAIT, Mucosal-Associated Invariant T cell; SHAP, SHapley Additive exPlanations; GPE, Glycerophosphoethanolamine; INFγ, Interferon Gamma; CD, Cluster of Differentiation; Th, T helper cell; TMAO, Trimethylamine N-oxide; KEGG, Kyoto Encyclopedia of Genes and Genomes. **Supporting Materials:** Supplemental Table 7-8, Supplemental Figure 6.

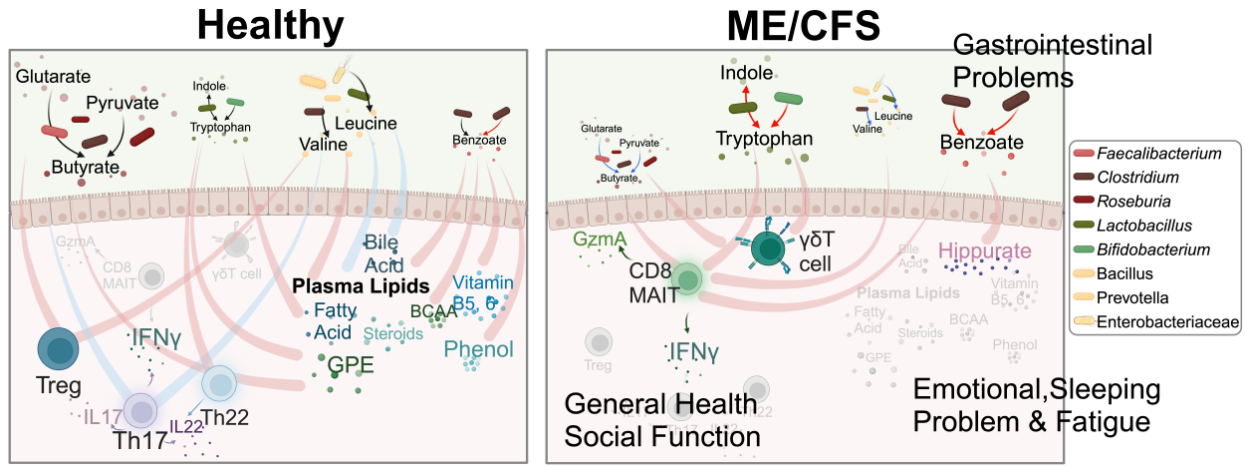

**Figure 5: Overview of Dysbiotic Host-Microbiome Interactions in ME/CFS.** This conceptual diagram visualizes the host-microbiome interactions in healthy conditions (left) and its disruption and transition into the disease state in ME/CFS (right). The base icons of the figure remain consistent, while gradients and changes in color and size visually represent the progression of the disease. Process of production and processing is represented by lines with arrows, where the color indicates an increase (red) or decrease (blue) in the pathway in disease; lines without arrows indicate correlations, with red representing positive and blue representing negative correlations. In healthy conditions, microbial metabolites support immune regulation, maintaining mucosal integrity and healthy inflammatory responses by positively regulating Treg and Th22 cell activity, and controlling Th17 activities, including the secretion of IL17 (purple cells), IL22 (blue), and IFNγ. These microbial metabolites also maintain many positive interactions with plasma metabolites like lipids, bile acids, vitamins, and phenols. In ME/CFS, there is a significant decrease in beneficial microbes and a disruption in metabolic pathways, marked by a decrease in the butyrate (brown-red dots) and BCAA (yellow) pathways and an increase in tryptophan (green) and benzoate (red) pathways. These changes are linked to gastrointestinal issues. In ME/CFS, the regulatory capacity of the immune system diminishes, leading to the loss of health-associated interactions with Th17, Th22, and Treg cells, and an increase in inflammatory immune activity. Pathogenic immune cells, including CD8 MAIT and γδT cells, show increased activity, along with the secretion of inflammatory cytokines such as IFNγ and GzmA, contributing

to worsened general health and social functioning. Healthy interactions between gut microbial metabolites and plasma metabolites weaken or even reverse in the disease state. A notable strong connection increased in ME/CFS is benzoate transformation to hippurate, associated with emotional disturbances, sleep issues, and fatigue.

**Abbreviations:** IFNγ, Interferon gamma; Th17, T helper 17 cells; Th22, T helper 22 cells; Treg, Regulatory T cells; GzmA, Granzyme A; MAIT, Mucosa-Associated Invariant T cells; γδT, Gamma delta T cells; BCAA, Branched-Chain Amino Acids; GPE, Glycerophosphoethanolamine.

**Methods**

**Study Design.** This was 4-year prospective study. All participants had a physical examination at the baseline visit that included evaluation of vital signs, BMI, orthostatic vital signs, skin, lymphatic system, HEENT, pulmonary, cardiac, abdomen, musculoskeletal, nervous system and fibromyalgia (FM) tender points. We enrolled a total of 153 ME/CFS patients (of which 75 had been diagnosed with ME/CFS <4 years before recruitment and 78 had been diagnosed with ME/CFS >10 years before recruitment) and 96 healthy controls. Among them, 110 patients and 58 healthy controls were followed one year after the recruitment as timepoint 2; 81 patients and 13 healthy controls were followed two years after the recruitment as timepoint 3; and 4 patients were followed four years after the recruitment as timepoint 4. Subject characteristics are shown in Supplemental Table 1 and Supplemental Figure 1A.

Medical history and concomitant medications were documented. Blood samples were obtained prior to orthostatic and cognitive testing. The 10-minute NASA Lean Test and cognitive testing were conducted after the physical examination and blood draw[151]. Cognitive efficiency was tested with the DANA Brain Vital, measuring three reaction time and information processing measurements[152]. The orthostatic challenge was assessed with the 10-minute NASA Lean Test (NLT). Participants rested supine for 10 minutes, and baseline blood pressure (BP) and heart rate (HR) were measured twice during the last 2 minutes of rest[153].

Participants were provided with an at-home stool collection kit at the end of each in-person visit. The following questionnaires were completed at baseline: DePaul Symptom Questionnaire (DSQ), Post-Exertional Fatigue Questionnaire, RAND-36, Fibromyalgia Impact Questionnaire-R, ACR 2010 Fibromyalgia Criteria Symptom Questionnaire, Pittsburgh Sleep Quality Index (PSQI), Stanford Brief Activity Survey, Orthostatic Intolerance Daily Activity Scale, Orthostatic Intolerance Symptom Assessment, Brief Wellness Survey, Hours of Upright Activity (HUA), medical history and family history. All but medical history and family history were administered again when participants came for their annual visit.

Approval was received before enrolling any subjects in the study (The Jackson Laboratory Institutional Review Board, 17-JGM-13). All participants were educated about the study prior to enrollment and signed all appropriate informed consent documents. Research staff followed Good Clinical Practices (GCP) guidelines to ensure subject safety and privacy.

**ME/CFS Cohort**. Beginning in January 2018, we enrolled ME/CFS patients who had been sick for <4 years or sick for >10 years. No ME/CFS patients with duration ≥4 years and ≤10 years were enrolled in order to have clear distinctions between short and long duration of illness with ME/CFS. All participants were 18 to 65 years old at the time of enrollment. ME/CFS diagnosis according to the Institute of Medicine clinical diagnostic criteria and disease duration of <4 years were confirmed during clinical differential diagnosis and thorough medical work up[154]. Additional inclusion criteria required, 1) a substantial reduction or impairment in the ability to engage in pre-illness levels of occupational, educational, social, or personal activities that persists for more than 6 months and less than 4 years and is accompanied by fatigue, which is often profound, is of new or definite onset (not lifelong), is not the result of ongoing excessive exertion, and is not substantially alleviated by rest, and 2) post-exertional malaise. Exclusionary criteria for the <4 year ME/CFS cohort were, 1) morbid obesity BMI>40, 2) other active and untreated disease processes that explain most of the major symptoms of fatigue, sleep disturbance, pain, and cognitive dysfunction, 3) untreated primary sleep disorders, 4) rheumatological disorders, 5) immune disorders, 6) neurological disorders, 7) infectious diseases, 8) psychiatric disorders that alter perception of reality or ability to communicate clearly or impair physical health and function, 9) laboratory testing or imaging are available that support an alternate exclusionary diagnosis, and 10) treatment with short-term (less than 2 weeks) antiviral or antibiotic medication within the past 30 days.

For the >10 year ME/CFS cohort, disease duration of >10 year and clinical criteria was confirmed to meet the Institute of Medicine criteria for ME/CFS during clinical evaluation and medical history review[154]. Other than disease duration, inclusion and exclusion criteria were the same as for <4 year ME/CFS cohort

**Healthy Control Cohort**. Healthy control participants were also between 18 to 65 years of age and in general good health. Enrollment began in 2018 and subjects were selected to match the <4 year ME/CFS cohort by age (within 5 years), race, and sex (~2:1 female to male ratio). Exclusion criteria for healthy controls included, 1) a diagnosis or history of ME/CFS, 2) morbid obesity BMI>40, 3) treatment with short-term (less than 2 weeks) antiviral or antibiotic medication within the past 30 days or 4) treatment long-term (longer than 2 weeks) antiviral medication or immunomodulatory medications within the past 6 months.

**Clinical Metadata and Scores.** Clinical symptoms and baseline health status were assessed on the day of physical examination and biological sample collection for both case and control subjects. For each participant, we collected demographic information (including age, gender, diet, race, BMI, family, work, and education), medical histories, clinical tests and questionnaires. From questionnaires and test as described above, we summarized 12 clinical scores to cover major symptoms of ME/CFS: Scores 1-8 were derived from the RAND36, following standardized rules [155] and summarized into eight categories: Physical Functioning (also referred to as Daily Activity in the main contents), Role Limitations due to Physical Health (Physical Limitations), Role Limitations due to Emotional Problems (Emotional Problems), Energy/Fatigue, Emotional Wellbeing (Mental Health), Social Functioning (Social Activity), Pain, and General Health (Health Perception). Cognitive Efficiency was summarized from the DANA Brain Vital test, Orthostatic Intolerance from the NLT test, Sleeping Problem Score from the Pittsburgh Sleep Quality Index (PSQI) questionnaire, and Gastrointestinal Problems Score from the Gastrointestinal Symptom Rating Scale (GSRS) questionnaire. Each score was transformed into a 0–1 scale to facilitate combination and comparison, where a score of 1 indicates maximum disability or severity and a score of 0 indicates no disability or disturbance.

**Plasma Sample collection and Preparation.** Healthy and patient blood samples were obtained from Bateman Horne Center, Salt Lake City, UT and approved by JAX IRB. One 4 mL lavender top tube (K2EDTA) was collected, and tube slowly inverted 8-10 times immediately after collection. Blood was centrifuged within 30 minutes of collection at 1000 x g with low brake for 10 minutes. 250 uL of plasma was transferred into three 1 mL cryovial tubes, and tubes were frozen upright at -80°C. Frozen plasma samples were batch shipped overnight on dry ice to The Jackson Laboratory, Farmington, CT, and stored at -80°C. One green top tube (Heparin) was collected, and tube slowly inverted 8-10 times immediately after collection. Heparinized blood samples were shipped overnight at room temperature. Peripheral blood mononuclear cells (PBMC) were isolated using Ficoll-paque plus (GE Healthcare) and cryopreserved in liquid nitrogen.

**Plasma untargeted metabolome by UPLC-MS/MS.** Plasma samples were sent to Metabolon platform and processed by Ultrahigh Performance Liquid Chromatography-Tandem Mass Spectroscopy (UPLC-MS/MS) following the CFS cohort pipeline. In brief, samples were prepared using the automated MicroLab STAR® system from Hamilton Company. The extract was divided into five fractions: two for analysis by two separate reverse phases (RP)/UPLC-MS/MS methods with positive ion mode electrospray ionization (ESI), one for analysis by RP/UPLC-MS/MS with negative ion mode ESI, one for analysis by HILIC/UPLC-MS/MS with negative ion mode ESI, and one sample was reserved for backup. QA/QC were analyzed with several types of controls were analyzed including a pooled matrix sample generated by taking a small volume of each experimental sample (or alternatively, use of a pool of well-characterized human plasma), extracted water samples, and a cocktail of QC standards that were carefully chosen not to interfere with the measurement of endogenous compounds were spiked into every analyzed sample, allowed instrument performance monitoring, and aided chromatographic alignment. Compounds were identified by comparison to Metabolon library entries of purified standards or recurrent unknown entities. The output raw data included the annotations and the value of peaks quantified using area-under-the-curve for metabolites

**Immune Profiling: Flow Cytometry Analysis.** Frozen PBMC aliquots were thawed, counted and divided into two parts, one part for day 0 surface staining, and the other part cultured in complete RPMI 1640 medium (RPMI plus 10% Fetal Bovine Serum (FBS, Atlanta Biologicals) and 1% penicillin/streptomycin (Corning Cellgro) supplemented with IL-2+IL15 (20ng/ml) for Treg subsets day 1 surface and transcription factors staining after culture with IL-7 (20ng/ml) for day 1 and day 6 intracellular cytokine staining, and a combination of cytokines (20ng/ml IL-12, 20ng/ml IL-15, and 40ng/ml IL-18) for day 1 intracellular cytokine staining (IL-12 from R&D, IL-7 and IL-15 from Biolegend). Surface staining was performed in staining buffer containing PBS + 2% FBS for 30 minutes at 4°C. When staining for chemokine receptors the incubation was done at room temperature. Antibodies used in the surface staining are 2B4, CD1c, CD14, CD16, CD19, CD25, CD27, CD31, CD3, CD303, CD38, CD4, CD45RO, CD56, CD8, CD95, CD161, CCR4, CCR6, CCR7, CX3CR1, CXCR3, CXCR5, γδ TCR bio, HLA-DR, IgG, IgM, LAG3, PD-1, TIM3, Va7.2, Va24Ja18 all were obtained from Biolegend.

For intracellular cytokine staining, cells were stimulated with PMA (40ng/ml for overnight cultured cells and 20ng/ml for 6 days cultured cells) and Ionomycin (500ng/ml) (both from Sigma-Aldrich) in the presence of GolgiStop (BD Biosciences) for 4 hours at 37°C. For cytokine secretion after stimulation with IL-12+IL-15+IL-18, GolgiStop was added to the culture on day 1 for 4 hours. For intracellular cytokine and transcription factor staining, PMA+Ionomycin stimulated cells of unstimulated cells were collected, stained with surface markers including CD3, CD4, CD8, CD161, PD1, 2B4, Va7.2, CD45RO, CCR6, and CD27 followed by one wash with PBS (Phosphate buffer Saline) and staining with fixable viability dye (eBioscience). After surface staining, cells were fixed and permeabilized using fixation/permeabilization buffers (eBioscience) according to the manufacturer's instruction. Permeabilized cells were then stained for intracellular

FOXP3, Helios, IL-4, IFNγ, TNFα, IL-17A, IL-22, Granzyme A, GM-CSF, and Perforin from Biolegend. Flow cytometry analysis was performed on Cytek Aurora (Cytek Biosciences) and analyzed using FlowJo (Tree Star).

**Fecal Sample Collection and DNA Extraction.** Stool was self-collected at home by volunteers using a BioCollector fecal collection kit (The BioCollective, Denver, CO) according to manufacturer instructions for preservation for sequencing prior to sending the sample in a provided Styrofoam container with a cold pack. Upon receipt, stool and OMNIgene samples were immediately aliquoted and frozen at –80°C for storage. Prior to aliquoting, OMNIgene stool samples were homogenized by vortexing (using the metal bead inside the OMNIgene tube), then divided into 2 microfuge tubes, one with 100µL aliquot and one with 1mL. DNA was extracted using the Qiagen (Germantown, MD, USA) QIAamp 96 DNA QIAcube HT Kit with the following modifications: enzymatic digestion with 50µg of lysozyme (Sigma, St. Louis, MO, USA) and 5U each of lysostaphin and mutanolysin (Sigma) for 30 min at 37 °C followed by bead-beating with 50 µg 0.1 mm of zirconium beads for 6 min on the Tissuelyzer II (Qiagen) prior to loading onto the Qiacube HT. DNA concentration was measured using the Qubit high sensitivity dsDNA kit (Invitrogen, Carlsbad, CA, USA).

**Metagenomic Shotgun Sequencing.** Approximately 50µL of thawed OMNIgene preserved stool sample was added to a microfuge tube containing 350 µL Tissue and Cell lysis buffer and 100 µg 0.1 mm zirconia beads. Metagenomic DNA was extracted using the QiaAmp 96 DNA QiaCube HT kit (Qiagen, 5331) with the following modifications: each sample was digested with 5µL of Lysozyme (10 mg/mL, Sigma-Aldrich, L6876), 1µL Lysostaphin (5000U/mL, Sigma-Aldrich, L9043) and 1µL oh Mutanolysin (5000U/mL, Sigma-Aldrich, M9901) were added to each sample to digest at 37°C for 30 minutes prior to the bead-beating in the in the TissueLyser II (Qiagen) for 2 x 3 minutes at 30 Hz. Each sample was centrifuged for 1 minute at 15000 x g prior to loading 200µl into an S-block (Qiagen, 19585) Negative (environmental) controls and positive (in-house mock community of 26 unique species) controls were extracted and sequenced with each extraction and library preparation batch to ensure sample integrity.

Pooled libraries were sequenced over 13 sequencing runs using both HiSeq (N=87) and NovaSeq (N=392) platforms. To address potential biases arising from varying read depths, all samples were down-sampled, using seqtk[156] (v1.3-r106), to 5 million reads. This threshold corresponds to the 95th percentile of the read count distribution across the dataset.

Sequencing adapters and low-quality bases were removed from the metagenomic reads using scythe (v0.994) and sickle (v1.33), respectively, with default parameters. Host reads were removed by mapping all sequencing reads to the hg19 human reference genome using Bowtie2 (v2.3.1), under 'very-sensitive' mode. Unmapped reads (i.e., microbial reads) were used to estimate the relative abundance profiles of the microbial species in the samples using MetaPhlAn4.

**Taxonomic Profiling (Specie Abundance) and KEGG Gene Profiling.** Taxonomic compositions were profiled using Metaphlan4.0[157] and the species whose average relative abundance > 1e-4 were kept for further analysis, giving 384 species. The gene profiling was computed with USEARCH[158] (v8.0.15) (with parameters: evalue 1e-9, accel 0.5, top_hits_only) to KEGG Orthology (KO) database v54, giving a total of 9452 annotated KEGG genes. The reads count profile was normalized by DeSeq2[159] in R. Genes with a prevalence of over 20% were selected for downstream analysis.

**Confounder Analysis.** Confounder analysis was done by R package MaAsLin2[106]. We considered demographic features (including age, gender, BMI, ethnicity, and race), diet records, medications (antivirals, antifungals, antibiotics, and probiotics), and self-reported IBS scores as potential confounders. The analysis followed the model formula:

$$expr \sim \text{age} + \text{gender} + \text{bmi} + \text{ethnic} + \text{race} + \text{IBS} + \text{diet\_meat} + \text{diet\_sugar} + \text{diet\_veg}$$
$$+ \text{diet\_grains} + \text{diet\_fruit} + \text{antifungals} + \text{antibiotics} + \text{probiotics}$$
$$+ \text{antivirals} + (1|\text{sample\_id\_tp1})$$

where $expr$ refers to the 'omics matrix. For each feature in the 'omics data, we ran this generalized linear model to identify multivariable associations between each 'omics feature and each metadata feature. Identified confounders were handled differently based on the type of data. For species and KEGG genes, any feature with a significant statistical association with any metadata feature was removed from all subsequent analyses, resulting in the removal of 21 species and 946 microbial genes. For immune profiling and plasma metabolomics, to remove the effects of identified confounders, each feature was adjusted by retaining the residuals[157], i.e., the part of the outcome not explained by the confounding factors, from a general linear model:

$$y' = (y \sim \text{predicted confounders})\$residual$$

Additionally, for network and patient subset analysis (Methods), age, gender, BMI, and IBS were not included as confounders since we analyzed different age groups, gender groups, weight groups, and IBS groups separately. However, other identified confounders were still considered in the residual models.

**BioMapAI.** The primary goal of BioMapAI is to connect high-dimensional biology data, $X$ to mixed-type output matrix, $Y$. Unlike traditional ML or DL classifiers that typically predict a single outcome, $y$, BioMapAI is designed to learn multiple objects, $Y = [y_1, y_2, ..., y_n]$, simultaneously within a single model. This approach allows for the simultaneous prediction of diverse clinical outcomes - including binary, categorical, continuous variables - with 'omics profiles, thus address disease heterogeneity by tailoring each patient's specific symptomology.

**1. BioMapAI Structure.** BioMapAI is a fully connected deep neural network framework comprising an input layer $X$, a normalization layer, three sequential hidden layers, $Z^1, Z^2, Z^3$, and one output layer $Y$.

**1) Input layer ($X$)** takes high-dimensional 'omics data, such as gene expression, species abundance, metabolome matrix, or any customized matrix like immune profiling and blood labs. **2) Normalization Layer** standardizes the input features to have zero mean and unit variance, defined as

$$X' = \frac{X - \mu}{\sigma}$$

where $\mu$ is the mean and $\sigma$ is the standard deviation of the input features.

**3) Feature Learning Module** is the core of BioMapAI, responsible for extracting and learning important patterns from input data. Each fully connected layer (hidden layer 1-3) is designed to capture complex interactions between features. **Hidden Layer 1 ($Z^1$)** and **Hidden Layer 2 ($Z^2$)** contain 64 and 32 nodes, respectively, both with ReLU activation and a 50% dropout rate, defined as:

$$Z^k = \text{ReLU}(W^k Z^{k-1} + b^k), \qquad k \in \{1,2\}$$

**Hidden Layer 3 ($Z^3$)** has $n$ parallel sub-layers for each object, $y_i$ in $Y$. Every sub-layer, $Z_i^3$, contains 8 nodes, represented as:

$$Z_i^3 = \text{ReLU}(W_i^3 Z^3 + b_i^3), \qquad i \in \{1,2,\dots,n\}$$

All hidden layers used ReLU activation functions, defined as:

$$\text{ReLU}(x) = max(0, x)$$

**4) Outcome Prediction Module** is responsible for the final prediction of the objects. **The output layer ($Y$)** has $n$ nodes, each representing a different object:

$$y_i = \begin{cases} \sigma(W_i^4 Z_i^3 + b_i^4) & \text{for binary object} \\ \text{softmax}(W_i^4 Z_i^3 + b_i^4) & \text{for categorical object} \\ W_i^4 Z_i^3 + b_i^4 & \text{for continuous object} \end{cases}$$

The loss functions are dynamically assigned based on the type of each object:

$$\mathcal{L} = \begin{cases} \dfrac{1}{N}\sum_{i=1}^{N}[y_i \log(\hat{y}_i) + (1-y_i)\log(1-\hat{y}_i)] & \text{for binary object} \\ -\dfrac{1}{N}\sum_{i=1}^{N}\sum_{j=1}^{C} y_{ij}\log(\hat{y}_{ij}) & \text{for categorical object} \\ \dfrac{1}{N}\sum_{i=1}^{N}\begin{cases} 0.5(y_i - \hat{y}_i)^2, & \text{if } |y_i - \hat{y}_i| \leq \delta \\ \delta|y_i - \hat{y}_i| - 0.5\delta^2, & otherwise \end{cases} & \text{for continuous object} \end{cases}$$

During training, the weights are adjusted using the Adam optimizer. The learning rate was set to 0.01, and weights were initialized using the He normal initializer. L2 regularizations were applied to prevent overfitting.

**5) Optional Binary Classification Layer** (not used for parameter training). An additional binary classification layer is attached to the output layer $Y$ to evaluate the model's performance in binary classification tasks. This layer is not used for training BioMapAI but serves as an auxiliary component to assess the accuracy of predicting binary outcomes, for example, disease vs. control. This $\text{ScoreLayer}$ takes the predicted scores from the output layer and performs binary classification:

$$y_{binary} = \sigma(W_{binary} Y + b_{binary})$$

The initial weights of the 12 scores are derived from the original clinical data, and the weights are adjusted based on the accuracy of BioMapAI's predictions:

$$w_{\text{new}} = w_{\text{old}} - \eta \nabla \mathcal{L}_{MSE}$$

where $\nabla\mathcal{L}_{MSE}$ refers to the mean squared error (MSE) between the predicted $y'$ and true $y$, then adjusts the weights to optimize the accuracy of the binary classification.

## 2. Training and Evaluation of BioMapAI for ME/CFS – BioMapAI::DeepMECFS.

BioMapAI is a framework designed to connect high-dimensional, sparse biological 'omics matrix $X$ to multi-output $Y$. While BioMapAI is not tailored to a specific disease, it is versatile and applicable to a broad range of biomedical topics. In this study, we trained and validated BioMapAI using our ME/CFS datasets. The trained models are available on GitHub, nicknamed DeepMECFS, for the benefit of the ME/CFS research community.

**1) Dataset Pre-Processing Module: Handling Sample Imbalance.** To ensure uniform learning for each output $y$, it is crucial to address sample imbalance before fitting the framework. We recommend using customized sample imbalance handling methods, such as Synthetic Minority Over-sampling Technique (SMOTE)[160], Adaptive Synthetic (ADASYN)[161], or Random Under-Sampling (RUS)[162]. In our ME/CFS dataset, there is a significant imbalance, with the patient data being twice the size of the control data. To effectively manage this class imbalance, we employed RUS as a random sampling method for the majority class. Specifically, we randomly sampled the majority class 100 times. For each iteration $i$, a different random subset $S_i^{majority}$ was used. This subset $S_i^{majority}$ of the majority class was combined with the entire minority class $S^{minority}$. For each iteration $i$:

$$S_i^{\text{majority}} \subseteq S^{majortiy}, \qquad S^{minority} = S^{minority}$$

$$S_i = S_i^{majority} \cup S^{minority}$$

where the combined dataset $S_i$ was used for training at each iteration. This approach allows the model to generalize better and avoid biases towards the majority class, improving overall performance and robustness.

**2) Cross-Validation and Model Training.** DeepMECFS is the name of the trained BioMapAI model with ME/CFS datasets. We trained on five preprocessed 'omics datasets, including species abundances (Feature N=118, Sample N=474) and KEGG gene abundances (Feature N=3959, Sample N=474) from the microbiome, plasma metabolome (Feature N=730, Sample N=407), immune profiling (Feature N=311, Sample N=481), and blood measurements (Feature N=48, Sample N=495). Additionally, an integrated 'omics profile was created by merging the most predictive features from each 'omics model related to each clinical score (SHAP Methods), forming a comprehensive matrix of 154 features, comprising 50 immune features, 32 species, 30 KEGG genes, and 42 plasma metabolites.

To evaluate the performance of BioMapAI, we employed a robust 5-fold cross-validation. Training was conducted over 500 epochs with a batch size of 64 and a learning rate of 0.0005, optimized through grid search. The Adam optimizer was used to adjust the weights during training, chosen for its ability to handle sparse gradients on noisy data. The initial learning rate was set to 0.01, with beta1 set to 0.9, beta2 set to 0.999, and epsilon set to 1e-7 to ensure numerical stability. Dropout layers with a 50% dropout rate were used after each hidden layer to prevent overfitting, and L2 regularization ($\lambda = 0.008$) was applied to the kernel weights, defined as:

$$L_{reg} = \frac{\lambda}{2} \sum_{i=1}^{N} w_i^2$$

**3) Model Evaluation.** To evaluate the performance of the models, we employed several metrics tailored to both regression and classification tasks. The Mean Squared Error (MSE) was used to evaluate the performance of the reconstruction of each object. For each $y_i$, MSE was calculated as:

$$MSE_i = \frac{1}{N} \sum_{j=1}^{N} \left( y_i^j - \hat{y}_i^j \right)^2, i = 1, 2, \ldots, n$$

where $y_i^j$ is the actual values, $\hat{y}_i^j$ is the predicted values, and $N$ is the number of samples, $n$ is the number of objects. For binary classification tasks (ME/CFS vs control), we utilized multiple metrics including accuracy, precision, recall, and F1 score to enable a comprehensive evaluation of the model's performance.

To evaluate the performance of BioMapAI, we compared its binary classification performance with three traditional machine learning models and one deep neural network (DNN) model. The traditional machine learning models included: 1) Logistic Regression (**LR**) (C=0.5, saga solver with Elastic Net regularization); 2) Support Vector Machine (**SVM**) with an RBF kernel (C=2); and 3) Gradient Boosting Decision Trees (**GBDT**) (learning rate = 0.05, maximum depth = 5, estimators = 1000). **DNN** model employed the same hyperparameters as BioMapAI, except it did not include the parallel sub-layer, $Z_3$, thus it only performed binary classification instead of multi-output predictions. The comparison between BioMapAI and DNN aims to assess the specific contribution of the spread-out layer, designed for discerning object-specific patterns, in binary prediction. Evaluation metrics are detailed in Supplemental Table 3.

**4) External Validation with Independent Dataset.** To validate BioMapAI's robustness in binary classification, we utilized 4 external cohorts[114,115,116,117] comprising more than 100 samples. For these external cohorts, only binary classification is available. A detailed summary of data collection for these cohorts is provided in Supplemental Table 4. For each external cohort, we processed the raw data (if available) using our in-house pipeline. The features in the external datasets were aligned to match those used in BioMapAI by reindexing the datasets. The overlap between the features in the external dataset and BioMapAI's feature set was calculated to determine feature coverage. Any missing features were imputed with zeros to maintain consistency across datasets. The input data was then standardized as BioMapAI. We loaded the pre-trained BioMapAI, GBDT, and DNN for comparison. LR and SVM were excluded because they did not perform well during the in-cohort training process. The performance of the models was evaluated using the same binary classification evaluation metrics. Evaluation metrics detailed in Supplemental Table 4.

**3. BioMapAI Decode Module: SHAP.** BioMapAI is designed to be explainable, ensuring that it not only reconstructs and predicts accurately but also is interpretable, which is particularly crucial in the biological domain. To achieve this, we incorporated SHapley Additive exPlanations (SHAP) into our framework. SHAP offers a consistent measure of feature importance by quantifying the contribution of each input feature to the model's output.[163]

We applied SHAP to BioMapAI to interpret the results, following these three steps:

**1) Model Reconstruction.** BioMapAI's architecture includes two shared hidden layers - $Z^1$, $Z^2$- and one parallel sub-layers - $Z_i^3$- for each object $y_i$. To decode the feature contributions for each object $y_i$, we reconstructed sub-models from single comprehensive model:

$$Model_i = Z^1 + Z^2 + Z_i^3, i = 1,2,\ldots,n$$

where $n$ is the number of learned objects.

**2) SHAP Kernel Explainer.** For each reconstructed model, $Model_i$, we used the SHAP Kernel Explainer to compute the feature contributions. The explainer was initialized with the model's prediction function and the input data $X$:

$$explainer_i = shap.KernelExplainer(Model_i.predict, X), i = 1,2,...,n$$

Then SHAP values were computed to determine the contribution of each feature to $y_i$:

$$\phi_i = explainer_i(X), i = 1,2,...,n$$

The kernel explainer is a model-agnostic approach that approximates SHAP by evaluating the model with and without the feature of interest and then assigning weights to these evaluations to ensure fairness. For each $model_i$, with each feature $j$:

$$\phi_i^j(f, x) = \sum_{S_i \subseteq N_i \backslash \{j\}} \frac{|S_i|! (m - |S_i| - 1)!}{m!} \left( Model_i(S_i \cup j) - Model_i(S_i) \right)$$

$$= \frac{1}{m} \sum_{S_i \subseteq N_i \backslash \{j\}} \binom{m-1}{m - |S_i| - 1}^{-1} \left( Model_i(S_i \cup j) - Model_i(S_i) \right), i = 1,2,...,n$$

where $n$ is the number of learned objects, $m$ is the total number of features, $\phi_i^j$ is the Shapley value for feature $j$ in $model_i$, $N_i$ is the full set of features in $model_i$, $S_i$ is the subset of features not including feature $j$, $Model_i(S_i)$ is the model prediction for the subset $S_i$. The SHAP value matrix, $\phi_i$, were further reshaped to align with the input data dimensions.

**3) Feature Categorization.** Analyzing the SHAP value matrices, $[\phi_1, \phi_2, ..., \phi_n]$, features can be roughly assigned to two categories: shared features - important to all outputs; or specific features - specifically important to individual outputs. We set the cutoff at 75%, where features consistently identified as top contributors in 75% of the models were classified as shared important features, termed disease-specific biomarkers. Features that were top contributors in only a few models were classified as specific important features, termed symptom-specific biomarkers.

By reconstructing individual models, $Model_i$, for each object, $y_i$, and applying SHAP explainer individually, we effectively decoded the contributions of input features to BioMapAI's predictions. This method allowed us to categorize features into shared and specific categories—termed as disease-specific and symptom-specific biomarkers—providing novel interpretations of the 'omics feature contribution to clinical symptoms.

**4. Packages and Tools.** BioMapAI was constructed by Tensorflow(v2.12.0)[164] and Keras(v2.12.0). ML models were from scikit-learn(v 1.1.2)[165].

**WGCNA and Network Analysis.** To identify co-expressed patterns of each 'omics, we employed the Weighted Gene Co-expression Network Analysis (WGCNA) using the WGCNA[166] package in R. The analysis was performed on preprocessed omics data (Methods): species abundances (Feature N=373, Sample N=479) and KEGG gene abundances (Feature N=4462, Sample N=479) from the microbiome, plasma metabolome (Feature N=395, Sample N=414), immune profiling (Feature N=311, Sample N=489). Network construction and module detection involved choosing soft-thresholding powers tailored to each dataset: 6 for species, 7 for KEGG, 5 for immune, and 6 for metabolomic. The adjacency matrices were transformed into topological overlap matrices (TOM) to reduce noise and spurious associations. Hierarchical clustering was performed using the TOM, and modules were identified using the dynamic tree cut method with a minimum module size of 30 genes. Module eigengenes were calculated, and modules with highly similar eigengenes (correlation > 0.75) were merged. Module-trait relationships were assessed by correlating module eigengenes with clinical traits, and gene significance (GS) and module membership (MM) were used to identify hub genes within significant modules.

Network analysis was conducted using igraph[167] in R. Module eigengenes from the WGCNA analysis were extracted for each dataset. A combined network was constructed by calculating Spearman correlation coefficients (corrected, Methods) between the module eigengenes of different datasets, and an adjacency matrix was created based on a threshold of 0.3 (absolute value) to include only significant associations. Network nodes represented module eigengenes and edges represented significant correlations. Degree centrality and betweenness centrality were calculated to identify highly connected and influential nodes. Networks in patient subgroups were displayed as the correlation differences from their healthy counterparts to exclude the influence of covariates. For example, correlations in female patients were compared with female healthy, and correlations in older patients were compared with older healthy.

**Statistical Analysis.** The dimensionality reduction analysis was conducted by Principal Correspondence Analysis (PCoA) using sklearn.manifold.MDS function for 'omics. For combined 'omics data, PCoA was applied to combined module eigengenes from WGCNA. Fold change of species, genes, immune cells, and metabolites were compared between patient and control groups, short-term and control groups, and long-term and control groups. P values were computed by Wilcoxon signed-rank test with False Discovery Rate (FDR) correction, adjusted for multiple group comparisons. Spearman's rank correlation was used to assess correlation covariant. P-values were adjusted using Holm's method, accounting for multiple group comparisons. P value annotations: ns: $p > 0.05$, *: $0.01 < p <= 0.05$, **: $0.001 < p <= 0.01$, ***: $p <= 0.001$.

**Longitudinal Analysis**. To capture statistically meaningful temporal signals, we employed various statistical and modeling methods, accounting for both linear and non-linear trends and intra-individual correlations:

**1. Interquartile Range (IQR) and Intraclass Correlation Coefficient (ICC)**. We initially assessed statistics at different time points by computing the IQR and ICC. Data were standardized to a mean of zero and a standard deviation of one to ensure comparability across features with different scales. The IQR quantified variability, while the ICC assessed the dependence of repeated measurements[168], indicating the similarity of measurements over time. Data showed no statistical dependence and no trend of stable variance across time points.

**2. Generalized Linear Models (GLMs)**. GLMs[169] were then used to analyze the effects of time points, considering age, gender, and their interactions. Time points were included as predictors to reveal changes in dependent variables over time, with interaction terms exploring variations based on age and gender. Random effects accounted for intra-individual correlations. Although 12 features out of 5000 showed weak trends over time (slopes < 0.2), they were not deemed sufficient to be potential longitudinal biomarkers, possibly due to individualized patterns.

**3. Repeated Measures Correlation (rmcorr)**. To better consider individual effects, we employed rmcorr[170] to assess consistent patterns of association within individuals over time. This method captured stable within-individual associations across different time points. However, only 30 features out of 5000 showed weak slopes (< 0.3), and these were not considered sufficient to conclude the presence of longitudinal signals.

4. **Smoothing Spline ANOVA (SS-ANOVA)**. We then considered the longitudinal trends could be non-linear and more complex. To model complex, non-linear relationships between response variables and predictors over time, SS-ANOVA[171] was used. SS-ANOVA uncovered non-linear trends and interactions in the omics data, however, no strong temporal signals were identified.

In conclusion, robust analysis of the longitudinal data, accounting for both linear and non-linear trends and intra-individual correlations, revealed the difficulty in extracting strong and statistically meaningful temporal signals. As Myalgic Encephalomyelitis/Chronic Fatigue Syndrome (ME/CFS) is a disease that usually lasts for decades with non-linear progression, the four-year tracking period with annual measurements is likely insufficient for capturing consistent temporal signals, necessitating longer follow-up periods.

**Data and Code**

Metagenomics data is being deposited under the BioProject submission number SUB14546737 and will be publicly available as of the date of publication. Accession numbers are listed in the key resources table. BioMapAI framework is available at https://github.com/ohlab/BioMapAI/codes/AI. All original code, analyzed data and trained model has been deposited at https://github.com/ohlab/BioMapAI. Other 'omics data and any additional information required to reanalyze the data reported in this paper is available from the lead contact upon request.

**Supplemental Figure**

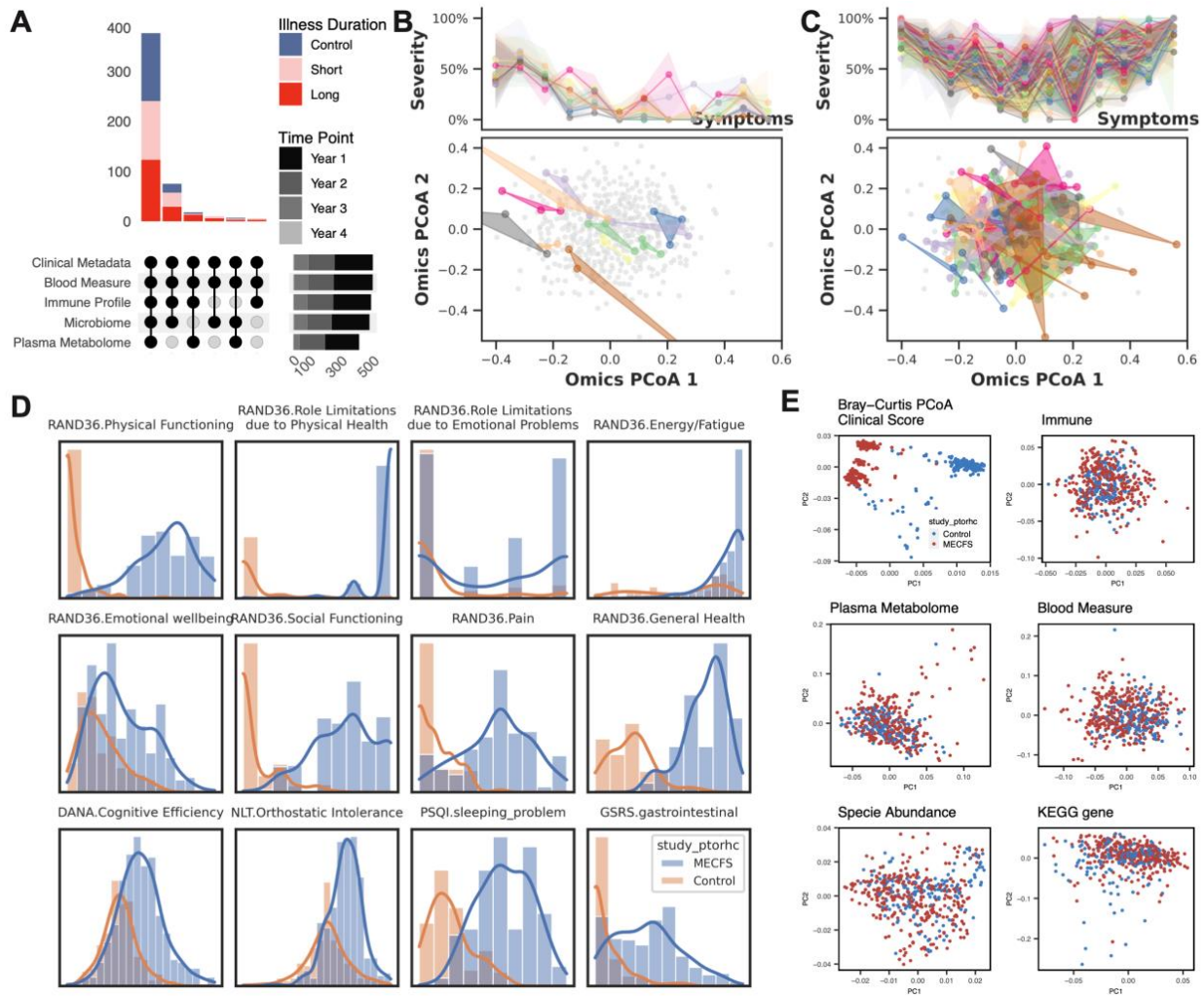

**Supplemental Figure 1: Data Pairedness Overview and Heterogeneity in Healthy and Patients.**

**A) Cohort Composition and Data Collection.** Over four years, 515 time points were collected: baseline year from all 249 donors (Healthy N=96, ME/CFS N=153); second year from 168 individuals (Healthy N=58, ME/CFS N=110); third year from 94 individuals (Healthy N=13, ME/CFS N=81); fourth year from N=4 ME/CFS patients. Nearly 400 collection points included complete sets of 5 'omics datasets, with others capturing 3-4 'omics profiles. Clinical metadata and blood measures were collected at all 515 points. Immune profiles from PBMCs were recorded at 489 points, microbiome data from stool samples at 479 points, and plasma metabolome data at 414 points. A total of 1,471 biosamples were collected. **B-C) Heterogeneity of B) Healthy Controls and C) All Patients in Symptom Severity and 'Omics Profiles.** Supplemental information for Figure 1B, which shows examples from 20 patients. Variability in symptom severity (top) and 'omics profiles (bottom) for all healthy controls and all patients with 3-4 time points. **D) Distribution of 12 Clinical Symptoms in ME/CFS and Control.** Density plots compare the distributions of 12 clinical scores between control (blue) and ME/CFS patients (orange) with the y-axis representing severity (scaled from 0% to 100%). Clinical scores include RAND36 subscales (e.g., Physical Functioning, Emotional Wellbeing), Cognitive Efficiency from the DANA test, Orthostatic Intolerance from the NLT test, Sleep Problems from the PSQI questionnaire, and Gastrointestinal Symptoms from the GSRS questionnaire. **E) Principal Coordinates Analysis (PCoA) of each 'Omics.** PCoA based on Bray-Curtis distance for clinical scores, immune profiles, plasma metabolome, blood measures, species abundance, and KEGG gene data. Control samples (blue) and ME/CFS patients (red) show distinct clustering. Here, except for the clinical scores, controls are indistinguishable from patients, highlighting the difficulty of building classification models. **Abbreviations:** ME/CFS, Myalgic Encephalomyelitis/Chronic Fatigue Syndrome; PCoA, Principal Coordinates Analysis; RAND36, 36-Item Short Form Health Survey; DANA, DANA Brain Vital; NLT, NASA Lean Test; PSQI, Pittsburgh Sleep Quality Index; GSRS, Gastrointestinal Symptom Rating Scale; KEGG, Kyoto Encyclopedia of Genes and Genomes. **Related to:** Figure 1-2.

**A**

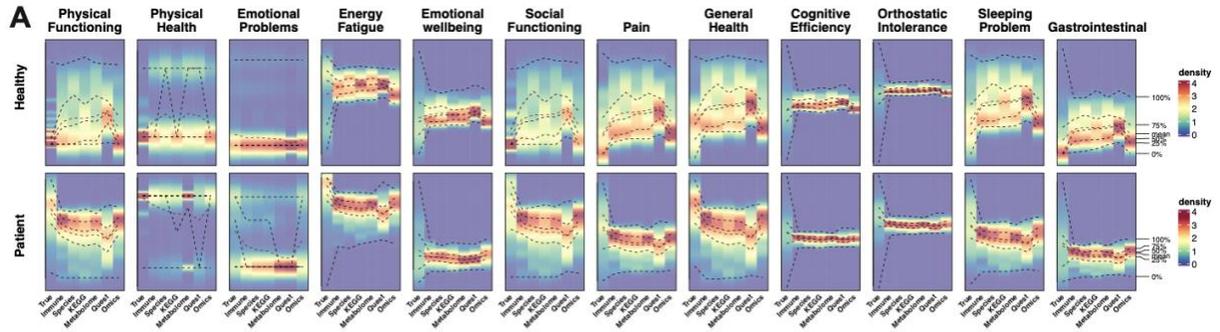

**B**

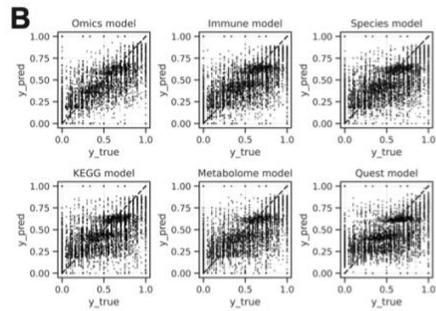

**C**

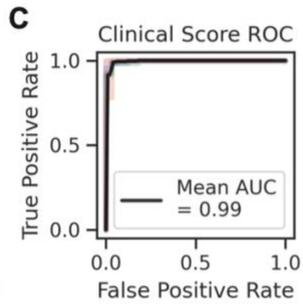

**D**

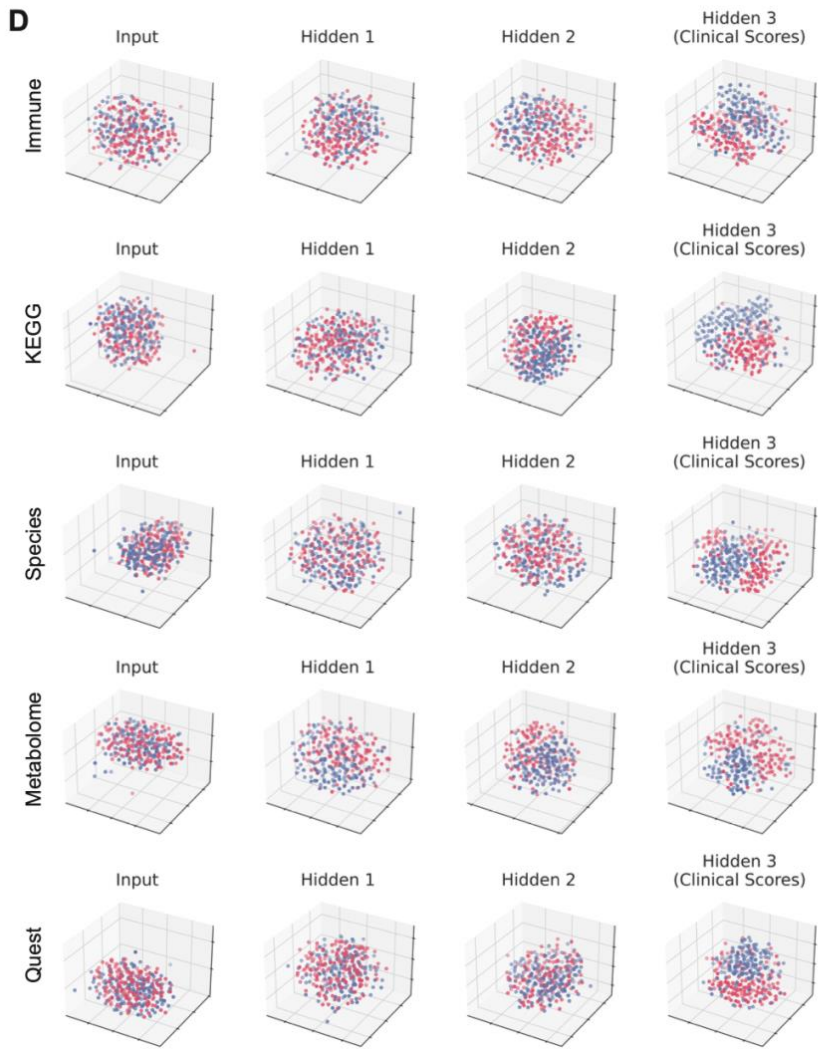

**Supplemental Figure 2: BioMapAI's Performance at Clinical Score Reconstruction and Disease Classification. A) Density map of True vs. Predicted Clinical Scores.** Supplemental information for Figure 2B, which shows three examples. Here, the full set of 12 clinical scores compares the true score, $y$ (Column 1), against BioMapAI's predictions generated from different 'omics profiles – $\hat{y}_{immune}, \hat{y}_{species}, \hat{y}_{KEGG},$ $\hat{y}_{metabolome}, \hat{y}_{quest}, \hat{y}_{omics}$ (Columns 2-7). **B) Scatter Plot of True vs. Predicted Clinical Scores.** Scatter plots display the relationship between true clinical scores (x-axis) and predicted clinical scores (y-axis) for six different models: Omics, Immune, Species, KEGG, Metabolome, and Quest Labs. Each plot demonstrates the clinical score prediction accuracy for each model. **C) ROC Curve for Disease Classification with Original Clinical Scores.** The Receiver Operating Characteristic (ROC) curve evaluates the performance of disease classification using the original 12 clinical scores. The mean Area Under the Curve (AUC) is 0.99, indicating high prediction accuracy, which aligns with the clinical diagnosis of ME/CFS based on key symptoms. **D) 3D t-SNE Visualization of Hidden Layers.** 3D t-SNE plots show how BioMapAI progressively distinguishes disease from control across hidden layers for five trained 'omics models: Immune, KEGG, Species, Metabolome, and Quest Labs. Each plot uses the first three principal components to show the spatial distribution of control samples (blue) and ME/CFS patients (red). The progression from the input layer (mixed groups) to Hidden Layer 3 (fully separated groups) illustrates how BioMapAI progressively learns to separate ME/CFS from healthy controls. **Abbreviations:** ROC, Receiver Operating Characteristic; AUC, Area Under the Curve; t-SNE, t-Distributed Stochastic Neighbor Embedding; PCs, Principal Components; $y$, True Score; $\hat{y}$, Predicted Score. **Related to:** Figure 2.

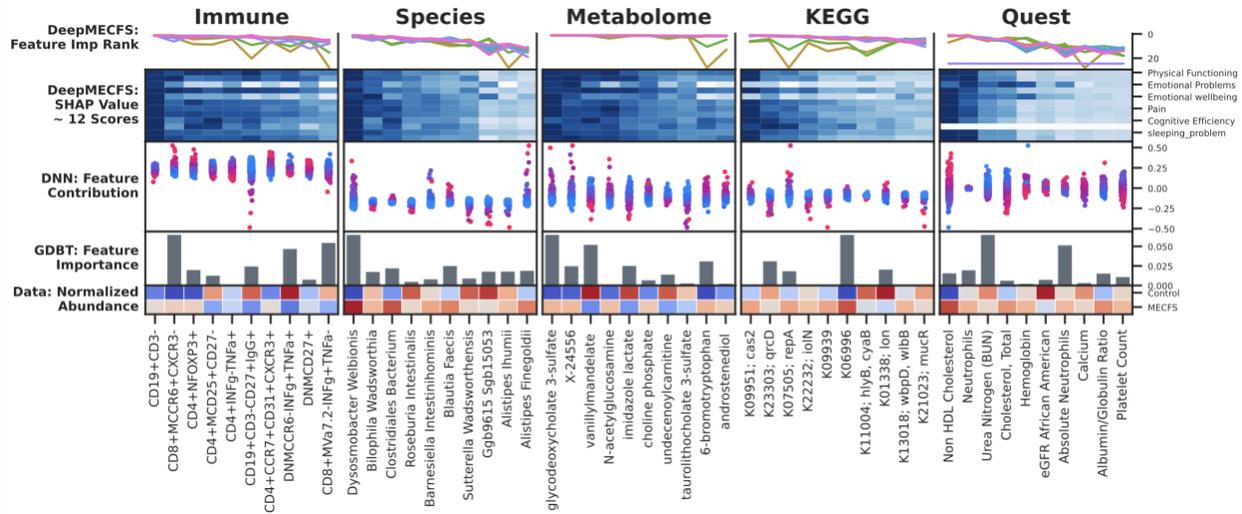

**Supplemental Figure 3: Disease-Specific Biomarkers - Top 10 Biomarkers Shared across Clinical Symptoms and Multiple Models.** Through the top 30 high-ranking features for each score, we discovered that the most critical features for all 12 symptoms were largely shared and consistently validated across ML and DL models, particularly the foremost 10. Here, this multi-panel figure presents the top 10 most significant features identified by BioMapAI across five 'omics profiles, highlighting their importance in predicting clinical symptoms and diagnostic outcomes across BioMapAI, DNN, and GBDT models, along with their data prevalence. Each vertical section represents one 'omics profile, with columns of biomarkers ordered by average feature importance from right to left. From top to bottom: *1. Feature Importance Ranking in BioMapAI.* Lines depict the rank of feature importance for each clinical score, color-coded by the 12 clinical scores. Consistency among the top 5 features suggests they are shared disease biomarkers crucial for all clinical symptoms; *2. Heatmap of SHAP Values from BioMapAI.* This heatmap shows averaged SHAP values with the 12 scores on the rows and the top 10 features in the columns. Darker colors indicate a stronger impact on the model's output; *3. Swarm Plot of SHAP Values from DNN.* This plot represents the distribution of feature contributions from DNN, which is structurally similar to BioMapAI but omits the third hidden layer ($Z^3$). SHAP values are plotted vertically, ranging from negative to positive, showing each feature's influence on prediction outcomes. Points represent individual samples, with color gradients denoting actual feature values. For instance, *Dysosmobacteria welbionis*, identified as the most

critical species, shows that greater species relative abundance correlates with a higher likelihood of disease prediction; *4. Bar Graphs of Feature Importance in GBDT*. GBDT is another machine learning model used for comparison. Each bar's height indicates a feature's significance within the GBDT model, providing another perspective on the predictive relevance of each biomarker; *5. Heatmap of Normalized Raw Abundance Data.* This heatmap compares biomarker prevalence between healthy and disease states, with colors representing z-scored abundance values, highlighting biomarker differences between groups. **Abbreviations:** DNN: Here refer to our deep Learning model without the hidden 3, 'spread out' layer; GBDT: Gradient Boosting Decision Tree; SHAP: SHapley Additive exPlanations. **Supporting Materials:** Supplemental Table 5. **Related to:** Figure 3.

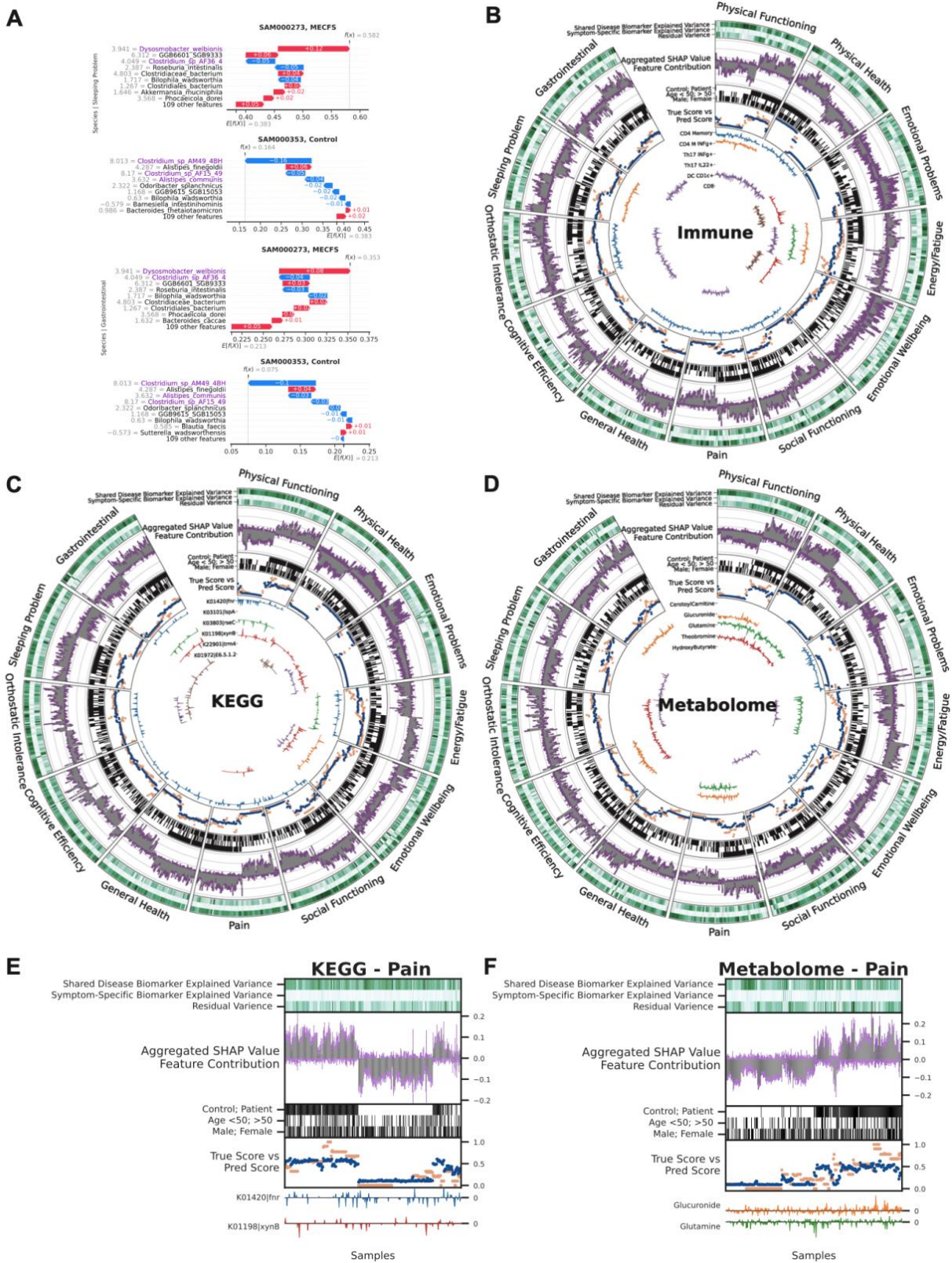

**Supplemental Figure 4: Symptom-Specific Biomarkers - Immune, KEGG and**

**Metabolome Models.** By linking 'omics profiles to clinical symptoms, BioMapAI identified unique symptom-specific biomarkers in addition to disease-specific biomarkers (Supplemental Figure 3). Each 'omics has a circularized diagram (Figure 3A, Supplemental Figure 4B-D) to display how BioMapAI use this 'omics profile to predict 12 clinical symptoms and to discuss the contribution of disease- and symptom-specific biomarkers. Detailed correlation between symptom-specific biomarkers and their corresponding symptoms is in Supplemental Figure 5. **A) Examples of Sleeping Problem-Specific Species' and Gastrointestinal-Specific Species' Contributions.** Supplemental information for Figure 3D, which shows the contribution of pain-specific species. **B-D) Circularized Diagram for Immune, KEGG and Metabolome Models.** Supplemental information for Figure 3A, which shows the species model. **E-F) Zoomed Segment for Pain in KEGG and Metabolome Model.** Supplemental information for Figure 3B, which shows the zoomed segment for pain in the species and immune models. **Abbreviations and Supporting Materials:** Supplemental Figure 5. **Related to:** Figure 3.

**Supplemental Figure 5: Symptom-Specific Biomarkers - Different Correlation Patterns of Biomarkers to Symptom**. Supplemental information for Figure 3C, which shows six pain biomarkers from multiple models. Here for each 'omics, we plotted the correlation of symptom-specific biomarkers (x-axis) to its related symptom (y-axis), colored by SHAP value (contribution to the symptom). **Abbreviations:** CD4, Cluster of Differentiation 4; CD8, Cluster of Differentiation 8; IFNg, Interferon Gamma; DC, Dendritic Cells; MAIT, Mucosal-Associated Invariant T; Th17, T helper 17 cells; CD4+ TCM, CD4+ Central Memory T cells; DC CD1c+ mBtp+, Dendritic Cells expressing CD1c+ and myelin basic protein; DC CD1c+ mHsp, Dendritic Cells expressing CD1c+ and heat shock protein; CD4+ TEM, CD4+ Effector Memory T cells; CD4+ Th17 rfx4+, CD4+ T helper 17 cells expressing RFX4; *F. prausnitzii, Faecalibacterium prausnitzii; A. communis, Akkermansia communis*; NAD, Nicotinamide Adenine Dinucleotide. **Related to:** Figure 3.

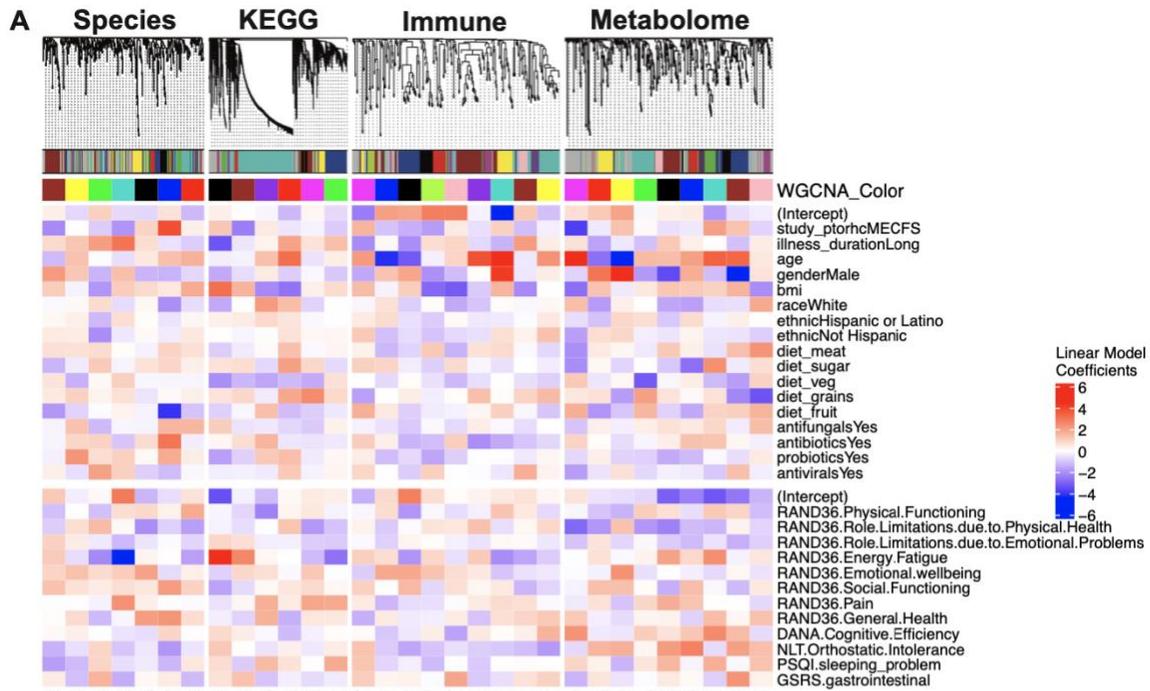

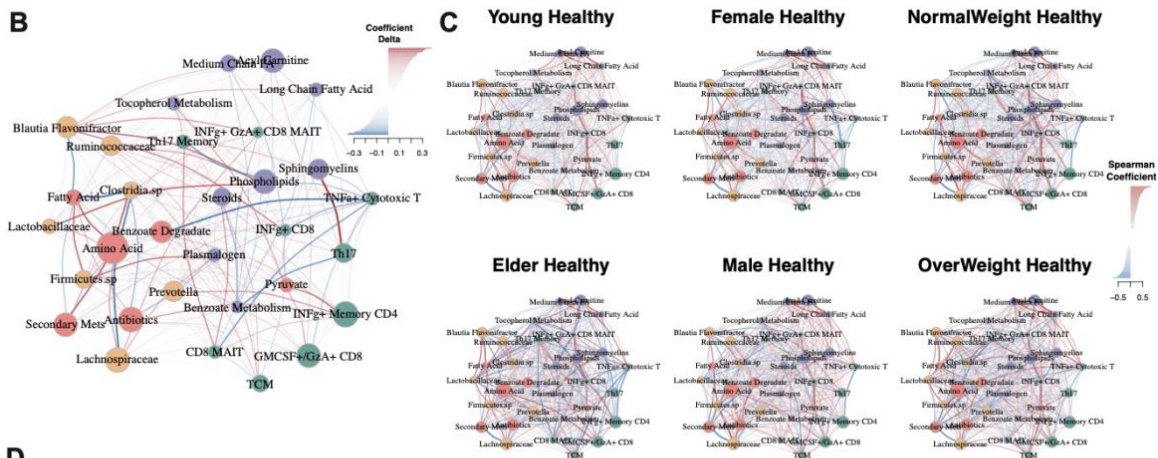

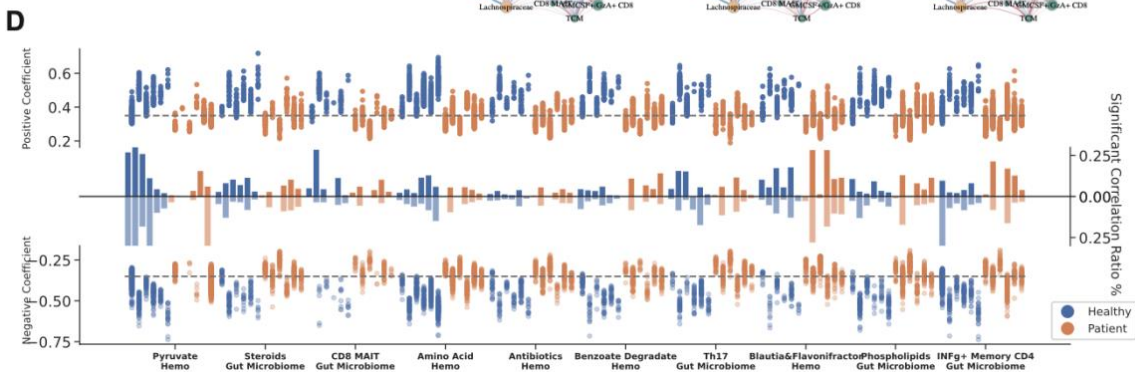

**Supplemental Figure 6: 'Omics WGCNA Modules and Host-Microbiome Network. A) Correlation of WGCNA Modules with Clinical Metadata.** Weighted Gene Co-expression Network Analysis (WGCNA) was used to identify co-expression modules for each 'omics layer: species, KEGG, immune, and metabolome. The top dendrograms show hierarchical clustering of 'omics features, with modules identified. The bottom heatmap shows the relationship of module eigengenes (colored as per dendrogram) with clinical metadata – including demographic information and environmental factors - and 12 clinical scores. General linear models were used to determine the primary clinical drivers for each module, with the color gradient representing the coefficients (red = positive, blue = negative). Microbial modules were influenced by disease presence and energy-fatigue levels, while metabolome and immune modules correlated with age and gender. **B-C) Microbiome-Immune-Metabolome Network in B) Patient and C) Healthy Subgroups.** Supplemental information for Figure 4A (Healthy Network) and 4B (Patient Subgroups). Figure 4A is the healthy network; here, Supplemental Figure 6B presented the shifted correlations in all patients. Figure 4B represented the network in patient subgroups; here, Supplemental Figure 6C is the corresponding healthy counterpart, for example, female patients were compared with female controls to exclude gender influences. **D) Differences in Host-Microbiome Correlations between Healthy and Patient Subgroups.** Selected host-microbiome module pairs are grouped on the x-axis (e.g., pyruvate to blood modules, steroids to gut microbiome). Significant positive and negative correlations (top and bottom y-axis) of module members pairs are shown as dots for each subgroup (blue = healthy, orange = patient) (Spearman, adjusted $p < 0.05$), from left to right: Young, Elder, Female, Male, NormalWeight, OverWeight Healthy and Young, Elder, Female, Male, NormalWeight, OverWeight Patient. The middle bars represent the total count of associations. This panel highlights the shifts in host-microbiome networks from health to disease, for example, in patients, the loss of pyruvate to host blood modules correlation and the increase of INFg+ CD4 memory correlation with gut microbiome. **Abbreviations:** WGCNA, Weighted Gene Co-expression Network Analysis; AA, Amino Acids; SCFA, Short-Chain Fatty Acids; IL, Interleukin; GM-CSF, Granulocyte-Macrophage Colony-Stimulating Factor. **Related to**: Figure 4.

**Supplemental Table**

**Supplemental Table 1** Sample Metadata and Clinical Scores

**Supplemental Table 2** Model Performance at Reconstructing Twelve Clinical Scores: Averaged Average Mean Squared Error by Model

**Supplemental Table 3** Model Performance in Diagnostic Comparison—Within-Cohort, Cross-Validated by Various ML and DL Models

**Supplemental Table 4** Model Performance in Diagnostic Comparison—Across Independent Cohorts

**Supplemental Table 5** Disease-Specific Biomarker: Averaged Feature Contribution of BioMapAI, DNN and GDBT

**Supplemental Table 6** Symptom-Specific Biomarker: Distinct Sets of Biomarkers for Each Symptom

**Supplemental Table 7** WGCNA Module Eigengene

**Supplemental Table 8** Targeted Pathways: Normalized Gene Read Counts and Their Correlation with Blood Responders

# Discussion

Democratization of AI technologies and large-scale multi-'omics has the promise of revolutionizing precision medicine[172,173,174,175]. This study generated among the richest, most extensive paired multi-'omics dataset to date[33,114,115,116,117,176,177,178], with new insights not only into ME/CFS, but potential other applications to heterogeneous and complicated diseases like fibromyalgia[179] and long COVID[180]. BioMapAI marks the first AI trained to systematically decode these complex, multi-system symptoms. Traditionally, diagnosing ME/CFS has been challenging, often relying heavily on self-reported questionnaires[181,182]. However, the crux for long-term post-viral infection syndromes like ME/CFS is not necessarily pinpointing an exact diagnosis or tracing disease origins[183,43] (typically infections[184]), but rather addressing the chronic, multifaceted symptoms that significantly impacts patients' quality of life[185,186]. Our study introduces a highly nuanced approach to link physiological changes in gut microbiome, plasma metabolome, and immune status, with host symptoms, moving beyond the initial causes of the disease[187,188]. Importantly, we validated key biomarkers in external cohorts[114,115,116,117], despite significant demographic and methodological differences between the studies.

In addition, by integrating these datatypes, we constructed complex new host-microbiome networks contrasted in health vs. ME/CFS. Networks constructed in healthy individuals revealed unique microbe-immune-metabolome connections and set a baseline for comparing numerous disease conditions while, critically, accounting for cohort covariates, including age, gender, and weight, as these factors reshape these networks by differing degrees, just as comorbid conditions like aging or obesity can complicate and individualize disease profiles. This approach enhanced the reliability of our findings in ME/CFS by rigorously accounting for potential confounders and solidified our proposed mechanisms exclusively to the disease itself[189,190]. For example, gut microbiome abnormalities were most relevant to ME/CFS, while changes in immune profiles and plasma metabolome were significant but influenced by factors like age and gender. Symptomatologically, the gut microbiome was expectedly linked to gastrointestinal issues and unexpectedly, to pain, fatigue, and mental health problems, possibly due to disruptions in the gut-brain axis from abnormal microbial metabolic functions, such as lost network connections with key plasma metabolites, particularly lipids. We previously noted immune abnormalities in ME/CFS[92]; in this study, we further analyzed activation of mucosal and inflammatory immunity, namely MAIT and γδ T cells, which linked to dysbiosis in gut microbial functions. These nuanced insights, while still premature for actual treatment applications, lay the groundwork for more precise controlled experiments and interventional studies. For instance, personalized treatment options could include supplementation of butyrate and amino acids for patients suffering from severe gastrointestinal and emotional symptoms, or targeted treatments for chronic inflammation for those experiencing significant pain and fatigue.

Taken together, our results underscore BioMapAI's particular suitability to complex datatypes that collectively, better explains the phenotypic heterogeneity of diseases such as ME/CFS than any one alone. BioMapAI's specialized deep neuron network structure with two shared general layers and one outcome-focused parallel layer is moreover generalizable and scalable to other cohort studies that aim to utilize 'omics data for a range of outputs (e.g., not just limited to clinical symptoms). For instance, researchers could employ our model to link whole genome sequencing data with blood or protein measurements. Constructed to automatically adapt to any input matrix $X$ and any output matrix $Y = [y_1, y_2, ..., y_m]$, BioMapAI defaults to parallelly align specific layers for each output, $y$. Currently, the model treated all 12 studied symptoms, $[y_1, y_2, ..., y_{12}]$, with equal importance due to the unclear symptom prioritization in ME/CFS[191]. We computed modules to assign different weights to symptoms to enhance diagnostic accuracy. While this approach was not particularly effective for ME/CFS, it may be more promising for diseases with more clearly defined symptom hierarchies[192,193]. In such cases, adjusting the weights of symptoms in the model's final layer could improve performance and help pinpoint which symptoms are truly critical.

Limitations of our study include that that our study population was comprised more females and older individuals, majorly Caucasian, though this is consistent with the epidemiology of ME/CFS[107,194,195], and was from a single geographic location (Bateman Horne Center). This may limit our findings to certain populations. In addition, previous RNA sequencing studies have suggested mitochondrial dysfunction and altered energy metabolism in ME/CFS[196,197,198,199,200]; thus, incorporating host PBMC RNA or ATAC sequencing in future research could provide deeper insights into regulatory changes. The typical decades-long disease progression of ME/CFS makes it challenging for our four-year longitudinal design to capture stable temporal signals - although separating our short-term (<4 years) and long-term (>10 years) provided valuable insights – ideally, tracking the same patients over a longer period would likely yield more accurate trends[201,202]. Long disease history also increases the likelihood of exposure to various diets and medications[203], which could influence biomarker identification, particularly in metabolomics. Finally, model-wise, BioMapAI was trained on < 500 samples with fivefold cross-validation, which is relatively small given the complexity of the outcome matrix; expanding the training dataset and incorporating more independent validation sets could potentially enhance its performance and generalizability[204,205].

# Acknowledgements

We are thankful to the Oh, Unutmaz, and Li laboratories for inspiring discussions and acknowledge the contribution of the Genome Technologies Service at The Jackson Laboratory for expert assistance with sample sequencing for the work described in this publication. We also thank the clinical support team at the Bateman Horne Center and all the individuals who participated in this study. This work was funded by 1U54NS105539.